%% file: toric.tex
\documentclass[paper, 12pt, letterpaper, epsf]{JHEP} 
\input{epsf.tex}
\usepackage{graphics} 

\def\Bbb{\bf} 

\def\C{{\Bbb C}} 
\def\R{{\Bbb R}}
\def\Z{{\Bbb Z}} 
\def\H{{\Bbb H}} 
\def\Q{{\Bbb Q}}
\def\P{{\Bbb P}}
\def\i{{\Bbb i}}\def\j{{\Bbb j}}\def\k{{\Bbb k}}
\def\I{{\Bbb I}}\def\J{{\Bbb J}}\def\K{{\Bbb K}}

\def\boldlambda{\mbox{\boldmath $\lambda$}}

 \font\mybb=msbm10 at 12pt
 \def\bb#1{\hbox{\mybb#1}}

\def\S {\bb{S}}
\def\T {\bb{T}} 

\font\gothics=ygoth at 10pt

\def \g{\hbox{\gothics g}}

\def\id{\protect{{1 \kern-.28em {\rm l}}}}

\newcommand{\be}{\begin{equation}} \newcommand{\ee}{\end{equation}}
\newcommand{\bea}{\begin{eqnarray}} \newcommand{\eea}{\end{eqnarray}}
\newcommand{\beann}{\begin{eqnarray*}} \newcommand{\eeann}{\end{eqnarray*}}
\newcommand{\bfig}{\begin{figure}} \newcommand{\efig}{\end{figure}}
\newcommand{\nn}{\nonumber}
\newcommand{\ba}{\begin{array}}\newcommand{\ea}{\end{array}}

\newtheorem{Proposition}{Proposition}[section]

\newtheorem{Theorem}{Theorem}[section]
\newtheorem{Lemma}{Lemma}[section]
\newtheorem{Corollary}{Corollary}[section]

\newcommand{\bp}{\begin{Proposition}} \newcommand{\ep}{\end{Proposition}} 
\newcommand{\bt}{\begin{Theorem}} \newcommand{\et}{\end{Theorem}} 
\newcommand{\bl}{\begin{Lemma}} \newcommand{\el}{\end{Lemma}} 
\newcommand{\bc}{\begin{Corollary}} \newcommand{\ec}{\end{Corollary}} 

\title{M-theory compactifications on certain `toric' 
cones of $G_2$ holonomy}

\author{L. Anguelova and  C. I. Lazaroiu 
\\C.~N.~Yang Institute for Theoretical Physics\\
SUNY at Stony Brook, NY11794-3840,
U.S.A.\\anguelov, calin @insti.physics.sunysb.edu}

\abstract{ We develop methods to study the singularities of certain 
$G_2$ cones related to toric hyperkahler spaces and Einstein selfdual 
orbifolds. This allows us to 
determine the low energy gauge groups of chiral $N=1$
compactifications of M-theory on a large family of such backgrounds, which 
includes the models recently studied by Acharya and Witten. 
All M-theory compactifications 
belonging to our family admit a $T^2$ of isometries, and therefore
T-dual IIA and IIB descriptions. We argue that 
reduction through such an isometry leads generically to 
systems of weakly and strongly coupled IIA 6-branes, T-dual to 
delocalized type IIB 5-branes. 
We find a simple criterion for the existence of 
a `good' isometry which leads to IIA models containing only weakly-coupled 
D6-branes, and construct examples of such backgrounds.
Some of the methods we develop may also 
apply to different situations, 
such as the study of certain singularities in the hypermultiplet moduli space
of $N=2$ supergravity in four dimensions.}

\preprint{YITP-SB-02-22}

\begin{document}

\tableofcontents

\pagebreak 

\vskip .6in

\section{Introduction} 

Compactifications of M-theory on spaces of $G_2$ holonomy have recently 
attracted renewed attention (see \cite{Achar, AW, Witten_Acharya, Witten_anom,
Gukov, Brandhuber, CGLP, Cvetic, CSU3, Gukov} for a very partial list of 
references). Perhaps the most important discovery from a 
physical perspective is that M-theory compactifications 
on certain singular $G_2$ spaces can lead to $N=1$ supersymmetric 
nonabelian gauge  theories in four dimensions with chiral matter
\cite{Witten_anom, Witten_Acharya}. This 
contrasts markedly with the smooth case, which upon compactification 
leads to Abelian vector multiplets and no net chirality 
\cite{PT}. The appearance of chiral matter 
in the singular case (more precisely, for $G_2$ spaces 
with singularities in codimension four) can be seen indirectly through 
an anomaly cancellation argument \cite{Witten_anom}. 
It can also be demonstrated in certain 
particular cases by reducing a local description of such singularities 
to a system of D6-branes in type IIA string theory \cite{Witten_Acharya}
(see also \cite{CSU1,CSU2}).

Since chiral nonabelian gauge theories are of obvious relevance to 
phenomenology, such compactifications may offer a way to study the strong 
coupling limit of various models whose physics has 
traditionally been accessible only at weak coupling. Unfortunately, progress 
in this direction is obstructed by our very poor understanding of 
$G_2$-holonomy spaces, which makes it very difficult to extract specific 
results. In fact, current knowledge provides extremely few constructions of 
$G_2$ spaces, both in the compact and non-compact set-up.  
Among these are compact examples of the type $(CY\times S^1)/\Z_2$ due 
to \cite{Joyce, Joyce_book} 
as well as two other constructions due to \cite{Joyce} 
and \cite{Kovalev}. 
In the non-compact case, our knowledge seems to be largely limited to 
certain $G_2$ lifts of the local geometry of the conifold 
\cite{Brandhuber, Cvetic, CGLP, Gukov}
(which are  known for the case of a `simple' conifold point, i.e. a 
point where a single rational curve or a 3-sphere is collapsed) as well 
as a construction of conical $G_2$ spaces due to 
Bryant and Salamon \cite{BS} and Gibbons, Page and Pope \cite{GP}.
The latter allows one to produce an infinite family of $G_2$ 
cones. The construction of \cite{BS,GP} starts with a four-dimensional, 
Einstein-self-dual space $M$ of positive scalar curvature 
and builds a $G_2$ metric on the real 
cone ${\cal C}(Y)$ over its twistor space $Y$. In the case considered in 
\cite{BS,GP} (namely when $M$ is smooth), 
this leads to only two conical $G_2$ metrics, since a smooth Einstein selfdual
four-manifold must coincide with either of $S^4$ or $\C\P^2$. $M$-theory 
physics on the $G_2$ cones obtained from these two choices was studied 
in \cite{AW}. 
However, the construction of \cite{BS,GP} generalizes to the case 
when $M$ is allowed to have orbifold singularities. As is by now well-known
\cite{GL,CP,AG, Abreu, BGMR, BG}, 
compact Einstein self-dual orbifolds of positive scalar curvature 
admit a rich geometry, and in particular
there exists an infinity 
of inequivalent examples of such spaces. A very small class 
among these is given by the celebrated 
models of \cite{GL}, which are orbifold 
equivalent with the weighted projective 
spaces $W\C\P^2_{p,q,r}$, but endowed with an Einstein self-dual metric  
(which differs from  the usual (toric) metric). These models are 
very special in many regards. For example, it is shown in 
\cite{AG} that the orbifold ESD metrics of \cite{GL} 
are Hermitian, and therefore  
conformal to certain Bochner-Kahler metrics \cite{Abreu, Bryant}.

In the paper \cite{Witten_Acharya}, the $G_2$ cone 
construction of \cite{BS,GP} 
was applied to the ESD orbifolds $M$ of \cite{GL} and used to
show that 
M-theory compactification on the associated  $G_2$ cone 
leads to chiral nonabelian gauge theories 
in four dimensions. The analysis of \cite{Witten_Acharya} relies on 
knowledge of the singularity structure of $M=W\C\P^2_{p,q,r}$, which is 
used in order to extract the singularities of $Y$ and thus of ${\cal C}(Y)$.
The location and type of such singularities is crucial for the physical 
analysis, since they allow one to identify the resulting low energy gauge 
group as well as the associated type IIA description.
For particular values of $p,q,r$, the authors of \cite{Witten_Acharya}  
show that performing the Kaluza-Klein reduction with respect to an 
appropriately chosen isometry of the $G_2$ metric leads to a system of three 
D6-branes in type IIA string theory; this gives an alternate 
explanation for the presence of chiral fermions. 

Since the choice of ESD orbifold $M$ used by \cite{Witten_Acharya} is very 
restrictive, this particular class of models does not allow for the 
construction of $G_2$ lifts of more generic D-brane systems. 
To extend it, one can use the construction of \cite{BS, GP} with a 
more general choice for the ESD orbifold $M$. The purpose of the present 
paper is to carry out part of this generalization, namely by considering 
ESD orbifolds $M$ which admit a two-torus of isometries.

While the most general ESD metric admitting two independent and 
commuting isometries is explicitly known due to the 
work of Calderbank and Pedersen\footnote{Particular metrics of this type 
were considered for example in \cite{Ivanov1} and \cite{Ivanov2}.} 
\cite{CP}, its expression is rather 
complicated and a direct analysis of singularities 
starting from the metric is quite involved. This makes it 
difficult to apply the simple methods of \cite{Witten_Acharya} to 
more general examples. Instead, we shall proceed in indirect fashion, 
by using a now well-known correspondence \cite{BGMR, BG, Swann, 
LeBrun, Salamon, Bergery}
between a quaternion-Kahler orbifold $M$, 
its twistor space $Y$ and its {\em hyperkahler cone} $X$\footnote{This 
can be completed by including a fourth geometry, namely
a 3-Sasaki space \cite{BG,BGMR} -- as we recall in Section 3.}. This 
allows us to determine the singularities of ${\cal C}(Y)$ by studying the 
singularities of the hyperkahler cone $X$ and the fixed points 
of a certain $U(1)$ action 
on $X$ whose Kahler quotient recovers the twistor space. 
To make the analysis
manageable, we shall require that the cone $X$  be 
{\em toric hyperkahler} \cite{BD}, i.e. a hyperkahler quotient 
(at zero moment map level) of some affine quaternion space by a torus action. 
This amounts to requiring the existence of two commuting, 
independent and `compact' isometries 
of $M$ and will allow us to use a slight adaptation of the 
results of \cite{BD} in order to identify
\footnote{It will turn out that a nice description can be obtained 
if one imposes a further technical condition, namely that $X$ be 
a `good' toric hyperkahler cone (see Sections 2 and 5).} 
the singularities of $X$; the singularities of $Y$ then follow from an 
analysis of the $U(1)$ reduction. In fact, we will be able to extract a 
simple algorithm for identifying such singularities, and 
therefore the low energy gauge group of $M$-theory on such $G_2$ 
holonomy backgrounds. 

The $G_2$ cones considered in this paper possess a two-torus of isometries,
induced by those of the Einstein self-dual space $M$. 
Upon choosing one of these,
one can perform the associated IIA reduction, thereby producing a certain
configuration of D6-branes. The remaining isometries of the $G_2$ cone 
descend to an $S^1$ of isometries of the IIA metric, which allows 
one to obtain a T-dual IIB description. The nature of the resulting IIA and 
IIB solutions depends on the choice of isometry used to reduce the 
$G_2$ metric. 

By characterizing such isometries in toric hyperkahler 
language, we argue that one will generically obtain 
a system of strongly coupled
6-branes in IIA and a T-dual system of strongly coupled, 
delocalized 5-branes in IIB. 
By varying the choice of the isometry used to perform
the reduction,  one is generally able to bring some of these branes 
to weak coupling. Whether a weakly-coupled description can be achieved 
for all D-branes is a characteristic of the model, 
and depends on the existence of a `good'
isometry, a property which can be tested through a simple criterion. 
In particular, we are able to construct models which {\em do} admit such 
an isometry, and therefore lead to IIA and IIB descriptions containing 
only weakly coupled branes. It will be 
clear from our description that models with this property are rather rare, 
and in particular that the generic models produced by our construction 
do not allow for an interpretation only in terms of  
weakly coupled branes. 
Among the models which admit a good isometry,
we shall find some whose type IIA description is given in terms of 
three D6-branes, carrying the same nonabelian gauge groups as those obtained 
from certain models considered in \cite{Witten_Acharya}. 
Since these are 
realized through different geometries, they must correspond to 
different choices of relative angles between the branes. 

The present paper is organized as follows. In Section 2, we give a short 
summary of our main results and illustrate them with one of the models
already analyzed in \cite{Witten_Acharya} with different methods. 
Section 3 recalls the construction of \cite{BS} and \cite{GP}, as
well as the relation between ESD spaces, twistor spaces, toric hyperkahler 
cones and 3-Sasaki spaces. It also discusses certain isometries of the 
$G_2$ cone of \cite{BS,GP}, which can be traced back to triholomorphic 
isometries of the hyperkahler cone of $M$. Section 4 introduces toric 
hyperkahler spaces following \cite{BD}, but under slightly more general 
assumptions. Section 5 discusses toric hyperkahler cones. After giving their 
description as an intersection of quadrics in a toric ambient space, we 
discuss a presentation of such cones as torus fibrations over a real affine 
space, which will play an important role in the remainder of the paper.
We also discuss the hyperkahler potential of such spaces, and give a criterion 
for identifying their singularities (this extends certain 
results of \cite{BD} to the case of hyperkahler cones). Much of the 
discussion of Section 5 is valid for $dim X=4n$ with arbitrary $n$, and we 
present it in this generality due to its relevance for other problems, such 
as the study of hypermultiplet moduli spaces in $N=2$ supergravity in four 
dimensions. Subsection 5.4.3. specializes our results to the eight-dimensional
case, which is relevant for the remainder of the paper. In Section 6, 
we apply the general construction reviewed in Section 3 to the case of toric 
hyperkahler cones. This allows us to describe the associated twistor space, 
quaternion-Kahler space and 3-Sasaki space in terms of certain 
explicit quotients. We also extract a description of the twistor 
space as an intersection of quadrics in a toric variety, and present a 
criterion for effectiveness of a certain 
$U(1)$ action on $X$ which leads to the 
twistor space upon performing the Kahler reduction. In Section 7, 
we use these 
preparations in order to extract the location and type of singularities of 
the twistor space $Y$, which immediately determine the singularities 
of the $G_2$ cones ${\cal C}(Y)$. We start by showing that all singularities 
of $Y$ must lie on a certain {\em distinguished locus}, which is naturally 
associated with a {\em characteristic polygon}. After presenting 
criteria for identifying the singularities of $Y$, we discuss the issue of 
good isometries. Finally, we mention an alternate, toric geometry approach to 
finding the singularities of $Y$.
In section 8, we apply the results of Section 7 to a few explicit models. 
This allows us to extract the associated low energy gauge groups and discuss 
the existence of good isometries. In particular, we present new 
models which admit such an isometry. Section 9 discusses M-theory reduction 
to type IIA on our backgrounds, as well as its T-dual, type IIB description. 
By using geometric arguments, we extract the interpretation of such models 
in terms of (generally strongly coupled) branes. Section 10 presents our 
conclusions. Some technical results are discussed in Appendices A-E.

\section{Summary of results}

As mentioned in the introduction, the singularities of the $G_2$ cone 
${\cal C}(Y)$ are immediately determined by the singularities of the twistor 
space 
$Y$. Recall that $Y$ is an $S^2$ fibration over a four-dimensional, 
Einstein selfdual space $M$. The twistor space 
is a compact, 3-dimensional complex variety, which 
turns out to have singularities both on points and along certain 
holomorphically embedded 
two-spheres. In algebraic geometry language, the later are 
compound du Val singularities \cite{Reid}, in our case families of 
$A_n$ singularities
depending on one complex parameter. 
In our models, there will be two types of spheres which 
may become singular:

(a) Vertical spheres, i.e. spheres which are fibers of $Y\rightarrow M$

(b)Horizontal spheres, which are lifts of spheres lying in  $M$.

As explained in the introduction, the twistor space  can be viewed
as a Kahler reduction $Y=X//_{\zeta}U(1)$ 
of a hyperkahler cone $X$ at some positive 
moment map level $\zeta$. 
Since the action of $U(1)$ on $X$ is uniquely determined,
$Y$ is completely specified by the choice of $X$.

The results of the present paper concern the particular case when $X$ is 
a {\em toric hyperkahler cone}, which means that it can be presented 
as a {\em hyperkahler} reduction (at zero level) 
of some affine quaternion space $\H^n$ through the 
Abelian group $T^{n-2}=U(1)^{n-2}$:
\be
X=\H^n///_{0}T^{n-2}~~.
\ee 
Such a quotient is completely specified by the choice of $U(1)^{n-2}$ action, 
which we characterize by the associated charge matrix $Q$. This is an 
$(n-2)\times n$ matrix whose entries $Q_{\alpha j}=q^{(\alpha)}_j$ 
appear in the 
transformation rules of the quaternion coordinates $u_1\dots u_n$ 
of $\H^n$:
\be
\label{uaction}
u_k\rightarrow \prod_{\alpha=1}^{n-2}{\lambda_\alpha^{q_k^{(\alpha)}}} u_k~~\
(k=1\dots n)~~,
\ee 
with $\lambda_\alpha$ some complex numbers of unit modulus. 
Note that (\ref{uaction})  maps $U(1)^{n-2}$  
into the torus $T^n=U(1)^n$ which acts 
diagonally on $u_k$:
\be
u_k\rightarrow \Lambda_k u_k~~(\Lambda_k\in U(1))~~.
\ee
We shall impose two technical conditions on the matrix $Q$:

(A) We require that $Q$ is `torsion-free', i.e. that the greatest 
common divisor of all of its $(n-2)\times (n-2)$ minor determinants 
equals one. Equivalently, the 
integral Smith form\footnote{Given an integral $r\times n$ matrix  
$F$ with $r\leq n$, 
one can find matrices $U\in GL(r,\Z)$ and $V\in GL(n,\Z)$ such that 
the matrix $F^{ismith}=U^{-1}FV$ has the {\em integral Smith form}
$F_{ismith}=[D,0]$, where $D=diag(t_1 \dots t_r)$, with $t_1 \dots t_r$ 
some non-negative integers satisfying the division relations 
$t_1|t_2|\dots |t_r$. These integers are called the {\em invariant factors}
(or `torsion coefficients')
of $F$, and their product $t_1\dots t_r$ coincides with the $r^{th}$ 
discriminantal divisor $\g (F)$ of $F$, which is the greatest common divisor 
of all $r\times r$ minors of $F$. It is clear that $\g(F)=1$ if and only 
if all $t_i$ equal one. A similar result holds for $r>n$.}
of $Q$ is $[I_{n-2},0]$, where $I_{n-2}$ is the $(n-2)\times (n-2)$ identity matrix
(in particular, $Q$ has maximal rank, so that its rows are linearly independent over 
the reals).
This condition assures that the action of $U(1)^{n-2}$ on $\H^n$ is 
effective.

(B) We also require that the matrix $Q$ is {\em good}, which means that 
none of its $(n-2)\times (n-2)$ minors can vanish.

If $X$ is such a hyperkahler cone, then one can describe the points of 
the twistor space as follows. Picking the first complex structure 
of $\H^n$, we decompose $u_k$ into complex coordinates:
\be
u_k=w_k^{(+)}+\j w_k^{(-)}~~.
\ee
Then a point of $Y$ is specified by $\{w_k^{(\pm)} \}$, taken 
to obey certain moment map constraints and considered 
modulo the action (\ref{uaction}):
\be
w_k^{(\pm)}\rightarrow 
\prod_{\alpha=1}^{n-2}{\lambda_\alpha^{\pm q_k^{(\alpha)}}} w^{(\pm)}_k~~
\ee
and the action of the `projectivising' 
$U(1)$ group by which one quotients to produce $Y$:
\be
\label{projaction}
w_k^{(\pm)}\rightarrow \lambda w_k^{(\pm)}~~.
\ee

Let us define a $2\times n$ integral matrix $G$ as the 
`transpose of the kernel 
of $Q$'. More precisely, the rows of $G$ are {\em integral} and 
{\em primitive}\footnote{We remind the reader that an 
integral vector is called primitive if the greatest common divisor 
of its components equals one.} 
vectors which form a basis for the real vector space $ker Q$.
The columns of $G$ are two-dimensional 
integral vectors $\nu_1\dots \nu_n$, which we shall call
{\em toric hyperkahler generators}. Note that we do {\em not } require that 
$\nu_j$ be primitive.
It will also be convenient to consider the $(n-1)\times (2n)$ matrix 
${\tilde Q}=\left[\begin{array}{ccc}Q&~&-Q\\1&\dots&~1\end{array}\right]$.

With the hypotheses (A),(B), we prove the following results:

(0a) A point $u$ of the hyperkahler cone $X$ can be singular only if
one of the quaternion coordinates $u_j$ vanishes.
If two quaternion coordinates vanish, then $u$ must coincide with 
the apex of $X$ (i.e. one has $u=0$). 
In particular, 
all singularities of $X$ must occur on one of the four-dimensional 
loci $X_j$ defined
by the equations $u_j=0$, and  two such loci intersect at a single point, namely the 
apex of $X$.

Suppose that $u$ is such that $u_j=0$
but $u\neq 0$. Then $u$ is  a singular point of $X$ if and only if the 
associated toric hyperkahler generator $\nu_j$ fails to be primitive. 
In this case, $X$ has a $\Z_{m_j}$ singularity at $u$, where $m_j$ is the 
greatest common divisor of the coordinates of $\nu_j$.

(0b) The trivially acting subgroup of the projectivising 
$U(1)$ 
is the trivial group or the $\Z_2$ subgroup $\{-1,1\}$.
The $\Z_2$ subgroup acts trivially on $X$ if and only if there exists a 
collection of rows of $Q$ whose sum is a vector all of whose entries 
are odd. Such a collection of rows is unique if it exists. 

(1) If $u$ is a singular point of the twistor space $Y$, then one of the 
following holds:

(1a) $u_j=0$ (i.e. $w_j^{(+)}=w_j^{(-)}=0$) for some $j=1\dots n$

or

(1b) There exists a choice of signs $\epsilon_j=\pm 1$ such that 
$w_1^{(-\epsilon_1)}=w_2^{(-\epsilon_2)}=\dots =w_n^{(-\epsilon_n)}=0$. 

Condition (1a) defines a locus $Y_j$ in $Y$, 
while (1b) defines a locus $Y_\epsilon$.
The union $Y_H$ of all $Y_\epsilon$ will be called {\em the horizontal locus}, 
while the union $Y_V$ of all $Y_j$ is the {\em vertical locus}.
The union $Y_D:=Y_H\cup Y_V$ will be called the {\em distinguished locus}.
We note that some horizontal components $Y_\epsilon$ may be void, and
that some of the  components $Y_j, Y_\epsilon$ may consist of smooth points 
of $Y$.

(2) Every component $Y_j$ is a vertical sphere of $Y$. 
A component $Y_\epsilon$ is either void
or a horizontal sphere (as defined on page 7).

(3) The planar polygon $\Delta\subset \R^2$ defined by:
\be
\Delta=\{a\in \R^2|\sum_{j=1}^n{|\nu_j\cdot a}|=\zeta\}
\ee
will be called the {\em characteristic polygon}. 
It has the following properties:

(3a) $\Delta$ is convex polygon on $2n$ vertices and is
symmetric with respect to the sign inversion $a\rightarrow -a$ of the plane.
Moreover, its principal diagonals $D_j$ (i.e. those diagonals of $\Delta$
which pass through the origin) lie on the lines $h_j$ defined 
by $\nu_j\cdot a=0$.

(3b) Every two-sphere $Y_j$ of the vertical locus can be written as 
an $S^1$ fibration over the principal diagonal $D_j$ 
of $\Delta$; the $S^1$ fiber
degenerates to a point above the opposite vertices of $\Delta$ connected by 
this diagonal. 

(3c) Every {\em non-void} component of the horizontal locus is associated 
with an edge of $\Delta$. If $e$ is such an edge, then the associated 
component of $Y_H$ is $Y_e:=Y_{\epsilon(e)}$, 
where $\epsilon(e)$ is the collection 
of signs defined by:
\be
\epsilon_k(e)=sign(\nu_k\cdot p_e)~~(k=1\dots n)~~,
\ee
with $p_e$ the vector from the origin to the middle point of the edge $e$.
Any component $Y_\epsilon$ of $Y_H$ for which $\epsilon$ cannot be written in 
this form must be void. In particular, $Y_H$ contains precisely $2n$ non-void 
components $Y_e$.

(3d) If $e$ is an edge of $\Delta$, then $Y_e$ is an $S^1$ fibration over 
$e$, whose fiber degenerates to a point at the two vertices of 
$\Delta$ connected by $e$.

(3e) Each vertex  $A$
of $\Delta$ corresponds to a common point $Y_A$ of two horizontal 
spheres $Y_e$ and $Y_{e'}$ and one vertical sphere $Y_j$, where $e, e'$ are 
the edges of $\Delta$ adjacent to the given vertex and $D_j$ is the principal 
diagonal passing through it. 

(3f) The antipodal map acting on the fibers of $Y\rightarrow M$ covers 
the sign inversion $\iota: a\rightarrow -a$ of $\Delta$ when restricted to the 
horizontal locus $Y_H$. This map takes each horizontal sphere 
$Y_e$ into the sphere $Y_{-e}$ associated with the opposite edge while 
preserving the vertical components $Y_j$.  

(3d) The ESD space $M$ is topologically 
a $T^2$ fibration over the compact planar convex  polytope 
bounded by the polygon $\Delta_M$ 
which results from $\Delta$ upon dividing through 
the sign inversion $\iota$. The $T^2$ fibers of $M\rightarrow \Delta_M$ 
degenerate to circles above its edges and to points above its vertices.

(4) Assume that the $\Z_2$ subgroup of the projectivising $U(1)$ acts 
non-trivially on $X$ (see criterion (0b)). Then 
the singularity type of $Y$ along $Y_j$ and $Y_e$ can be determined as 
follows:

(4a) Given a horizontal sphere $Y_e$, consider the integral vector: 
\be
\nu_e=\sum_{k=1}^n{\epsilon_k(e)\nu_k}~~.
\ee
(It can be shown that this vector cannot vanish).
Let $m_e$ be the greatest common 
divisor of the two components of $\nu_e$. 
Then $Y$ has a $\Gamma_e=\Z_{m_e}$ singularity at every point of $Y_e$, 
possibly with the exception of the two points 
lying above the vertices connected by the edge $e$. 
The generator of $\Z_{m_e}$ acts as follows 
on the complex 
coordinates $w^{(-\epsilon_k(e))}_k$ transverse to the locus $Y_e$:
\be
w_k^{(-\epsilon_k(e))}\rightarrow e^{\frac{2\pi i}{m_e}}~
w_k^{(-\epsilon_k(e))}~~.
\ee
The 
horizontal spheres $Y_e$ and $Y_{-e}$ associated with opposite edges 
have the same singularity type (since $\epsilon_k(-e)=-\epsilon_k(e)$).

(4b) Given a vertical sphere $Y_j$, consider the matrix ${\tilde Q}_j$ 
obtained by 
deleting the $j^{th}$ and $(j+n)^{th}$ columns of ${\tilde Q}$.
Then singularity group  $\Gamma_j$ 
of $Y$ along $Y_j$ coincides with $\Z_{m_j}$ or $\Z_{2m_j}$. 
To find which of these cases occurs, one computes the integral Smith 
form of the matrix ${\tilde Q}_j$, which is assured to be of the type:
\be
{\tilde Q}_j^{ismith}=[diag(1\dots 1, t_j),0]~~,
\ee
where $t_j=m_j$ or $t_j=2m_j$. The singularity group $\Gamma_j$ coincides with 
$\Z_{t_j}$. The generator of $\Z_{t_j}$ acts on the transverse 
coordinates as:
\be
\label{u_j_action}
u_j\rightarrow e^{\frac{2\pi i}{t_j}}u_j\Leftrightarrow 
w_j^{(\pm)}\rightarrow e^{\pm \frac{2\pi i}{t_j}}w_j^{(\pm)}~~.
\ee

(4c) Given a vertex $A$ of $\Delta$, let $e,e'$ and $D_j$ be the two 
edges and the principal diagonal passing through this vertex. 
Then we have $\epsilon_k(e)=\epsilon_k(e'):=\epsilon_k$ for all $k\neq j$ 
and $\epsilon_j(e)=-\epsilon_j(e')$.
Consider the $(n-1)\times (n-1)$ matrix:
\be
{\bar Q}_A=\left[\begin{array}{cccccc}
\epsilon_1col(Q,1)&\dots &
\epsilon_{j-1}col(Q,j-1)~, &\epsilon_{j+1}col(Q,j+1)&\dots&
\epsilon_n col(Q,n)\\
1&\dots &~~1 &1&\dots& 1
\end{array}
\right]~~.
\ee 
If its integral Smith form is 
${\bar Q}_A^{ismith}=diag(t_1\dots t_{n-1})$, then 
$Y$ has a singularity of type 
$\Gamma_A=\Z_{t_1}\times\dots \times \Z_{t_{n-1}}$ 
at the point $Y_A$ (the common point of $Y_e,Y_{e'}$ and $Y_j$)
\footnote{This group is expected to be cyclic in our models (and it is 
cyclic in the models we investigated). However, 
we shall not attempt to give an independent proof of this statement.}.
The transverse action of this group can be determined as explained in 
Appendix A. 

(5) If the $\Z_2$ subgroup of the projectivising $U(1)$ acts trivially on 
$X$, then the singularity 
group along $Y_j$ is $\Z_{m_j}$, with transverse action given 
by (\ref{u_j_action}) (with $t_j=m_j$). The singularity groups 
along the loci $Y_e$ and $Y_A$ are given by the quotients
of the groups $\Gamma_e, \Gamma_A$ 
determined at (4) through $\Z_2$ (which is assured to be a subgroup 
of $\Gamma_e$ and $\Gamma_A$). 
The action of $\Gamma_e/\Z_2$ and $\Gamma_A/\Z_2$ 
on the coordinates transverse to $Y_e$ and $Y_A$ is induced by the 
actions determined at (4) upon taking this quotient.

(6) The toric hyperkahler cone $X$ has a $T^2$ of isometries which 
preserve its hyperkahler structure. These descend to isometries of 
the ESD space $M$ and induce isometries of the $G_2$ cone ${\cal C}(Y)$.
Such isometries of the $G_2$ cone will be called {\em special}.

(6a) Given an edge $e$ of $\Delta$, the associated locus ${\cal C}(Y_e)$
in the $G_2$ cone is (point-wisely) fixed precisely by that 
$U(1)$ subgroup of the special isometry 
group whose Lie algebra equals $\R\nu_e$ (viewed as a subalgebra 
of the Lie algebra $\R^2$ of $T^2$).

(6b) The locus ${\cal C}(Y_j)$ (the cone over $Y_j$ in ${\cal C}(Y)$) 
is (point-wisely) fixed by that $U(1)$ of 
special isometries whose Lie algebra equals $\R \nu_j$. 

Let $E_{sing}$ be the set of edges $e$ of $\Delta$ for which $Y_e$ consists of 
singular points of $Y$ and $J_{sing}$ be the subset of indices $j\in 
\{1\dots n\}$ for which $Y_j$ consists of singular points. 
A {\em good 
isometry} of ${\cal C}(Y)$ is a {\em special isometry} whose fixed 
point set contains the union 
${\cal C}(Y_{sing}):=
\left[\cup_{e\in E_{sing}}{{\cal C}(Y_e)}\right]\cup  
\left[\cup_{j\in J_{sing}}{{\cal C}(Y_j)}\right]$. 
Good isometries form a (possibly trivial) 
Lie subgroup of the two-torus of special isometries.

(6c) The Lie algebra of the group of good isometries coincides with
the intersection 
$\left[\cap_{e\in E_{sing}}{\R\nu_e}\right]\cap \left[
\cap_{j\in J_{sing}}{\R\nu_j}\right]$, 
when viewed as a subalgebra of the Lie algebra $\R^2$ of $T^2$. In particular,
${\cal C}(Y)$ admits a good isometry if and only if this intersection does 
not vanish. In this case, the intersection is a one-dimensional 
space and ${\cal C}(Y)$ admits precisely
a $U(1)$ group of good isometries.

It is clear from this result that models admitting good isometries are 
non-generic in our class. If the model admits a good isometry, then we shall 
argue that its IIA reduction through this isometry 
can be described in terms of weakly-coupled
D6-branes. In this case, the T-dual IIB solution describes a system 
of delocalized D5-branes.  Reduction through isometries which are not good
leads to strongly coupled IIA and IIB descriptions. 

\subsection{Example: A model with one hyperkahler charge}

Consider the toric hyperkahler cone $X=\H^3///U(1)$, with the $U(1)$ action given by:
\be
Q=\left[\begin{array}{ccc} 1 &2 &2\end{array}\right]~~.
\ee
This is one of the models studied in \cite{Witten_Acharya}, whose 
ESD base is $M=W\C\P_{4,3,3}$ .
The matrix $Q$ has trivial invariant factors, 
since its integral Smith form is $\left[\begin{array}{ccc}1&0&0\end{array} \right]$. 
An integral basis for the kernel of $Q$ is given by the rows of the matrix:
\be
G=ker(Q)^t=\left[\begin{array}{ccc}2&0&-1\\0&1&-1\end{array}\right]~~
\ee
whose columns $\nu_1, \nu_2$ and $\nu_3$ are the toric hyperkahler generators.
Since the last two generators are primitive, the loci $X_2:u_2=0$ and 
$X_3:u_3=0$ associated with the flats 
$H_2,H_3$ are smooth in the hyperkahler cone $X$. 
For the non-primitive vector $\nu_1$ we have 
$m_1=gcd(\nu_1^1, \nu_1^2)=gcd(2,0)=2$, 
which gives $\Z_2$ singularities along the locus $X_1:u_1=0$. The singularity is worse at the apex
$u=0$. It is clear that the projectivising $U(1)$ acts effectively on 
$X$ (since the row vector $Q$ contains even entries).

The twistor space $Y=X//_{\zeta}U(1)$ can be realized as the 
quadric hypersurface: 
\be
w^{(+)}_1w^{(-)}_1+2w^{(+)}_2w^{(-)}_2+2w^{(+)}_3w^{(-)}_3=0
\ee
in the four-dimensional toric variety $\T=(\C^{6}-\{0\})/(\C^*)^2=
\C^{6}//T^2$ defined by the charge matrix:
\be
{\tilde Q}=
\left[\begin{array}{cccccc}1&2&2&-1&-2&-2
\\1&1&1&1&1&1\end{array}\right]~~,
\ee
whose integral Smith form equals:
\be
{\tilde Q}^{ismith}=\left[\begin{array}{cccccc}1&0&0&0&0&0
\\0&1&0&0&0&0\end{array}\right]~~.
\ee
Since ${\tilde Q}$ has trivial invariant factors, 
the associated $T^2$ action is effective on $\C^6$. 

\begin{figure}[hbtp]
\begin{center}
\mbox{\epsfxsize=7truecm \epsffile{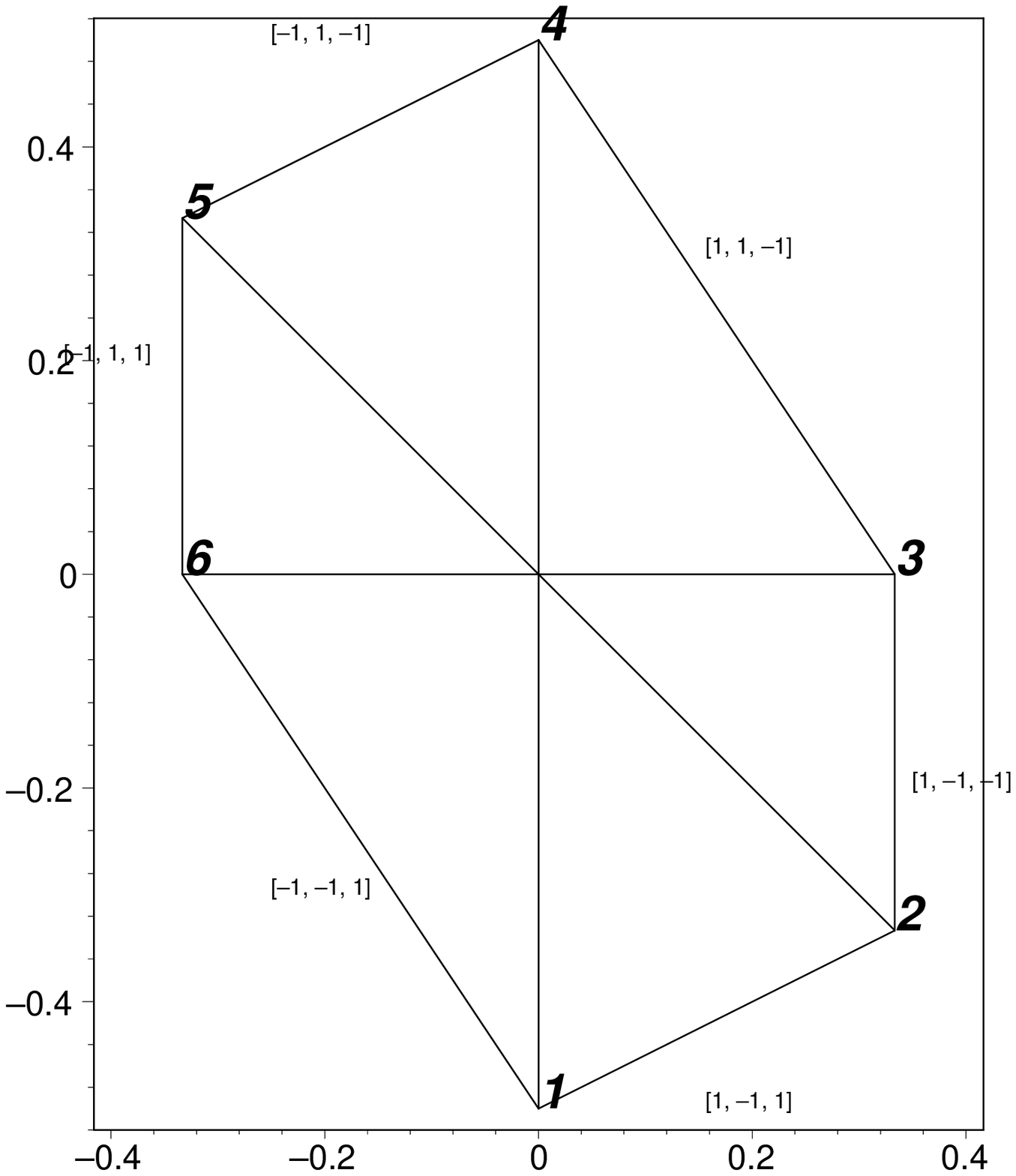}}
\end{center}
\caption{The polygon $\Delta$ for $\zeta=1$.\label{Delta_witten}}
\end{figure}

The distinguished locus is described by the planar polygon:
\be
\Delta:|\nu_1\cdot a|+|\nu_2\cdot a|+|\nu_3\cdot a|=\zeta~~,
\ee
where $a$ is a vector in $\R^2$. This is a hexagon whose vertices
(numbered $1\dots 6$ in figure \ref{Delta_witten}) 
correspond to the columns of the matrix:
\be 
\left[ \begin {array}{cccccc} 0&1/3&1/3&0&-1/3&-1/3
\\-1/2&-1/3&0&1/2&1/3&0\end {array} \right]~~.
\ee

Let us consider the sign vectors 
$\epsilon=(\epsilon_1,\epsilon_2,\epsilon_3)$ associated with the various 
edges as shown in the figure:
\bea
&&[12]:(+,-,+)~~,~~[23]:(+,-,-)~~,~~[34]:(+,+,-)~~\\
&&[45]:(-,+,-)~~,~~[56]:(-,+,+)~~,~~[61]:(-,-,+)~~.\nn
\eea
These give the vectors 
$\nu_e:=\sum_{j=1}^3{\epsilon_j(e)\nu_j}$:
\be
\nu_{12}=-\nu_{45}=\left[\begin{array}{c}~~1\\-2\end{array}\right]~~,~~
\nu_{23}=-\nu_{56}=\left[\begin{array}{c}3\\0\end{array}\right]~~,~~
\nu_{34}=-\nu_{61}=\left[\begin{array}{c}3\\2\end{array}\right]~~.
\ee
Since $\nu_{12}=-\nu_{45}$ and $\nu_{34}=-\nu_{61}$ 
are primitive, the associated edges of $\Delta$ correspond to smooth 
two-spheres in $Y$.
The vectors $\nu_{23}=-\nu_{56}$ 
give $\Z_3$ singularities along the horizontal 
spheres associated with the opposite edges $[23]$ and $[56]$; 
these are related by the antipodal map of the fibration $Y\rightarrow M$. 

The locus $Y_V$ consists of two-spheres (fibers of $Y\rightarrow M$) 
associated with the principal diagonals 
$D_1=[14]$, $D_2=[36]$ and $D_3=[25]$ of $\Delta$, which 
lie along the lines 
$h_1:\nu_1\cdot a=0$, $h_2:\nu_2\cdot a=0$ and $h_3:\nu_3\cdot a=0$. 
To identify the singularity type along these spheres, we compute the 
integral Smith forms of the matrices:
\bea
&&{\tilde Q}_1=\left[\begin{array}{cccccc}2&2&-2&-2
\\1&1&1&1\end{array}\right]\Rightarrow 
{\tilde Q}_1^{ismith}=\left[ \begin {array}{cccc} 1&0&0&0\\
0&4&0&0\end {array} \right]\\
&&{\tilde Q}_2={\tilde Q}_3=\left[\begin{array}{cccccc}1&2&-1&-2
\\1&1&1&1\end{array}\right]\Rightarrow 
{\tilde Q}_2^{ismith}={\tilde Q}_3^{ismith}=
\left[ \begin {array}{cccc} 1&0&0&0\\0&1&0&0\end {array} \right]~~.\nn
\eea
Thus the $S^2$ fibers associated with $D_2$ and $D_3$ are smooth in $Y$,
while the fiber associated with $D_1$ 
gives a sphere's worth of $\Z_4$ singularities. 

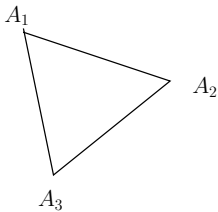
\begin{figure}[hbtp]
\begin{center}
\scalebox{0.5}{\input{deltaM0.pstex_t}}
\end{center}
\caption{ The polygon $\Delta_M$.\label{deltaM0}}
\end{figure}

\paragraph{Conclusion} The twistor space contains two $S^2$'s 
worth of $\Z_3$ singularities and one $S^2$ of $\Z_4$ singularities.
The $\Z_3$ spheres are horizontal (lifts of spheres lying in the selfdual base $M$), while the $\Z_4$ sphere is vertical (a fiber of 
$Y\rightarrow M$). The reduced polygon $\Delta_M$ is a triangle, which is covered by $\Delta$ through the sign inversion in $\R^2$
(figure \ref{deltaM0}). These conclusions clearly agree with those of 
\cite{Witten_Acharya}. While models of this type are well-understood 
from the work just cited, the methods of the present paper apply
more generally, and lead to many new examples, some of which are discussed 
in Section 8.  

\section{The basic set-up}

\subsection{Quaternion-Kahler spaces, twistor spaces and hyperkahler cones}

\begin{figure}[hbtp]
\begin{center}
\scalebox{0.9}{\input{geometries.pstex_t}}
\end{center}
\caption{Relation between the four geometries. \label{geometries}}
\end{figure}
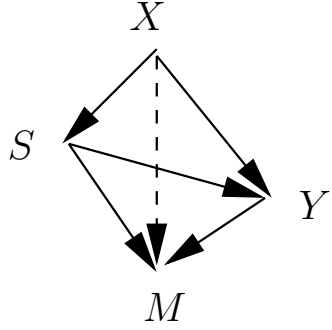

There exists a well-known relation \cite{BG} between 
four types of Riemannian geometries, namely quaternion-Kahler, 
twistor, 3-Sasaki and conical hyperkahler. This connects a 
$4d$-dimensional hyperkahler cone $X$ with a
$4d-4$-dimensional
quaternion-Kahler space $M$, a $4d-2$ -dimensional twistor space $Y$ and a 
$4d-1$ dimensional 3-Sasaki space $S$. The various relations are summarized
in figure \ref{geometries}, whose arrows are explained below\footnote{In the 
present paper we only consider `the positive case'. Thus $M$ will have 
positive scalar curvature and $X$ will carry a positive-definite metric.
The `negative case' is relevant for applications to 
hypermultiplet moduli spaces in supergravity, and 
part of our considerations can be extended to that situation.}.

We remind the reader that a {\em hyperkahler cone} is a hyperkahler space 
$X$ (of real dimension $4d$) which can be written as the metric cone over 
a compact, $4d-1$ dimensional Riemannian space $S$. 
Up to a possible finite cover $X\rightarrow {\tilde X}$, 
this amounts \cite{Swann} to considering a (complete) 
hyperkahler space ${\tilde X}$  
which admits a {\em hyperkahler potential}, 
i.e. a real-valued function ${\cal K}$ which is 
a Kahler potential with respect to {\em all} compatible complex structures. 
In this case, the 
vector field $\Xi$ defined through $d{\cal K}(.)=-g(\Xi,.)$ (where $g$ is 
the hyperkahler metric) induces 
a one-parameter semigroup of homotheties. 
One can choose the integration constant in 
${\cal K}$ such that the homothety action 
rescales it according to 
${\cal K}\rightarrow \alpha^2 {\cal K}$, where 
$\alpha>0$ is the homothety parameter. Accordingly, the hyperkahler potential
defines a {\em radial distance function} $r:={\cal K}^{1/2}$, 
which scales as $r\rightarrow \alpha r$ and identifies $\Xi$ with 
$r\frac{\partial}{\partial r}$. 
This presents (a cover\footnote{Taking a finite cover of $X$ may be
required in order to insure compactness of $S$.}  $X$ of ) ${\tilde X}$
as the metric cone 
over a $4d-1$-dimensional compact space $S$ obtained by restriction to a 
level $r=\zeta^{1/2}>0$.
Since its metric cone is hyperkahler, 
$S$ is a {\em 3-Sasaki} space \cite{BG}; 
in particular, the metric induced on $S$ 
by restriction is Einstein and of positive scalar curvature ---
the choice of $\zeta$ fixes its overall scale.
The arrow $X\rightarrow S$ in figure \ref{geometries} stands for this 
restriction.

The hyperkahler cone admits an isometric (but not triholomorphic) 
$Sp(1)$ action which preserves 
the hyperkahler potential and thus the distance function $r$; its generators
are the vector fields $I\Xi=I(r\partial_r)$, where $I$ are the 
compatible complex structures of $X$. Existence of the hyperkahler 
potential implies \cite{Swann} that this 
action rotates (i.e. acts transitively on) the compatible 
complex structures of $X$. Upon fixing some 
$I$, one obtains a group 
isomorphism $Sp(1)\approx SU(2)$ 
which is dependent of this choice; the totality of 
such isomorphisms is parameterized by a two-sphere. Then 
the diagonal $U(1)$ subgroup of 
$SU(2)$ acts on $X$ with generator $I\Xi$; 
this action preserves the Kahler form of $I$. 
The hyperkahler potential ${\cal K}=r^2$
coincides with the Kahler moment map of this $U(1)$ action.
In fact, the $U(1)$ may have a trivially acting $\Z_2$ subgroup (the diagonal 
subgroup $\{1,-1\}$ of $Sp(1)$), 
so the effectively acting group $U(1)_{eff}$ is either $U(1)$ or 
$U(1)/\Z_2$. The quotient 
$Y:=S/U(1)_{eff}={\cal K}^{-1}(\zeta)/U(1)_{eff}$ is 
the Kahler reduction of $X$ at level $\zeta$. 
The map $X\rightarrow Y$ in figure 
(\ref{geometries}) stands for this quotient, while the map $S\rightarrow Y$ 
stands for the associated $U(1)_{eff}$ quotient of $S$. Using the fact that 
$S$ is 3-Sasaki (or that the cone $X$ is hyperkahler), one 
shows \cite{BG} that $Y$ is a {\em twistor space} 
\cite{Salamon, Bergery, LeBrun, LeBrun_finite}, i.e. a projective complex 
Fano\footnote{We remind the reader that a complex variety $Y$ is {\em Fano} 
if $c_1(TY)$ is positive, i.e. the anticanonical line bundle 
$K_Y^{-1}$ is ample.} 
variety admitting a so-called {\em complex contact structure}
\footnote{ A {\em complex contact structure} \cite{Salamon, Bergery, LeBrun} 
on a complex variety 
$Y$ is a maximally non-integrable holomorphic Frobenius 
distribution on $Y$, i.e. 
a corank one holomorphic subbundle $D$ of the holomorphic tangent bundle
$TY$, with the property that the Frobenius obstruction map 
$D\times D\stackrel{[.,.]}{\longrightarrow}TY/D$ is nondegenerate everywhere.
In this case, the holomorphic line bundle $L:=TY/D$ is called the 
{\em contact line bundle} of $D$. If $Y$ has complex dimension $2d-1$, then 
it easy to see \cite{LeBrun} that $L^{d}$ is isomorphic with the 
anticanonical line bundle of $Y$, so that $L$ is a $d^{th}$ root of 
$K^{-1}_Y$.} \cite{Salamon, Bergery, LeBrun}
and a Kahler-Einstein metric of positive curvature. The Kahler-Einstein metric 
on $Y$ is induced from the hyperkahler metric of $X$ by the Kahler reduction.
Such a variety can always be written as the twistor space $tw(M)$ 
(in the sense of \cite{Salamon, Bergery}) of a 
quaternion-Kahler space $M$, which is uniquely determined by this property
\cite{LeBrun}. 
For this, one uses the complex contact structure in order to 
build a fibration of $Y$ through holomorphically embedded 2-spheres, 
whose normal bundle in $Y$ has the 
form $O_{\P^1}(1)^{\oplus d}$. The space $M$ is recovered as the base of this 
fibration, i.e. the image of the map $p$ which contracts the $\P^1$ fibers.
Then the contact distribution of $Y$ coincides with the horizontal 
distribution of this fibration, 
and the Kahler-Einstein metric of the twistor space induces a 
quaternion-Kahler metric on $M$ with respect to which $p$ becomes  
locally a Riemannian submersion. The arrow $Y\rightarrow M$ in figure 
(\ref{geometries}) describes this process of passing from $Y$ to $M$.

The arrow $S\rightarrow M$ can be described as follows. As mentioned above, 
the hyperkahler cone $X$ admits an isometric $Sp(1)$ action which preserves 
the distance function $r$. This action restricts to the 3-Sasaki space $S$. 
It turns out that the 
diagonal $\Z_2$ subgroup $\{1,-1\}$ of $Sp(1)$ may act trivially, 
which means that the effectively acting group 
$Sp(1)_{eff}$ is either $Sp(1)$ or 
$Sp(1)/\Z_2=SO(3)$. The quotient of $S$ through this action
coincides metrically and topologically with the quaternion-Kahler space $M$.
In particular, the $Sp(1)=S^3$ or 
$SO(3)$ fibration map $S\rightarrow M$ is locally a 
Riemannian  submersion. 

Since this procedure leads to the same space $M$, the full arrows in figure 
(\ref{geometries}) commute. The dashed vertical arrow $X\rightarrow M$ 
is defined 
as the composition of the two arrows on the left, which equals the composition 
of the two arrows on the right. This describes the so-called 
{\em conformal quotient} of \cite{martin,conf_quotient,martin_review}, 
which presents $M$ as ${\cal K}^{-1}(\zeta)/Sp(1)_{eff}=S/Sp(1)_{eff}$. 

The correspondences shown in figure \ref{geometries} admit certain 
inverses, which can be described as follows. 
Given the 3-Sasaki space $S$, 
one recovers $X$ as the metric cone over $S$. Given the twistor space $Y$, 
one obtains $X$ by a procedure due to LeBrun 
\cite{Salamon,LeBrun, Swann}, which can 
be described as follows. If $K_Y$ is the canonical line bundle of $Y$, then 
one considers the {\em contact line bundle} $L:=K_Y^{-1/d}$ (the precise 
root is determined by the contact structure of $Y$) and the space 
$(L^{-1})^\times=(K_Y^{1/d})^\times$ obtained from the total space of the 
dual line bundle $L^{-1}$ by removing its zero section. 
LeBrun shows that this 
carries a hyperkahler structure, which covers the well-known 
Ricci-flat Kahler structure on $K_Y^\times$ due to Calabi. The hyperkahler 
cone $X$ is obtained from $(L^{-1})^\times$ by completing this hyperkahler 
metric, which topologically has the effect of replacing the zero section 
of $L^{-1}$ with the apex of $X$ (thus identifying all points of this zero 
section to a single point). The 3-Sasaki space $S$ is recovered as the  
sphere bundle associated with $L^{-1}$, taken with respect to the metric
induced on $L^{-1}$ by the Kahler-Einstein metric of $Y$. If $||.||_{L^{-1}}$ 
is the fiberwise norm with respect to this metric, then the radial distance 
function of $X$ is $r=||.||_{L^{-1}}$, and $S$ is recovered by imposing the 
condition $r^2=\zeta$ in each fiber. The homothety action on $X$ is given 
by the standard fiberwise rescaling $u\rightarrow \alpha u$, 
while the $U(1)$ action on $X$ 
associated with the complex structure induced by $L^{-1}$ is 
given by the fiberwise action $u\rightarrow e^{2\pi i\phi}u$. 
The 3-Sasaki space and hyperkahler cone 
constructed in this manner carry an effective 
action of $SO(3)$ (so that $Sp(1)_{eff}=SO(3)$, $U(1)_{eff}=U(1)/\Z_2$ 
and $Y=S/U(1)_{eff}=S/(U(1)/\Z_2)$). Whether this lifts to an effective 
$Sp(1)$  action on some other 3-Sasaki space $S'$ or hyperkahler
cone $X'$ depends on whether the contact line bundle admits a square root $L'$.
In this case, one can obtain $S'$ and $X'$ by repeating the construction 
with $L$ replaced by $L'$. 

\begin{figure}[hbtp]
\begin{center}
\scalebox{0.9}{\input{double.pstex_t}}
\end{center}
\caption{Diagram of `inverse' geometries when $M$ has vanishing 
Marchiafava-Romani class.\label{double}}
\end{figure}
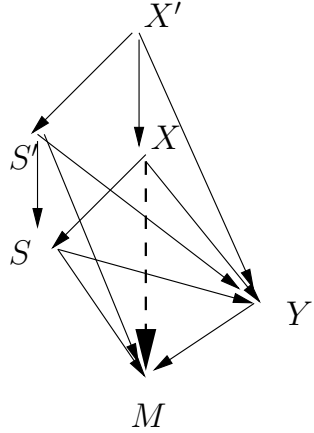

Given the quaternion-Kahler space $M$, its twistor 
space is recovered through the standard construction of Salamon and Bergery 
\cite{Salamon, Bergery}, while the 3-Sasaki space $S$ can be extracted as explained in 
\cite{Konishi,BG}. The latter construction recovers $S$ as the principal 
$SO(3)$ bundle associated with the subbundle ${\cal G}\subset End(TM)$
which specifies the quaternion-Kahler structure of $M$.
Finally, the hyperkahler cone $X$ can be recovered from $M$ through the 
construction of \cite{Swann}. This constructs a hyperkahler 
metric on a principal $\H^*/\Z_2$ bundle (the Swann bundle) over $M$, 
which can be completed to a metric on the hyperkahler cone $X$. The total 
space of the Swann bundle ${\cal U}(M)$ coincides with the space 
$(L^{-1})^\times$ in LeBrun's construction, and the hyperkahler metric 
on ${\cal U}(M)$ constructed in \cite{Swann} agrees with the hyperkahler 
metric on $(L^{-1})^\times$. Whether the $SO(3)$ bundle $S$ lifts to an 
$Sp(1)$ bundle (equivalently, whether Swann's bundle lifts to an $\H^*$-bundle)
depends on the topology of $M$ and is decided by vanishing of the so-called 
Marchiafava-Romani class \cite{MR, Salamon}. This is essentially the same 
obstruction as the existence of a square root of the contact line bundle 
$L$ mentioned above. If this obstruction vanishes, then one has {\em two}
3-Sasaki spaces $S,S'$ and {\em two} hyperkahler cones $X,X'$
associated with $M$, even though the twistor space $Y$ is 
uniquely determined. The two choices differ in whether the $\Z_2$ subgroup
of $Sp(1)$ acts trivially ($S,X$) or nontrivially ($S',X'$), 
with $S'$ being a double cover of $S$ and $X'$ a double cover of $X$
(figure \ref{double}).
This $\Z_2$ ambiguity will be important in Section 6.

\subsection{$G_2$ cones from quaternion-Kahler spaces}

\subsubsection{The construction} 

In the case $d=2$ (so that $M$ is a $4$-dimensional quaternion-Kahler
space), a construction due to \cite{BS} and \cite{GP} allows one to
obtain a metric of $G_2$ holonomy from the metric on a cone ${\cal C}(Y)$
built over the twistor space $Y=tw(M)$. The metric of \cite{BS,GP} is
obtained as follows. First, let us recall that in the four-dimensional
(positive) case, a four-manifold $M$ is quaternion-Kahler if
it admits a self-dual metric which is Einstein and of positive scalar
curvature. In this situation, the twistor space of $M$ can be
(topologically) identified with the sphere bundle associated with
the bundle $\Lambda^{2,-}(T^*M)$ of antiselfdual two-forms on $M$: \be
Y=S(\Lambda^{2,-}(T^*M))~~.  \ee Since the anti-selfduality condition
is invariant under conformal transformations of the metric on $M$, it
follows that $Y$ depends only on the conformal equivalence class of
this metric. This is a special property of twistor spaces associated
with quaternion-Kahler four-manifolds. If $d\sigma^2$ is the self-dual
Einstein metric on $M$, then the Kahler-Einstein metric on the twistor
space has the form: 
\be
\label{rho}
d\rho^2=|d\sigma|^2+|d_Au|^2~~, 
\ee
where $u=(u_1,u_2,u_3)$ is a local
frame of sections of $\Lambda^{2,-}(T^*M)$ with $\sum_{i=1}^3 u_i^2 = 1$ and $A$ is the connection
induced on this bundle by the Levi-Civita connection of $M$.

The construction of \cite{BS,GP} proceeds as follows. First, one
considers the following {\em modified} metric on $Y$: \be
\label{rho'}
d\rho'^2=\frac{1}{2}\left[d\sigma^2+\frac{1}{2}|d_Au|^2\right]~~.  \ee

The $G_2$ cone ${\cal C}(Y)$ of \cite{BS,GP} 
is simply the metric cone built over the Riemannian space
$(Y,d\rho'^2)$: 
\be
\label{G2metric}
ds^2=dr^2+r^2d\rho'^2=dr^2+\frac{r^2}{2}(d\sigma^2+
\frac{1}{2}|d_Au|^2)~~.
\ee The topological 
space obtained by removing the apex of ${\cal C}(Y)$
can be identified with $(\Lambda^{2,-}(T^*Y))^\times$, the space
obtained from the total space of $\Lambda^{2,-}(T^*Y)$ by removing its zero 
section. In this case, the coordinate $r$ is
identified with the radial coordinate inside each fiber.  Then ${\cal C}(Y)$
is obtained by completing the metric (\ref{G2metric}) on
$(\Lambda^{2,-}(T^*Y))^\times$.  The
metric (\ref{G2metric}) admits a one-parameter family of deformations
of the form: \be
\label{G2deformed}
ds^2=\frac{1}{1-(r_0/r)^4}dr^2+\frac{r^2}{2}(d\sigma^2+\frac{1}{2}(1-(r_0/r)^4)|d_Au|^2)~~,~~r_0\geq
0 \ee whose members have $G_2$ holonomy; the conical limit
is obtained for $r_0=0$.  The conical singularity is
smoothed out in the metrics (\ref{G2deformed}), though
other singularities are still present. These partially
resolved $G_2$ metrics are complete on the bundle $\Lambda^{2,-}(T^*M)$
(with its zero section included).

\subsubsection{Special isometries}

It is clear from (\ref{rho},\ref{rho'}) that any isometry of the
Einstein self-dual metric on $M$ will lift to an isometry of both the
Kahler-Einstein metric $\rho$ and the modified metric $\rho'$ on
$Y=tw(M)$.  Through the explicit constructions (\ref{G2metric}) and
(\ref{G2deformed}), such a symmetry of $M$ also lifts to an isometry
of the conical and deformed $G_2$ metrics. This observation will be
important in later sections, when we will study isometries of $M$
which are induced from its hyperkahler cone $X$; according to the
discussion above, such symmetries will automatically translate into
isometries of the associated $G_2$ metric, which can therefore be used
to extract the type II interpretation of our models.

\subsection{From hyperkahler cones to $G_2$ cones}

As recalled above, the study of quaternion-Kahler spaces can be
reduced to that of hyperkahler cones. In particular, if one starts
with an 8-dimensional hyperkahler cone, one has an associated Einstein
self-dual space, and therefore an associated $G_2$ cone. This allows
us to produce (and study) large families of $G_2$ cones by starting
with eight-dimensional hyperkahler cones. Since we are interested in
understanding M-theory physics on the former, we shall pay particular
attention to their singularities. It is clear from the explicit
construction of (\ref{G2metric}) that the singularities of the $G_2$
cone are completely determined from the singularities of the twistor
space of $M$, and that the nature of the former does not depend on
which of the metrics $\rho$ and $\rho'$ one uses on $Y$. This allows
us to determine the singularities of ${\cal C}(Y)$ by studying
the singularities of the twistor space $Y$, without worrying about the
metric change $\rho\rightarrow \rho'$ involved in the construction
(\ref{G2metric}). The singularities of $Y$ will be 
determined by studying its presentation
$Y=X//_{>0}U(1)$ as a Kahler quotient of the hyperkahler cone $X$. 
With arbitrary hyperkahler cones, the study of singularities 
is rather involved. To 
simplify the problem, we shall limit ourselves to a particular class
of cones $X$, namely those which are {\em toric
hyperkahler}. This will allow us to study the singularities of $Y$ in
a systematic manner, and therefore extract the singularities of the
associated $G_2$ cone, by applying methods reminiscent of those familiar
from toric geometry \cite{Oda, Fulton, Danilov, Audin, Cox, Cox_review}.

\section{Toric hyperkahler spaces}

Toric hyperkahler spaces form a hyperkahler analogue of the well-known 
toric varieties which play a central role in many 
subjects of algebraic geometry. To define them, we start with a brief 
description of torus actions on affine quaternion spaces.

\subsection{Torus actions on affine quaternion spaces}

Let $\H$ denote the field of quaternions, with quaternion units $1,\i,\j,\k$. 
The field $\C$ of complex numbers embeds into $\H$ upon 
identifying the complex units $1,i$ with the 
quaternion units $1, \i$. This is the only 
embedding of $\C$ into $\H$ which will be used in this paper.
Given a quaternion $u=u^0+\i u^1+\j u^2+\k u^3$ 
(with $u^0\dots u^3$ some real numbers), one defines its conjugate by 
${\overline u}=u^0-\i u^1-\j u^2-\k u^3$ and its norm by 
$||u||^2={\overline u}u=(u^0)^2+(u^1)^2+(u^2)^2+(u^3)^2$.

Consider the quaternion affine space $\H^n$, with quaternion coordinates $u_1\dots u_n\in \H$. 
This carries the standard flat metric:
\be
\label{metric}
ds^2=\sum_{k=1}^n{d{\overline u}_kdu_k}~~,
\ee 
which induces the norm $||u||^2=\sum_{k=1}^n{||u_k||^2}$. 
When endowed with this metric, $\H^n$ becomes a 
hyperkahler manifold, whose three compatible complex structures are induced 
by component-wise multiplication from the {\em right} 
with the quaternion units $\i,\j,\k$:
\be
\label{IJK}
\I (u)=(u_1 \i \dots u_n \i)~~,~~\J(u)=(u_1\j\dots u_n\j)~~,~~\K(u)=(u_1\k\dots u_n\k)
\ee

\subsubsection{The diagonal $T^n$ action}

The space $\H^n$ admits an obvious action by the $n$-dimensional torus $T^n=U(1)^n$:
\be
\label{Tn}
u=(u_1\dots u_n)\rightarrow u'=(\Lambda_1 u_1\dots \Lambda_n u_n)~~,
\ee
where $\Lambda_k$ are complex numbers of unit 
modulus. In (\ref{Tn}), we view $\Lambda_k$
as elements of $\H$, and use juxtaposition to denote quaternion multiplication. 
This action preserves the metric (\ref{metric}) and commutes with the 
three complex structures (\ref{IJK}). 
Therefore, it admits a hyperkahler moment map 
$\eta:\H^n\rightarrow ({\rm Im} \H)^n$ with components:
\be
\label{eta}
\eta^{(j)}(u)={\overline u}_j\i u_j \in 
{\rm Im} \H~~,~~j=1\dots n~~.
\ee
We remind the reader that the space ${\rm Im} \H$ of imaginary quaternions 
is the real vector subspace of $\H$ spanned by the imaginary quaternion 
units $\i,\j,\k$. This 3-dimensional space can be identified with 
$\R^3$ (i.e. ${\rm Im} \H = \langle ~\i,~\j,~\k~\rangle_\R\approx \R^3={\vec \R}$) through the map which sends 
$\i,\j,\k$ into the canonical basis $e_1,e_2,e_3$ of $\R^3$. 
Upon writing $\eta^{(j)}=\i\eta^{(j)}_1+\j \eta^{(j)}_2+\k \eta^{(j)}_3$, 
one obtains a map ${\vec \eta}^{(j)}=(\eta^{(j)}_1,\eta^{(j)}_2,\eta^{(j)}_3)
:\H^n\rightarrow \R^3$ and a map ${\vec \eta}:=({\vec \eta}^{(1)}\dots 
{\vec \eta}^{(n)}):\H^n\rightarrow (\R^3)^n=\R^{3n}$.
It will be convenient to use the following notation.
We let ${\vec \R}:=\R^3$, and 
${\vec \R}^n=(\R^3)^n\approx \R^{3n}$. The arrow superscript 
indicates that we use the particular presentation of $\R^{3n}$ as a direct 
product of $n$ copies of ${\vec \R}$, and 
amounts to thinking of $\R^{3n}$ as the tensor product $\R^3\otimes \R^n$.
We also let ${\vec \eta}_s=(\eta^{(1)}_s\dots \eta^{(n)}_s)\in \R^n$, 
for $s=1,2, 3$. Similar notations and conventions will be used for all 
hyperkahler moment maps appearing in this paper.

\subsubsection{Description of subtori of $T^n$ through maps of lattices}

Any {\em injective} integral linear map $q^*:\Z^r\rightarrow \Z^n$ induces a map of tori $T^r\rightarrow T^n$, 
by tensoring from the right with the Abelian group $U(1)$ (indeed, one has $T^m=\Z^m\otimes_{\Z}U(1)$ for all $m$). 
When combined with (\ref{Tn}), this gives a $T^r$ action on $\H^n$.
If the injective map $q^*$ is otherwise arbitrary, then 
the induced map of tori need not be injective (in technical language, tensoring with $U(1)$
is not a left exact operation). In this case, $T^r$ can not be regarded as a sub-torus of $T^n$, and the induced $T^r$ action on 
$\H^n$ will fail to be effective. To understand 
when this occurs, let us consider the short exact sequence:
\be
\label{QAsequence}
0\longrightarrow \Z^r\stackrel{q^*}{\longrightarrow}\Z^n\stackrel{g}{\longrightarrow} A\longrightarrow 0~~
\ee
obtained by computing the cokernel of $q^*$ (thus $A=\Z^n/q^*(\Z^r)$). As explained in Appendix A, this 
induces an exact sequence of the form:
\be
\label{GAsequence}
0\longrightarrow \Gamma\longrightarrow T^r\longrightarrow T^n\longrightarrow A\otimes_\Z U(1)\longrightarrow 0~~,
\ee
where the group $\Gamma$ coincides with the torsion subgroup of $A$. Therefore, the map $T^r\rightarrow T^n$ will be an 
embedding if and only if $A$ contains no torsion; in this case, $A$ coincides with the lattice $\Z^d$, where 
$d=n-r$. Moreover, this happens (See Appendix A) 
if and only if the charge matrix 
$Q$ associated with the transpose map 
$q:\Z^n\rightarrow \Z^r$ has trivial `torsion coefficients', which means 
that its integral Smith form equals:
\be
Q^{ismith}=\left[\begin{array}{cc}I_r &0_{r\times d}\end{array}\right]~~,
\ee
where $I_r$ denotes the $r\times r$ identity matrix. Equivalently, the 
$r^{th}$ {\em discriminantal divisor}\footnote{The $r^{th}$ discriminantal 
divisor $\g$ 
of $Q$ is defined as the greatest common divisor of all of its
$r\times r$ minor determinants. One has $t_1\dots t_r=\g$ (Appendix A).} 
of $Q$ equals one.

Throughout this paper, we shall assume that this condition is satisfied. Then (\ref{QAsequence}) reduces to a short exact sequence of 
lattices:
\be
\label{quat_sequence}
0\longrightarrow \Z^r\stackrel{q^*}{\longrightarrow}\Z^n\stackrel{g}{\longrightarrow} \Z^d\longrightarrow 0~~,
\ee
and (\ref{GAsequence}) collapses to a short exact sequence of tori:
\be
0\longrightarrow T^r\longrightarrow T^n\longrightarrow T^d\longrightarrow 0~~.
\ee
\noindent 
The $d\times n$ matrix $G$ of the cokernel map 
$g$ will be called `the matrix of generators'. Its columns $\nu_j$ are integral 
vectors belonging to the lattice $\Z^d$ and 
will be called `toric hyperkahler generators'. We note that the {\em rows} of $G$ are 
primitive and form a basis for the kernel of the matrix $Q$.

\subsection{Subtorus actions}

With these assumptions, the induced $T^r$ action on $\H^n$:
\be
\label{quat_action}
u_j\rightarrow u'_j=\prod_{\alpha=1}^r{\lambda_\alpha^{q_j^{(\alpha)}}}u_j~~,
\ee
is effective. In this relation, $q_j^{(\alpha)}$ stands for the entry $Q_{\alpha j}$ 
of the matrix $Q$.
The map $q^*$ describes the embedding:
\be
\Lambda_j=\prod_{\alpha=1}^r{\lambda_\alpha^{q_j^{(\alpha)}}}
\ee
of $T^r$ into $T^n$.

It is obvious that the $T^r$ action (\ref{quat_action}) is tri-holomorphic on $\H^n$ (since so is the $T^n$ action (\ref{Tn})).
Accordingly, it has a hyperkahler moment map $\mu:
\H^n\rightarrow {\vec \R}^r$ whose components are given by:
\be
\label{mu}
\mu^{(\alpha)}:=\sum_{j=1}^n{q^{(\alpha)}_j{\overline u}_j\i u_j}\in 
{\rm Im} \H~~.
\ee
As in Subsection 4.1.1, we let ${\vec \mu}:\H^n\rightarrow {\vec \R}^r$
be the map resulting from the identification ${\rm Im} \H =\R^3={\vec \R}$.
If we let ${\vec q}$ be the map from 
$\R^n$ to ${\vec \R}^r$ induced by $q$ upon tensoring with 
${\vec \R}=\R^3$, then we have ${\vec \mu}={\vec q}\circ {\vec \eta}$.

\subsection{Toric hyperkahler spaces}

With the notations of the previous subsection, a {\em toric hyperkahler space} is defined \cite{BD} as the hyperkahler reduction 
of $\H^n$ through a subtorus $T^r$ of $T^n$ at some level ${\vec \xi}\in {\vec \R}^r$ of the associated 
hyperkahler moment map:
\be
\label{Xdef}
{\bf X}=\H^n///_{\vec \xi} T^r={\vec \mu}^{-1}({\vec \xi})/T^r~~.
\ee
This description immediately shows that ${\bf X}$ is endowed with a hyperkahler structure induced via the reduction process
\cite{HKLR}. In general, the space ${\bf X}$ will be singular. 

In spite of the formal similarity of their definition, the geometry of toric hyperkahler spaces differs qualitatively from that 
of toric varieties (which result upon performing {\em Kahler} torus quotients of some {\em complex} affine space $\C^n$). 
For example, it can be shown \cite{BD} that the topology of $X$ is independent of the choice of hyperkahler moment map levels
${\vec \xi}$, as long as the latter lie in the complement of a codimension {\em three} subset of $\R^{3r}$ 
(i.e. the resulting spaces for different values of ${\vec \xi}$ are homeomorphic). This is in marked contrast with the behavior
of toric varieties, for which topology changing transitions occur on walls (i.e. subspaces of codimension {\em one}) 
in the space of Kahler moment map levels.

\paragraph{Observation 1}

In this paper, we do {\em not} assume that the toric hyperkahler generators $\nu_j$ are  
primitive. In fact, these vectors fail to be primitive even for the simplest examples one wishes to consider, 
namely the models based on the construction of \cite{GL}, 
which were studied in \cite{Witten_Acharya} by methods different from ours. 
This means that our hyperkahler spaces 
are {\em more general} than those studied in the work of \cite{BD}, which assumes
primitivity of generators for the most part. 
Due to this, we will have to modify and adapt some basic results of 
\cite{BD}. Some of the facts we shall use
require a direct proof, which can be found in the appendices.

\paragraph{Observation 2} We warn the reader that the term 
`toric hyperkahler spaces' is used somewhat ambiguously in the literature. 
In \cite{GGPT, G,GR}, this language is used to describe
quotients of $\H^n$ by Abelian groups which are {\em not} subtori of $T^n$; 
the resulting spaces are `toric' inasmuch as they admit a triholomorphic 
torus action, but they do not satisfy our definition. The prototypical 
example of this type is the Euclidean Taub-Nut space, which can be obtained 
upon dividing $\H^2$ through the non-compact group $(\R,+)$, 
acting through transrotations (see \cite{GR}).
In particular, this construction allows the authors of \cite{GGPT,GR} 
to give a hyperkahler quotient description of 
the generalizations of Taub-Nut metrics 
studied in \cite{LWY}. We stress that the spaces 
considered in \cite{GGPT, G, GR, LWY} are {\em not} toric 
hyperkahler according to our definition (which agrees with that of \cite{BD}).

\section{Toric hyperkahler cones}

Since we are interested in hyperkahler cones, we shall henceforth concentrate on the case of {\em vanishing} moment map level 
${\vec \xi}=0$ 
in definition (\ref{Xdef}). Then $X$ admits a one-parameter semigroup of 
homotheties induced by the obvious rescaling of coordinates of $\H^n$:
\be
\label{homothety}
u_j\rightarrow u_j\alpha~~,~~\alpha>0~~.
\ee 
It is clear that these transformations descend to homotheties of 
$X=\H^n///_0T^r$ in such a way that $X$ becomes a hyperkahler cone. 
We note that such cones have been studied in \cite{martin} from a local 
perspective.

\subsection{Description as a Kahler quotient and embedding in a toric variety} 

\subsubsection{Complex coordinates and Kahler quotient description}
 
Toric hyperkahler cones can be presented as Kahler quotients of affine 
algebraic varieties. 
For this, let us write $u_j=u^0_j+\i u^1_j+\j u^2_j +\k u^3_j$ and introduce 
complex coordinates  
through $w^{(+)}_j=u^0_j+\i u^1_j$ and $w^{(-)}_j=u^2_j-\i u^3_j$. Then: 
\be
\label{complex_coords}
u_j=w^{(+)}_j+\j w^{(-)}_j~. 
\ee
This amounts to endowing $\H^n$ with the complex 
structure $\I$. The flat hyperkahler metric (\ref{metric}) becomes the 
standard Kahler metric on $\C^{2n}$:
\be
ds^2=\sum_{j=1}^n{\left(d{\overline w}^{(+)}_jdw^{(-)}_j+d{\overline w}^{(-)}_jdw^{(+)}_j\right)}~~,
\ee
while the triholomorphic $T^r$ action (\ref{quat_action}) takes the form: 
\be
\label{complex_action}
w_j^{(+)}\rightarrow \prod_{\alpha=1}^r \lambda_\alpha ^{q^{(\alpha)}_j}w_j^{(+)}~~,~~
w_j^{(-)}\rightarrow \prod_{\alpha=1}^r \lambda_\alpha ^{-q^{(\alpha)}_j}w_j^{(-)}~~.  
\ee 
The minus sign in the second exponent follows from the 
anticommutation relation 
$\i \j=-\j \i$, which implies
$\lambda_\alpha \j=\j {\overline \lambda}_\alpha= \j \lambda_\alpha^{-1}$.
This action preserves the Kahler structure of $\C^{2n}$.
The hyperkahler moment map separates as ${\vec \mu}=
2[\i \mu_r+ \k \mu_c]$, with: 
\bea
\label{mu_rc}
\mu_r^{(\alpha)}(w)&=&\frac{1}{2}\mu^{(\alpha)}_1(w)=
\frac{1}{2}\sum_{j=1}^n{q^{(\alpha)}_j
\left(|w_j^{(+)}|^2-|w_j^{(-)}|^2\right)}\in \R \nn\\
\mu_c^{(\alpha)}(w)&=&
\frac{1}{2}[\mu^{(\alpha)}_3(w)+i\mu^{(\alpha)}_2(w)]
=\sum_{j=1}^n{q^{(\alpha)}_jw_j^{(+)}w_j^{(-)}}
\in \C~~.  
\eea 
The levels decompose accordingly as ${\vec \xi}=(\xi_r,\xi_c)\in \R^r\oplus \C^r\approx \R^{3r}$.
The real component $\mu_r$ is the Kahler moment map for the action (\ref{complex_action}) on $\C^{2n}$. 

These decompositions allow us to view the hyperkahler cone $X$ as a Kahler quotient: 
\be
\label{XKahler}
X={\cal Z}//_0T^r~~,
\ee
where ${\cal Z}$ is the affine algebraic variety defined through:
\be
\label{Z}
{\cal Z}=\mu_c^{-1}(0)\subset \C^{2n}~~.
\ee
The presentation (\ref{XKahler}) also gives 
$X$ as the solution set of the quadric 
equations: 
\be
\label{quadrics}
\sum_{k=1}^n{q_k^{(\alpha)}w_k^{(+)}w_k^{(-)}}=0~~(\alpha=1\dots r)~~  
\ee
in the toric variety $\S=\C^{2n}//_{0}T^r\approx \C^{2n}/(\C^*)^r$. 
This shows that hyperkahler cones are algebraic varieties of a 
familiar type. 

\subsubsection{Toric description of the ambient space}

It is straightforward to extract the 
toric description of $\S$, which we give below for the sake of completeness.
First notice that the $T^r$ action (\ref{complex_action}) on the complex 
coordinates has the $r\times (2n)$ charge matrix:
\be 
{\hat Q}=\left[\begin{array}{cc}Q, -Q\end{array}\right]~~, 
\ee 
which defines the lattice map ${\hat q}^*=(q^*,-q^*):\Z^r\rightarrow \Z^{2n}$.
It is shown in Appendix E that the cokernel of ${\hat
q}^*$ has no torsion, so one obtains an exact sequence:
\be
\label{q0sequence}
0\longrightarrow \Z^r\stackrel{{\hat
q}^*}{\longrightarrow}\Z^{2n}\stackrel{{\hat g}}{\longrightarrow}
\Z^{2d+r}\longrightarrow 0~~.
\ee 
This completely specifies the ambient toric variety 
$\S$. The columns of the $(2d+r)\times (2n)$ matrix ${\hat G}$ (the matrix
of the map ${\hat g}$ with respect to the canonical bases of $\Z^{2n}$ and 
$\Z^{2d+r}$) are the toric generators of $\S$.

Considering (\ref{quat_sequence}) and (\ref{q0sequence}), the 3-lemma 
shows that there exists a uniquely-determined and injective 
map $f:\Z^d\rightarrow \Z^{2d+r}$ which makes the diagram of figure 
\ref{3lemma} commute.

\begin{figure}[hbtp]
\begin{center}
\scalebox{0.5}{\input{3lemma.pstex_t}}
\end{center}
\caption{\label{3lemma} Embedding a hyperkahler cone in a toric variety.}
\end{figure}

Thus there exists a unique $(2d+r)\times d$
matrix $F$ satisfying the constraint: \be
\label{F0}
FG={\hat G}J~~,  
\ee 
where $J=(I_n,-I_n)$ is the $(2n)\times n$ matrix of the map 
$j=(id,-id):\Z^n\rightarrow \Z^{2n}$. Since $f$ is injective, 
the matrix $F$ has maximal rank. 
Its columns span a $d$-dimensional subspace of $\R^{2d+r}$,
which can be identified with the Lie algebra of the triholomorphic
$T^d$ action on $X$.  In particular, $f$ embeds the toric hyperkahler
generators $\nu_j$ as integral vectors $f(\nu_j)$ lying 
in this subspace.

\subsection{Description as a $T^d$ fibration}

It is clear from the hyperkahler quotient construction that the torus 
$T^d=T^n/T^r$ will act on $X$ preserving its
hyperkahler structure. In particular, one has an associated hyperkahler moment map ${\vec \pi}:X\rightarrow \R^{3d}$. It is shown 
in \cite{BD} that this map is surjective and descends to a homeomorphism between $X/T^d$ and $\R^{3d}$. In particular, 
$X$ is connected and can be viewed as a $T^d$ fibration over the {\em entire}\footnote{
This should be contrasted with the situation for toric varieties. In that case, 
the image of the Kahler moment map of the densely embedded torus is only a convex subset of real affine space, 
namely the Delzant polytope of the variety.} space $\R^{3d}$.

\subsubsection{Construction of the fibration}

To see this explicitly, we notice that the level set ${\cal N}:={\vec \mu}^{-1}(0)\subset \H^n$ can 
be described as follows. Since ${\vec \mu}={\vec q}\circ {\vec \eta}$, 
a point $u\in \H^n$ belongs to ${\cal N}$  
if and only if ${\vec \eta}(u)$ belongs to the kernel of ${\vec q}$. By dualizing (\ref{quat_sequence}), we
obtain a short exact sequence:
\be
0\longrightarrow \Z^d\stackrel{g^*}{\longrightarrow}\Z^n\stackrel{q}{\longrightarrow} \Z^r\longrightarrow 0~~,
\ee
which shows that $ker q=im g^*$. Since $g^*$ is injective, we find  that there exists a unique map 
${\vec \pi}_0:{\cal N}\rightarrow {\vec \R}^d$ such that ${\vec \eta}|_{\cal N}={\vec g}^*\circ {\vec \pi}$.
This map presents ${\cal N}$ as a $T^n$ fibration over ${\vec \R}^d$.
Moreover, it descends to a 
well-defined map ${\vec \pi}:X={\cal N}/T^r\rightarrow {\vec \R}^d$, since ${\cal N}$ and ${\vec \mu}$ 
are $T^r$-invariant; this amounts to writing $X$ as a $T^d$ fibration over $\R^{3d}$, obtained from ${\cal N}$ by quotienting out 
its $T^n$ fibers through the subtorus $T^r\subset T^n$.
It is easy to see that  
${\vec \pi}$ is surjective and coincides with the hyperkahler moment 
map of the induced $T^d$ action on $X$. This argument is summarized in figure 
\ref{moment}. 

\begin{figure}[hbtp]
\begin{center}
\scalebox{0.5}{\input{moment.pstex_t}}
\end{center}
\caption{Construction of the moment map for the induced $T^d$ action on $X$.\label{moment}}
\end{figure}

In words, a point $u=(w^{(+)}, w^{(-)})\in \H^n$ 
satisfies the moment map equations ${\vec \mu}(u)=0$ (and thus 
belongs to ${\cal N}$) if and only if there exists a vector 
${\vec v}=(v_1,v_2,v_3)$ in 
$\R^d\times \R^d\times \R^d$ 
such that ${\vec \eta}(u)={\vec g}^*({\vec v})$, i.e.:
\be
\label{ab}
\eta_r(u)=\frac{1}{2}(|w^{(+)}_j|^2-|w^{(-)}_j|^2)=
\nu_j\cdot a~~,~~
\eta_c(u)=
w^{(+)}_jw^{(-)}_j=\nu_j\cdot b~~,~{\rm~for~all~}j=1\dots n~~, \label{X}
\ee
where $a=\frac{v_1}{2}$, $b=\frac{v_3+iv_2}{2}$ 
and $\cdot$ stands for the standard scalar product. 
In this case, we have ${\vec \pi}_0(u)={\vec v}$ and similarly for the induced
map $\pi$,  so that:
\be 
\pi_r(u)=a {\rm ~and~} \pi_c(u)=b~~.
\ee
Passage to the 
`dual variables' ${\vec v}$ 
allows one to `solve' the moment map constraints ${\vec \mu}=0$. 
This is a well-known observation familiar from the work of \cite{HKLR,martin},
which we have simply reformulated in `invariant' language. One can describe
this more physically in the language of \cite{HKLR, martin} by introducing 
a four-dimensional $N=2$ nonlinear $\sigma$-model, in which case one identifies the quaternion coordinates
$u$ with hypermultiplets and the dual variables $a$ and $b$ with linear and  hyper- multiplets respectively. However, we wish to stress that this 
interpretation is purely formal, since such a fictitious $N=2$
theory need not (and will not) have any direct 
physical relevance in our situation. 

We end by noting that the 
fiber $X({\vec v})={\vec \pi}^{-1}({\vec v})$ is given by:
\bea
\label{Xu}
X({\vec v})=\{u=[w_+,w_-]~|~(w^{(+)}_j,w^{(-)}_j){\rm~satisfy~(\ref{X})} \}~~,
\eea 
where the square brackets indicate that  
$u$ is considered modulo the $T^r$ action.

\subsubsection{Degenerate fibers}

The degenerations of the fibers of 
${\vec \pi}$ can be described as follows \cite{BD}.
For every toric hyperkahler generator $\nu_j$, define a codimension three 
linear subspace 
$H_j=h_j\times h_j\times h_j$ of $\R^{3d}=\R^d\times \R^d\times \R^d$, where $h_j$ is the following hyperplane in $\R^d$:
\be
\label{h}
h_j=\{v\in \R^d|v\cdot \nu_j=0\}~~.
\ee 
Following the terminology of \cite{BD} we shall call $H_j$ a {\em flat}; 
in terms of the real-complex coordinates $(a,b)\in \R^d\times \C^d$, it 
corresponds to the locus $a\cdot \nu_j=0$, $b\cdot \nu_j=0$, which 
by virtue of (\ref{ab}) amounts to $w_j^{(+)}=w_j^{(-)}=0
\Leftrightarrow u_j=0$.
It thus follows from (\ref{Xu}) 
that the preimage ${\vec \pi}^{-1}(H_j)$ is the sublocus of $X$ defined 
by vanishing of the quaternion coordinate $u_j$:
\be
\label{Xj}
X_j:={\vec \pi}^{-1}(H_j)=\{u\in X|u_j=0\}~~.
\ee
It is clear that all flats have the origin of $\R^{3d}$ as 
a common point; the associated point in $X$ is the apex 
${\vec \pi}^{-1}(0)$, which corresponds to $u=0$. 
On the other hand, two flats $H_i, H_j$ intersect
outside of the origin if and only if the associated hyperplanes 
$h_i$ and $h_j$ intersect outside the origin in $\R^d$. 
Since $X_i\cap X_j={\vec \pi}^{-1}(H_i)\cap {\vec \pi}^{-1}(X_j)=
{\vec \pi}^{-1}(H_i\cap H_j)$, we find that the loci $X_i$ and $X_j$
can intersect outside the apex of $X$ if and only if $h_i$ and $h_j$ 
intersect outside the origin of $\R^d$.

\paragraph{Observation 1}
Because $h_i\cap h_j$ has codimension at least 
two in $\R^d$, intersections outside the 
origin will always occur for $d>2$. From this point of view, the case 
$d=2$ of eight-dimensional toric hyperkahler cones is rather special; in
that case, the hyperplanes $h_j$ are simply lines through the origin 
in $\R^2$, and two such lines can intersect outside of the origin 
if and only if they coincide. As a consequence, two loci 
$X_i$ and $X_j$ (which in this case have real dimension four) 
will either coincide or intersect at the apex of $X$ only.
It is this special case which will be of interest in the remaining sections.

\paragraph{Observation 2} Recall that $X$ can be viewed as an algebraic 
variety upon choosing the fist complex structure on $\H^n$. Since $X_j$ are 
defined by the equations $w_j^{(+)}=w_j^{(-)}=0$, each such locus 
is a sub-variety of $X_j$ of complex codimension two.

The configuration of flats $H_1\dots H_n$ can be used to describe the 
fixed points of the $T^d$ action on $X$.
Indeed, it is shown in \cite{BD} that the $T^d$-stabilizer of a point 
$u\in X$ is a subgroup $Stab_{T^d}(u)$ of $T^d$ whose Lie algebra 
$stab_{T^d}(u)$ is given by:
\be
\label{Tdstab}
stab_{T^d}(u)=
\langle \{\nu_j|{\vec \pi}(u)\in H_j\}\rangle_\R\subset \R^d~~.
\ee
In this relation, $\langle \dots \rangle_\R$ indicates the real linear span 
of the given set of vectors and $\R^d$ is viewed as the Lie algebra of $T^d$.
Thus $stab_{T^d}(u)$ is spanned by those toric hyperkahler generators 
$\nu_j$ which have the property ${\vec \pi}(u)\cdot \nu_j=0$, 
i.e. such that the associated flat $H_j$ contains ${\vec \pi}(u)$. 
It is clear from this that all degenerate fibers sit above 
the flats. Above each flat $H_j$, a certain one-cycle of the $T^d$ fiber 
collapses to zero length, 
with the Lie algebra of the collapsing $S^1$ lying in the direction $\nu_j$.
Above the intersection of $k$ flats, the $S^1$ cycles associated with each flat collapse simultaneously.

\subsection{The hyperkahler potential}

The hyperkahler potential of the 
metric induced on $X$ by the hyperkahler quotient construction
can be obtained \cite{BD} by applying the 
general methods of \cite{HKLR}. To describe the result, it is convenient to 
use the coordinate ${\vec x}=\frac{\vec v}{2}=\frac{1}{2}{\vec \pi}(u)$. 
We have 
${\vec x}=(x_1,x_2,x_3)$ with 
$x_s=(x_s^{(1)}\dots x_s^{(d)})\in \R^d$ for each $s=1,2,3$. 
Then the real-complex coordinates $a,b$
have the form $a=x_1=(x^{(1)}_1 \dots x^{(d)}_1)=\frac{\pi_1(u)}{2}$
and 
$b=x_3+ix_2=(x_3^{(1)}+ix_2^{(1)}\dots x_3^{(d)}+i x_2^{(d)})=
\frac{\pi_3(u)+i\pi_2(u)}{2}$.

We also consider coordinates 
$\phi^{(1)}\dots \phi^{(d)}$ 
on the $T^d$ fiber of $X\rightarrow \R^{3d}$, associated 
with the moment map components ${\vec \pi}^{(1)}\dots 
{\vec \pi}^{(d)}$. It is convenient to define 
$r_k=2\sqrt{(a\cdot \nu_k)^2+ 
(b\cdot \nu_k) ({\overline b}\cdot \nu_k)}=
2||\nu_j\cdot {\vec x}||_{\R^3}$, where $\nu_j\cdot {\vec x}=
(\nu_j\cdot x_1, \nu_j\cdot x_2, \nu_j\cdot x_3)$ is a vector with three 
components.
Note that $r_k$ is proportional 
to the distance between ${\vec x}$ and $H_k$; in particular, the flat $H_k$
corresponds to $r_k=0$ (to understand this correctly, one must keep in mind
that the flats have codimension {\em three} in $\R^{3d}$).
It is shown in \cite{BD} that $X$ admits the hyperkahler 
potential:
\be
{\cal K}=\frac{1}{2}\sum_{k=1}^n{r_k}=
\sum_{k=1}^n{||\nu_k\cdot {\vec x}||_{\R^3}}~~,
\ee
which, as expected, 
is independent of the $T^d$ fiber coordinates $\phi^{(1)}\dots \phi^{(d)}$.
Our normalization\footnote{This differs from the normalization of 
\cite{BD} by a factor of two.} is such that 
$\omega_A=i\partial_A {\overline \partial}_A{\cal K}$, where $\omega_A$ is 
the Kahler form associated with the compatible complex structure $I_A$.
Note that $a,b$, ${\vec x}_j$ and $r_k$ 
scale as $\alpha^2$ under the homothetic action (\ref{homothety}).
Thus ${\cal K}$ has the $\alpha^2$ scaling discussed in Section 2, and 
the radial distance function $r={\cal K}^{1/2}$ scales linearly.
Using relations (\ref{ab}), one finds that $r_k=||u_k||^2$ and thus: 
\be
\label{HKpot}
{\cal K}=\frac{1}{2}\sum_{k}{||u_k||^2}=\frac{1}{2}||u||^2~~.
\ee
The hyperkahler metric on $X$ can be obtained \cite{BD} from (\ref{HKpot}) 
by using the general methods of \cite{PP}.

\subsection{Singularities}

A toric hyperkahler cone $X=\H^n///_0 T^r={\cal N}/T^r$ 
will generally have two types of singularities, namely those inherited from 
the variety ${\cal N}={\vec \mu}^{-1}(0)$ and those 
due to fixed points of the $T^r$ action.
A singularity of ${\cal N}$ may become worse after taking the 
quotient, since the torus $T^r$ may have a subgroup acting trivially on a singular locus of ${\cal N}$.

\subsubsection{Good toric hyperkahler cones}

It is possible 
to identify a class of toric hyperkahler cones for which 
${\cal N}$ is smooth except at the origin. 
We show in Appendix B that the following conditions are equivalent: 

(a) All $d\times d$ minor determinants of the matrix $G$ are nonzero. 
Equivalently, any $d$ of the vectors $\nu_1\dots \nu_n$ are 
linearly independent over $\R$ (and thus form a basis of $\R^d$),

(b) All $r\times r$ minor determinants of $Q$ are nonzero,

\noindent and that, if they are satisfied, then the origin of 
$\H^n$ is the only singular point of ${\cal N}$.

\noindent Toric hyperkahler cones satisfying these conditions will be called {\em good}.

\subsubsection{Singularities of good toric hyperkahler cones}

Given a {\em good} toric hyperkahler cone $X$, the set ${\cal N}$ is smooth 
outside the origin and thus 
all singularities of $X-\{0\}$ arise from fixed points of the $T^r$ action. 
On the other hand,  (\ref{Tdstab}) shows that such fixed points 
(and thus singularities of $X-\{0\}$)  
can only occur on the subloci $X_j={\vec \pi}^{-1}(H_j)$ obtained by setting 
some quaternion coordinate $u_j$ to zero. If $u$ is a point of $X$, 
we let $V(u)$ be the set of indices $j$ such that $u_j=0$ and 
$N(u)$ be the set of indices $j$ such that $u_j\neq 0$. This provides 
a partition of the index set $\{1\dots n\}$, which characterizes the 
collection of flats which contains the point ${\vec \pi}(u)$. 
Indeed, one has ${\vec \pi}(u)\in H_j\Leftrightarrow u\in 
X_j={\vec \pi}^{-1}(H_j)$ if and only if $j\in V(u)$. Therefore, 
the stabilizer (\ref{Tdstab}) can be written:
\be
stab_{T^d}(u)=\langle \{\nu_j|j\in V(u)\}\rangle_{\R}~~.
\ee
It is not hard to see (Appendix B) that, for 
a {\em good} toric hyperkahler cone, the set $V(u)$ has at most $d-1$ 
elements (and thus $N(u)$ has at least $r+1$ elements), 
unless $u$ coincides with the apex of $X$ 
(in which case $V(u)=\{1\dots n\}$). This happens because 
any $d$ of the toric hyperkahler generators  are linearly independent, 
which implies that no more than $d-1$ of the flats $H_j$ can intersect 
outside of the origin in $\R^{3d}$. 

It turns out that the partition $\{1\dots n\}=V(u)\cup N(u)$ also 
characterizes the singularity type of $X$ at the point $u$.
Indeed, it is clear that 
$u$ will be fixed by an element $\boldlambda=(\lambda_1\dots \lambda_r)
\in T^r$ under the 
action (\ref{quat_action}) if and only if $\boldlambda$ is a solution of
the system:
\be
\label{fpsys}
\prod_{\alpha=1}^r{\lambda^{q_j^{(\alpha)}}}=0~~{\rm~for~}j\in N(u)~~.
\ee
Let us assume that $u\neq 0$. Since the cone $X$ is good and $N$ contains 
at least $r+1$ elements, the $r$ rows of exponents $q^{(\alpha)} 
~(\alpha=1\dots r)$ appearing in (\ref{fpsys}) are linearly independent.  
As discussed in Appendix A, the solution set of such a system forms 
a multiplicative subgroup $\Gamma_u$ of $T^r$, whose structure can 
be determined by computing the integral Smith form of the matrix 
$Q_N$ obtained from $Q$ by deleting all columns associated with 
indices belonging to the set $V(u)$. Using the fact that $X$ is good, 
one can in fact show that the structure of $\Gamma_u$
can also be determined from the integral Smith form of the matrix $G_V$ 
obtained from $G$ by deleting all columns associated with the index set 
$N(u)$. This result follows by chasing a certain diagram of lattices.
We refer the reader to Appendix A for the precise formulation and 
proof of these statements. 

\subsubsection{The eight-dimensional case}

\begin{figure}[hbtp]
\begin{center}
\scalebox{0.6}{\input{flats.pstex_t}}
\end{center}
\caption{For a good, eight-dimensional toric hyperkahler cone, 
two four-dimensional subloci $X_i$ and $X_j$ can intersect only at the apex. 
Equivalently, two flats $H_i$ and $H_j$ and two lines $h_i$ and $h_j$ can 
intersect only at the origin. The figure shows two loci $X_i,X_j$ and 
the associated lines $h_j,h_j\subset \R^2$.\label{flats}}
\end{figure}
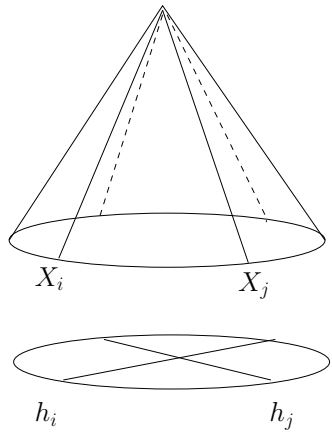

Let us consider the case of eight-dimensional toric hyperkahler cones,
which are of interest for the remainder of this paper. In this case,
one has $d=2$ and $r=n-2$. Let $X=\H^n///_0 T^{n-2}$ be such a cone.
As remarked in Observation 1 of Subsection 5.2.2, the case $d=2$ is
special due to dimension constraints which force the flats $H_j$ to
intersect only at the origin unless they coincide. Using this
observation, it is not hard to see that the cone $X$ will be good if
and only if no two of the flats coincide. In this case, the flats
intersect only at the apex, which means that the only degenerations
allowed for the $T^2$ fibers of $X-\{0\}$ correspond to the collapse
of a single cycle, a phenomenon which occurs along the
four-dimensional loci $X_j$.  Equivalently, at most one quaternion
coordinate of a point $u$ of $X-\{0\}$ can vanish (figure \ref{flats}).  
By the results mentioned above, this means that the singularity
type is constant above each flat (with the exception of the apex of
$X$). On the locus $X_j-\{0\}$, one has a single vanishing coordinate
$u_j$, which means that $V(u)=\{j\}$ and the matrix $G_V$ coincides
with the $j^{th}$ column of $G$, i.e.  with the toric hyperkahler
generator $\nu_j$. Therefore, the second description of singularities
mentioned in the previous subsection becomes particularly simple. Combining
these observations, one can show that the following statements are
equivalent:

(a) $X$ is a good toric hyperkahler cone, i.e. any two of the toric hyperkahler generators 
$\nu_1\dots \nu_n$ are linearly independent over $\R$. Equivalently, 
all $(n-2)\times (n-2)$ minor determinants 
of the $(n-2)\times n$ charge matrix $Q$ are non-vanishing. 

(b) No two of the three-dimensional flats $H_1\dots H_n$ coincide in $\R^6$

(c) No two of the lines $h_1\dots h_n$ coincide in $\R^2$.

In this case, the singularities of $X$ can be described as follows:

(1) All singularities of $X$ lie in one of the four-dimensional loci 
$X_j=\{u\in X|u_j=0\}=\pi^{-1}(H_j)$. 
Two such loci intersect at precisely one point, namely the apex of $X$.

(2) The locus $X_j-\{0\}$ is smooth if and only if the associated toric 
hyperkahler generator $\nu_j\in \Z^2$ is a primitive vector.

(3) If $\nu_j$ is not primitive, then each point on the locus 
$X_j-\{0\}$ is a $\Z_{m_j}$ quotient singularity of $X$, where $m_j$ is the 
greatest common divisor of the coordinates of $\nu_j$.

The formal proof of these statements can be found in Appendix B. 
The singularity group $\Z_{m_j}$ appears as the multiplicative group of 
solutions $\boldlambda=(\lambda_1\dots \lambda_{n-2})\in T^{n-2}$ 
to the system (\ref{fpsys}) for $V(u)=\{j\}$:
\be
\label{Gamma_j_sys}
\prod_{\alpha=1}^{n-2}{\lambda^{q_k^{(\alpha)}}}=1~~{\rm~for~}~~k\neq j~~.
\ee
In Appendix B, we show that such a solution will automatically 
also satisfy the equation:
\be
\prod_{\alpha=1}^{n-2}{\lambda^{q_j^{(\alpha)}}}=e^{\frac{2\pi i}{m_j}s}~~
\ee
for some element $s\in \Z_{m_j}$. The isomorphism between $\Gamma_j$
and $\Z_{m_j}$ is given by the map which takes a solution of 
(\ref{Gamma_j_sys}) into the element $s$.

\pagebreak

\section{Quaternion-Kahler spaces and twistor spaces from 
toric hyperkahler cones}

Given a toric hyperkahler cone $X=\H^n///_0T^r$, the associated twistor space $Y$ and 
quaternion-Kahler space $M$ 
can be recovered as follows.

\subsection{Construction of $M$ as a conformal quotient} 

Let us consider the group $\H^*=\H-\{0\}$ of invertible quaternions, which 
acts on $\H^n$ through: 
\be
\label{Haction}
u_j\rightarrow u_j \, t^{-1}~~,~~t\in \H^*~~.
\ee 
Since $t$ acts from the right in (\ref{Haction}), 
this commutes with the $T^r$ action (\ref{quat_action}), and thus 
descends to an $\H^*$ action on the hyperkahler cone $X$.

The subgroup of $\H^*$ consisting of unit norm quaternions is the symplectic 
orthogonal group $Sp(1)$, whose action (\ref{Haction}) rotates the complex 
structures of $\H^n$; this descends to an action on $X$ which rotates its 
complex structures and obviously preserves the hyperkahler potential 
(\ref{HKpot}). As discussed in Section 3, 
restriction from $X$ to a level set ${\cal K}=\zeta$ defines the associated 
3-Sasaki space $S$, and the quotient:
\be
M=S/Sp(1)_{eff}={\cal K}^{-1}(\zeta)/Sp(1)_{eff}~~
\ee
is a quaternion-Kahler space of positive scalar curvature. This is 
the presentation of $M$ as a conformal quotient \cite{martin, conf_quotient}.
As mentioned in Section 3, the $\Z_2$ subgroup $\{-1,1\}$ of $Sp(1)$ 
may act trivially on $X$, so the effectively acting group 
is $Sp(1)_{eff}=Sp(1)$ or $Sp(1)/\Z_2=SO(3)$. A criterion for deciding when 
the $\Z_2$ subgroup acts trivially is given in Subsection 5.4. below.

\subsection{Description of $M$ as a quaternionic quotient}

Since our cone is toric hyperkahler, it is also possible to present $M$
as a quaternionic quotient in the sense of \cite{GL}.
For this, one considers the quotient of $\H^n$ through (\ref{Haction}), 
which is the quaternion 
projective space $\H\P^{n-1}$. This is a 
quaternion-Kahler (but not hyperkahler) manifold.  
The quaternion projective space carries a quaternion analogue of the 
Fubini-Study metric:
\be
ds^2=\zeta\left[\frac{1}{||u||^2}{\sum_{j=1}^n{d{\overline u}_jdu_j}}-
\frac{1}{||u||^4}{\sum_{j,k=1}^n{{\overline u}^j du^jd{\overline u}^k u^k}}
\right]
~~,
\ee
where the scale factor $\zeta>0$ fixes the volume of $\H\P^{n-1}$.
The $T^r$ action (\ref{quat_action}) descends to an action on $\H\P^{n-1}$ which
preserves its quaternion-Kahler structure. Then the 
work of \cite{Galicki2, GL} implies that 
$M$ can also be described as the quaternionic quotient
\footnote{Since $\H\P^{n-1}$ is only quaternion-Kahler, 
its reduction is only possible at level zero \cite{GL}. }:
\be
\label{M}
M:=\H\P^{n-1}///T^r~~.
\ee
In this presentation, the quaternion structure of $M$ is inherited 
from that of $\H\P^{n-1}$ by quaternionic reduction \cite{GL}.
The scale of the resulting metric is fixed by the choice 
of $\zeta$.

\subsection{Description of the twistor space as a Kahler quotient}
Returning to the action (\ref{Haction}) on 
the quaternion affine space, let us pick the first complex structure 
${\bf I}$ and write the elements of $\H^*$ in the form: 
\be 
t=\alpha({\overline \lambda} +\j\kappa)~~,~~{\rm~with~}~~\alpha=||t||>0~~,
~~ \lambda, \kappa \in \C~~,~~|\lambda|^2+|\kappa|^2=1~~. 
\ee 
Then (\ref{Haction}) becomes:
\bea
\label{wtfs}
w_k^{(+)}&\rightarrow& \frac{1}{\alpha}(\lambda w^{(+)}_k+\kappa {\overline w}^{(-)}_k)~~\nn\\
{\overline w}_k^{(-)}&\rightarrow& 
\frac{1}{\alpha}(-{\overline \kappa} w^{(+)}_k+{\overline \lambda} {\overline w}^{(-)}_k)~~.
\eea
The subgroup of $\H^*$ consisting of unit norm quaternions $\tau={\overline \lambda}+\j \kappa$ 
coincides with $Sp(1)$. It acts on the vector $w_k:=\left[\begin{array}{c}w_k^{(+)}\\ 
{\overline w}_k^{(-)}\end{array}\right]$ through:
\be
\label{Aaction}
w_k\rightarrow A(\tau)w_k~~,
\ee
where $A(\tau)$ is the $SU(2)$ matrix: 
\be
A(\tau)=\left[\begin{array}{cc}\lambda &\kappa
\\-{\overline \kappa}& {\overline \lambda}\end{array}\right]~~.
\ee
The map $\tau\rightarrow A(\tau)$ gives the standard group isomorphism
$Sp(1) \approx  SU(2)$. Since the vector $w_k$ appearing in (\ref{Aaction}) 
contains the complex conjugate of $w_k^{(-)}$, this action rotates 
(i.e. acts transitively on) the complex structures. 

Fixing $\kappa=0$ gives a $U(1)$ subgroup $\tau=\lambda$ 
of $Sp(1)$ $(|\lambda|=1)$, 
which is identified 
with the diagonal $U(1)$ subgroup of $SU(2)$ given by $A(\lambda)=
\left[\begin{array}{cc}\lambda &0\\0& {\overline \lambda}\end{array}\right]$.
This $T^1$ subgroup acts on $w^{(+)}$ and $w^{(-)}$ through:
\be
\label{T1}
w^{(\pm)}\rightarrow \lambda w^{(\pm)}~~,
\ee
and in particular it {\em does} preserve the complex structure $\I$
(this is why the $Sp(1)$ orbit of the complex structures is $Sp(1)/U(1)=S^2$).
The action (\ref{T1}) on $\H^n$ admits the {\em Kahler} moment map:
\be
\label{mu0}
\mu_0(u)=\frac{1}{2}\sum_{j=1}^n{(|w_j^{(+)}|^2+|w_j^{(-)}|^2)}=
\frac{1}{2}\sum_{j=1}^n{||u_j||^2}~~.
\ee
The induced $T^1$ action on the hyperkahler cone preserves 
the first complex structure ${\bf I}$ of $X$ and has a moment map 
induced from (\ref{mu0}), which obviously coincides with the hyperkahler 
potential ${\cal K}$ of (\ref{HKpot}). 

As explained in Section 3, the twistor space is given 
by the quotient:
\be
\label{Yquot}
Y={\cal K}^{-1}(\zeta)/T^1_{eff}=X//_{\zeta}T^1_{eff}~~,
\ee
where $T^1_{eff}$ is the effectively acting subgroup of $T^1$.

\paragraph{Observation}
Under the action (\ref{Aaction}), 
the  projection $(\pi_r(u),\pi_c(u))=(a,b)$
of Subsection 5.2. transforms as:
\bea
\label{abtfs}
a&\rightarrow& (|\lambda|^2-|\kappa|^2)a+2 {\rm Re}(\lambda {\overline \kappa} b)~~\nn\\
b&\rightarrow& |\lambda|^2b -|\kappa|^2 {\overline b} 
-2 \lambda \kappa a~~.
\eea
In particular, the  element $\j\in Sp(1)$ (which corresponds to $\lambda=0$ and $\kappa=1$)
acts on complex coordinates as:
\be
\label{amap}
w_k^{(+)}\rightarrow {\overline w}_k^{(-)}~~,~~w_k^{(-)}\rightarrow 
-{\overline w}_k^{(+)}~~
\ee
and induces the transformations:
\be
\label{proj_amap}
a \rightarrow  -a~~,~~b\rightarrow -{\overline b}~~\Leftrightarrow v_1\rightarrow -v_1~~, 
~~v_2\rightarrow v_2~~,~~v_3\rightarrow -v_3~~.
\ee
On the other hand, the fibers of $Y\rightarrow M$ are the $S^2=Sp(1)/U(1)$ orbits 
of the induced $Sp(1)$ action on $Y$. It is clear that (\ref{amap}) acts along these fibers, and 
therefore induces an involution of $Y$ which commutes with the projection $Y\rightarrow M$.
This is the so-called `antipodal map' of the $S^2$ fibration $Y\rightarrow M$. Equations 
(\ref{proj_amap}) give the projection of this involution through the $T^d$ fibration 
map ${\vec \pi}:Y\rightarrow \R^{3d}$.

\subsection{Embedding of the twistor space in a toric variety} 

The last equation in (\ref{Yquot}) can be used to embed $Y$ in a toric variety.
For this, remember from Subsection 5.1. that $X=\mu_c^{-1}(0)//_0 T^r$ and notice that 
the $T^1$ action (\ref{T1}) preserves the level set 
$\mu_c^{-1}(0)$. Since ${\cal K}$ is induced by the 
moment map (\ref{mu0}), we find that $Y$ coincides with the Kahler quotient:
\be
Y={\cal Z}//T^{r+1}_{eff}~~,
\ee
where ${\cal Z}\in \H^n=\C^{2n}$ is the affine variety of Subsection 5.1. and 
$T^{r+1}_{eff}$ is the effectively acting subgroup of $T^{r+1}$. 
In this presentation, the first $r$ reductions are performed at zero 
moment map levels,  while the last is performed at positive level $\zeta$. 
The metric induced by the reduction is the Kahler-Einstein metric of $Y$.
Using the Kahler-quotient--holomorphic quotient correspondence, we 
obtain $Y=({\cal Z}-\{0\})/(\C^*)^{r+1}$, which allows us to view $Y$ as the intersection 
of quadrics (\ref{quadrics}) in the ambient toric variety $\T=\S//_{\zeta}U(1)=
(\C^{2n}-\{0\})/(\C^*)^{r+1}$. In the Kahler quotient description of $\T$, the first $r$ 
quotients are performed at zero moment map levels, while the last
quotient is performed at an arbitrary (but fixed) positive level.

The torus $T^r$ maps into $T^{r+1}$ according to the map $s:\Z^r\rightarrow \Z^{r+1}$ 
given by $s(v)=(v,0)$. The $T^{r+1}$ action on $\C^{2n}$ is described by the 
map ${\tilde q}^*:\Z^{r+1}\rightarrow \Z^{2n}$, whose transpose corresponds to 
the toric charge matrix ${\tilde Q}$ obtained by augmenting 
${\hat Q}$ with an $(r+1)^{th}$ row: 
\be
{\tilde Q}=\left[\begin{array}{cc}Q&-Q\\1\cdots 1&1\cdots 1\end{array}\right]~~.
\ee 
The maps ${\hat q}^*, {\tilde q}^*$ and $s$ satisfy:
\be
{\tilde q}^* \circ s={\hat q}^*~~.
\ee
The toric ambient space $\T$ is described by a short exact sequence:
\be
\label{fullsequence0}
0\longrightarrow \Z^{r+1}\stackrel{{\tilde q}^*}
{\longrightarrow}\Z^{2n}\stackrel{{\tilde g}}{\longrightarrow} A\longrightarrow 0~~,
\ee
where the group $A$ will generally contain torsion. 
This corresponds to the fact that the projectivising $U(1)$ action need 
not be effective on $\S$.  

\begin{figure}[hbtp]
\begin{center}
\scalebox{0.4}{\input{3lemma2.pstex_t}}
\end{center}
\caption{ Exact sequences for the toric embedding of $Y$.\label{3lemma2}}
\end{figure}

The situation is described by the commutative diagram of figure \ref{3lemma2}. 
Applying the 3-lemma gives a unique and surjective 
linear map $p:\Z^{2d+r}\rightarrow \Z^{2d+r-1}$ 
satisfying the constraint $p\circ {\hat g}={\tilde g}$. 

\subsection{Quotient description of the 3-Sasaki space} 

The form (\ref{HKpot}) of the hyperkahler potential shows that the 3-Sasaki 
space $S$ is the {\em 3-Sasaki reduction} \cite{BG} of the sphere 
$S^{4n-1}=\{u\in \H^n \, |\, \, \frac{1}{2}||u||^2=\zeta\}$ through the action of $T^r$:
\be
S=[S^{4n-1}\cap {\vec \mu}^{-1}(0)]/T^r~~.
\ee
In fact, $S^{4d-1}$ admits a 3-Sasaki structure determined by its hyperkahler 
cone $\H^n$, and the restriction of the hyperkahler moment map 
${\vec \mu}$ to this sphere is a so-called {\em 3-Sasakian moment 
map} \cite{BG}. Certain classes of 
torus reductions of 3-Sasakian spheres were studied in 
\cite{BGMR}, though their singularities were not determined there. 

\subsection{On effectiveness of the $Sp(1)$ and $U(1)$ actions on $X$}

In this subsection, we give a criterion for deciding when the $\Z_2$ 
subgroup $\{-1,1\}$ of $Sp(1)$ (and of $T^1$) acts trivially on $X$.
For this, consider the integral Smith form $Q_{ismith}$ 
of the $r\times n$ charge matrix $Q$ and matrices $U\in SL(r, \Z)$, 
$V\in SL(n,\Z)$ such that $Q=U^{-1}Q_{ismith}V$. As explained in Section 4, 
$Q_{ismith}$ has the form $[I, 0]$, where $I$ is the $r\times r$ identity 
matrix. 

Since the $\Z_2$ subgroup 
acts through sign inversion of the quaternion coordinates 
$u_j$, it will have trivial action on $X$ if and only if this transformation 
is realized by the $T^r$ action, i.e. if and only if the system:
\be
\label{Z2syst}
\prod_{\alpha=1}^{n-2}{\lambda_\alpha^{q_j^{(\alpha)}}}=-1~{\rm~for~all~}
j=1\dots n~~.
\ee
admits a solution $\boldlambda=(\lambda_1\dots \lambda_{n-2})
\in U(1)^r$. This system is analyzed in 
Appendix $C$, where we prove the following:

{\bf Proposition} The $\Z_2$ subgroup $\{1,-1\}$ of  $Sp(1)$ acts trivially 
on $X$ if and only if there exists $1\leq m\leq r$
and $1\leq \alpha_1<\alpha_2<\cdots \alpha_m\leq r$ such that all components 
of the $n$-vector $w$ defined as the sum of the rows 
$\alpha_1\dots \alpha_m$ of $Q$ are odd. The indices $\alpha_k$
with this property are uniquely determined.

This allows us to determine the effectively acting subgroups of $Sp(1)$ and 
$U(1)$ and is related to nonvanishing of the Marchiafava-Romani class of 
the quaternion-Kahler space $M$.

\section{Twistor space singularities in the six-dimensional case}

Throughout this section we consider a {\em good} eight dimensional toric 
hyperkahler cone $X=\H^n///T^{n-2}$, so that 
$d=2$ and $r=n-2$. We shall present a method for identifying the 
singularities of the 
six-dimensional twistor space $Y=X//_{\zeta}T^1$, where $\zeta$ is 
a fixed positive number. 
Our approach combines ideas from toric geometry with the description of $X$ as a 
$T^2$ fibration over $\R^6$. 

The basic idea is as follows. Since $X$ is a good toric hyperkahler
cone, its singularities outside the apex can be identified by the
methods of Section 5. These are the singularities along the loci
$X_j-\{0\}$, and are the only singularities of $X$ which can descend
to $Y$, since the apex of $X$ is removed when performing the
projectivising $U(1)$ quotient at a positive level. The 
corresponding loci in the twistor space are holomorphically embedded
two-dimensional spheres $Y_j=X_j//U(1)_{eff}$, 
which turn out to coincide with certain 
fibers of the $S^2$ fibration $Y\rightarrow M$. 
The union $Y_V$ of all $Y_j$ will be called the {\em vertical locus}; we 
define this union to contain all components $Y_j$, even though some 
of them could in fact be smooth in $Y$. 
The singularity type along $Y_j$ can be computed
by toric methods. 
This singularity type may be enhanced with respect to $X_j$ --- 
in particular, a locus $X_j$ which
happens to be smooth in $X$ may project to a sphere of $\Z_2$
singularities in $Y$.

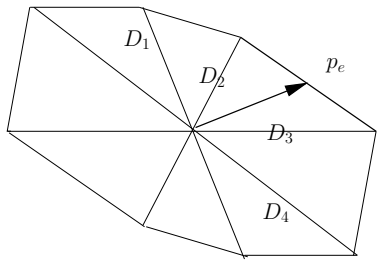
\begin{figure}[hbtp]
\begin{center}
\scalebox{0.5}{\input{Delta_gen.pstex_t}}
\end{center}
\caption{Shape of the characteristic polygon $\Delta$ for the case $n=4$.
\label{Delta_gen}}
\end{figure}

A second class of singularities arises from smooth points of $X$ which
have nontrivial stabilizer under the projectivising action.
Presenting $X$ as a $T^2$ fibration over $\R^6$ as in Section 5, the 
$\pi$--projection of such points must be invariant under the action of
this stabilizer. This observation will allow us to extract a 
locus $X_H$  which
we define as the union of those $T^2$ fibers of $X$ whose projection to $\R^6$
is fixed by a nontrivial subgroup of $U(1)_{eff}$. It will turn out that 
$X_H$ consists of  those points $u\in X$ having the property
$\pi_c(u)=0$, i.e. it is the union of the $T^2$ fibers which lie above 
the two-plane $b=0$ in $\R^2\times \C^2=\R^6$. The real part $\pi_r$ of the 
moment map $\pi$ induces a $T^2$ fibration of $X_H$ over this plane. 
When descending 
to the twistor space, one obtains the {\em horizontal locus} $Y_H=X_H//U(1)$. 
Since this requires imposition of the moment map constraint 
$\frac{1}{2}\sum_{j=1}^n{||u_j||^2}=\zeta$, we shall find that $Y_H$ 
is an $S^1$ fibration over a one-dimensional locus $\Delta$ 
lying inside the plane $b=0$. It will turn out that $\Delta$  is 
a convex polygon with $2n$ vertices,  which is entirely
determined  by $\zeta$ and by the matrix of generators $G$. We will call it the {\it characteristic polygon}. It 
is symmetric under reflection through the origin of the
plane, and in particular contains the origin in its interior 
(figure \ref{Delta_gen}).

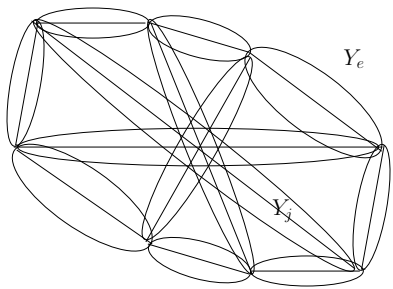
\begin{figure}[hbtp]
\begin{center}
\scalebox{0.5}{\input{dist.pstex_t}}
\end{center}
\caption{The distinguished locus (the union of the horizontal and vertical loci in the text) is an $S^1$ fibration over the edges and 
principal diagonals of $\Delta$. The vertical spheres touch the horizontal 
spheres at single points, in spite of our inability to draw this in two 
dimensions.
\label{dist}}
\end{figure}

The $S^1$ fibers of $Y_H$ degenerate to points above the vertices of $\Delta$, 
which means that $Y_H$ is a union of two-spheres $Y_e$ associated with 
its edges $e$; two spheres $Y_{e}$ and $Y_{e'}$ associated with 
adjacent edges intersect at a point $Y_A$ in $Y$ lying above their
common vertex $A$. Each horizontal sphere $Y_e$ 
turns out to be holomorphically embedded in $Y$, and is 
the lift of a locus 
lying in the base $M$ through the $S^2$ fibration map $Y\rightarrow M$.
Therefore, the intersection of $Y_e$ with a given $S^2$ fiber consists of 
at most one point. Once again, $Y_H$ is 
defined to contain all spheres $Y_e$, even though some of them may 
be smooth in $Y$.
It will also turn out that two horizontal spheres $Y_e, Y_{e'}$ have the
same projection on $M$ if they are associated with opposite edges
$e'=-e$ of $\Delta$. In this case, they are either both smooth or
singular in $Y$, with the same singularity type, which can be determined 
by a simple criterion.  
In fact, the sign inversion
of $\R^2$ is covered by the antipodal map of the fibration
$Y\rightarrow M$. 
Those diagonals $D_j$ of $\Delta$ which connect opposite vertices (and thus 
pass through the origin) will be called {\em principal}. 
It will turn out that 
the restriction of $\pi_r$ to $X_j$ can be used to present 
each vertical sphere $Y_j$ as an $S^1$ fibration over the principal 
diagonal $D_j$; its $S^1$ fibers collapse above the opposite vertices 
which this diagonal connects. The corresponding points of $Y_j$
coincide with the points of the horizontal locus which lie
above these vertices. Each of these points corresponds to the intersection 
of $Y_j$ with two horizontal spheres (figure \ref{dist}).

\subsection{The distinguished locus}

Recall from Section 5 that the hyperkahler moment map 
${\vec \pi}$ presents $X$ as a $T^2$ fibration over 
$\R^6=\R^2\times \R^2\times \R^2$. 
The fiber above a point ${\vec v}=(v_1,v_2,v_3)$ 
(with $v_j$ in $\R^2$) is given by the solutions of the system:
\be
\label{fibers}
\frac{1}{2}(|w_j^{(+)}|^2-|w_j^{(-)}|^2)=\nu_j\cdot a~~,~~
w_j^{(+)}w_j^{(-)}=\nu_j\cdot b~~,
\ee 
where $a=v_1/2\in \R^2$ and $b=(v_3+iv_2)/2\in \C^2$. 
The variables $w_j^{(\pm)}$ are subject to identifications 
induced by the triholomorphic $T^r$ action:
\be
\label{taction}
w_j^{(+)}\rightarrow e^{2\pi i \sum_{\alpha=1}^r{q_j^{(\alpha)}\phi_\alpha}}w_j^{(+)}~~,~~w_j^{(-)}\rightarrow 
e^{-2\pi i \sum_{\alpha=1}^r{q_j^{(\alpha)} \phi_\alpha}}w_j^{(-)}~~.
\ee
As explained in Section 5, solutions to (\ref{fibers}) 
will automatically satisfy the hyperkahler moment map constraints
${\vec \mu}={\vec 0}$.

The twistor space results from $X$ by performing the projectivising 
quotient at a fixed level $\zeta>0$, 
which amounts to imposing the moment map condition:
\be
\label{pmoment}
{\cal K}(w^{(+)},w^{(-)})
=\frac{1}{2}\sum_{j=1}^{n}{(|w_j^{(+)}|^2+|w_j^{(-)}|^2)}=\zeta ~~
\ee 
and quotienting by the action:
\be
\label{Uproj_action}
w_j^{(+)}\rightarrow \lambda w_j^{(+)}~~,
~~w_j^{(-)}\rightarrow \lambda w_j^{(-)}~~, \label{QA}
\ee
where $\lambda=e^{2\pi i \phi}$.
We let ${\cal M}_\zeta:={\cal K}^{-1}(\zeta)\subset X$ 
be the seven-dimensional locus defined by (\ref{pmoment}).
Since ${\cal K}$ is invariant under the $T^n$ action, 
the subspace ${\cal M}_\zeta$ 
is also a $T^2$ fibration, obtained by restricting 
the image of ${\vec \pi}$ from $\R^6$ to the 5-dimensional locus 
$\Sigma_\zeta:={\vec \pi}({\cal M}_\zeta)$. 

Remember that $X$ is smooth outside 
$X_j={\vec \pi}^{-1}(H_j)=\{u\in X |u_j=0\}$.
Hence singularities of $Y$ can only occur on one of the loci
$Y_j:=(X_j\cap {\cal M}_\zeta)/U(1)$ or at points outside these loci 
whose projection $(a,b)$ to $\R^2\times \C^2$ 
is stabilized by a nontrivial subgroup of 
$U(1)_{eff}$. Let $Y_V:=\cup_{j=1}^n{Y_j}$.
{\em We shall show that a point $u$ lying outside $Y_V$ 
can have a nontrivial stabilizer in $U(1)_{eff}$ only if $b=0$.} 
The argument is as follows.

Using equations (\ref{fibers}) and the fact that 
the vectors $\nu_j$ generate $\R^2$, 
it is easy to see  that the projectivising $U(1)$ action 
descends to a well defined action on $\Sigma$ 
through the projection ${\vec \pi}$. 
From (\ref{fibers}) and (\ref{QA}) it follows that a point 
$(\pi_r(u),\pi_c(u))=(a,b)$ transforms as:
\be
\label{paction}
a\rightarrow a~~,~~b\rightarrow \lambda^2 b~~.
\ee
Hence the 
$\Z_2$ subgroup $\lambda=\pm 1$ of $U(1)$ acts trivially on $\Sigma$. 
In particular, since the trivially acting subgroup $G_0$ 
of the projectivising $U(1)$ must preserve $(a,b)$, it follows that 
$G_0$ is a subgroup of this $\Z_2$, i.e. is either trivial or coincides with 
$\Z_2$. 
As explained in Section 6, one has $G_0=\Z_2$ if and only if the 
following system admits  a solution $\lambda\in U(1)^r$:
\be
\label{Z2system}
\prod_{\alpha=1}^{n-2}{\lambda_\alpha^{q_j^{(\alpha)}}}=-1
~{\rm~for~all~}j=1\dots n~~.
\ee
We distinguish two cases:

(a) If (\ref{Z2system}) has a solution, then $G_0=\Z_2$ 
and the symmetry $u\rightarrow -u$ is not part of the effectively acting 
group $U(1)_{eff}=U(1)/G_0=U(1)/\{-1,1\}$. In this case, the transformation 
rule (\ref{paction}) shows that a point $u$ can be stabilized by a nontrivial 
subgroup of $U(1)_{eff}$ only if $b=0$. 

(b) If (\ref{Z2system}) has no solutions, then $G_0$ is the trivial group, 
and the projectivising $U(1)$ acts effectively on $X$. Let us assume that 
$b\neq 0$ (otherwise, there is nothing left to prove). 
In this case, relation (\ref{paction}) shows that the $U(1)$ 
stabilizer must be a subgroup of $\{-1,1\}$. This stabilizer is nontrivial 
precisely when there exists a $U(1)^r$ transformation which implements 
the sign inversion $w^{(\pm)}_k\rightarrow -w_k^{(\pm)}$ on all complex 
coordinates of $u$. To show that the stabilizer is trivial, we proceed
in two steps: 

(b1) Show that at most one complex coordinate of $u$ can vanish.

To understand why, let us assume that two complex coordinates of $u$ 
equal zero. Since $u$ does not belong to $Y_V=\cup_{j=1}^n{Y_j}$, we 
cannot have $w_k^{(+)}=w_k^{(-)}=0\Leftrightarrow u_k=0$, 
since this would imply $u\in Y_k$. Therefore, we must have 
$w_j^{(+)}=w_k^{(+)}=0$ or $w_j^{(-)}=w_k^{(-)}$ for some $k\neq j$. 
In both cases, equations (\ref{fibers}) give $\nu_j\cdot b=\nu_k\cdot b=0$, 
which implies\footnote{Since we assume that the toric hyperkahler cone 
is good, the two-vectors $\nu_j$ and $\nu_k$ are linearly independent.}
$b=0$, thereby contradicting our assumption. 

(b2) Show that no $U(1)^r$ transformation can implement the 
sign inversion $w_i^{(\pm)}\rightarrow -w_i^{(\pm)}$.

According to $(b1)$, only one of the complex coordinates of $u$ can vanish. 
Without loss of generality, we can assume that $w_k^{(+)}=0$. In this case, 
a $U(1)^r$ transformation implementing the desired sign inversion exists if 
an only if the system:
\bea
\prod_{\alpha=1}^{n-2}{\lambda_\alpha^{q_j^{(\alpha)}}}~~&=&-1
~{\rm~for~all~}j\neq k~~\nn\\
\prod_{\alpha=1}^{n-2}{\lambda_\alpha^{-q_j^{(\alpha)}}}&=&-1
~{\rm~for~all~}j=1\dots n~~
\eea 
admits a solution $\lambda=(\lambda_1\dots \lambda_r)\in U(1)^r$. 
Now, 
it is clear that the second set of equations in this system implies
the first, which means that the entire system is equivalent with
(\ref{Z2system}), which has no solutions by the hypothesis of case 
(b). It follows that $(b2)$ holds.
Combining everything, we see once again 
that the $U(1)$ stabilizer of $u$ must 
be trivial unless $b=0$. This finishes the proof of our claim.

Let us define $Y_H$ to be the set of points $u$ in $Y$ for which 
$\pi_c(u)=0\Leftrightarrow b=0$. According to the discussion above, 
a singular point of $Y$ must belong to one of the loci $Y_V$
or $Y_H$. We conclude that {\em all} singularities of $Y$ lie along the 
{\em distinguished locus}:
\be
Y_D:=Y_V\cup Y_H \subset Y ~~,
\ee
where $Y_V:=\cup_{j=1}^n Y_j$ with $Y_j=\{u\in Y|u_j=0\}=(X_j\cap 
{\cal M}_\zeta)/T^1$ 
and $Y_H:=X_H/U(1)$ where $X_H={\vec \pi}^{-1}(\{(a,b)\in \Sigma |b=0\})$.

\subsection{Geometry of the component $Y_H$ and the characteristic 
polygon $\Delta$}

To characterize $Y_H$, consider equations (\ref{fibers}) for $b=0$:
\be
\label{sfibers}
\frac{1}{2}(|w_j^{(+)}|^2-|w_j^{(-)}|^2)=
\nu_j\cdot a~~,~~w_j^{(+)}w_j^{(-)}=0~~.
\ee
Since the second set of conditions requires either 
$w_j^{(+)}=0$ or $w_j^{(-)}=0$ for each $j$, we have $2^n$ possible 
branches $X_\epsilon\subset X_H$, parameterized by a `sign vector' 
$\epsilon=(\epsilon_1\dots \epsilon_n)$, with 
$\epsilon_j=\pm 1$. 
The branch $X_\epsilon$ is defined by choosing the solutions 
$w^{(-\epsilon_j)}=0$ for the second set of equations in (\ref{sfibers}). 
Thus $X_\epsilon$ is given by the constraints:
\be
\label{Xe}
w_j^{(-\epsilon_j)}=0~~,~~
\frac{1}{2}|w_j^{(\epsilon_j)}|^2=\epsilon_j\nu_j\cdot a~~,~~
\frac{1}{2}\sum_{j=1}^{n}{|w_j^{(\epsilon_j)}|^2}=\zeta~~
\ee
and the action:
\be
\label{qepsilon}
w_j^{(\epsilon_j)}\rightarrow e^{2\pi i \epsilon_j \sum_{\alpha=1}^r{q_j^{(\alpha)}\phi_\alpha}}w_j^{(\epsilon_j)}~~.
\ee
The second equation in (\ref{Xe}) requires 
$\epsilon_j\nu_j\cdot a\geq 0$ for all $j=1\dots n$.

The third constraint in (\ref{Xe}) results from the moment map 
condition (\ref{pmoment}). 
When combined with the second equation in 
(\ref{Xe}), it becomes:
\be
\label{Delta}
\sum_{j=1}^{n}{|\nu_j\cdot a|}=\zeta~~,
\ee
a condition for $a$ whose solution set forms the union of edges of a
convex polygon $\Delta$ in $\R^2$. Since (\ref{Delta}) is invariant
under the sign inversion $a\rightarrow -a$, this polygon is symmetric
with respect to the origin.

The map ${\vec \pi}$ presents $X_\epsilon$ as a $T^2$ fibration over
the following subset of $\Delta$:
\be
\Delta_\epsilon={\vec \pi}(X_\epsilon)=
\{a\in \Delta|\epsilon_j\nu_j\cdot a\geq 0\}~~.
\ee
We have 
$\Delta=\cup_{\epsilon}{\Delta_\epsilon}$ and $X_H=
\cup_{\epsilon}{X_\epsilon}$. 
Moreover, $X_\epsilon$ will be non-void precisely when the system of equations:
\be
\epsilon_j\nu_j\cdot a\geq 0~~,~~\sum_{j=1}^{n}{|\nu_j\cdot a|}=\zeta~~
\ee
admits a solution, in which case $\Delta_\epsilon$ coincides with an 
edge of $\Delta$. Thus one can 
index the {\em non-void} components among $X_\epsilon$ by the edges 
$e$ of $\Delta$: $X_\epsilon=X_e$ for some edge $e$, 
if $X_\epsilon\neq \emptyset$. 
Given an edge $e$, the associated signs $\epsilon_j$ are obtained as follows.
If $p_e$ is a vector lying in the interior of $e$ (for example the 
middle point of $e$), we let
$\epsilon_j(e)$ be the sign of the scalar product $\nu_j\cdot p_e$. Then
$X_e=X_{\epsilon(e)}$. Thus $X_e$ is the sublocus in $X$ which
corresponds to vanishing of the variables
$w_j^{(-\epsilon_j(e))}$; in particular, this shows that $X_e$ and $Y_e$ 
are complex subvarieties of $X$ and $Y$. Eliminating the void branches gives
$X_H=\cup_{e}{X_e}$.

We conclude that $X_H$ is obtained by restricting to those $T^2$
fibers of $X$ which lie above the edges of the polygon $\Delta$ (note that 
this automatically implements the $U(1)$ moment map constraint 
${\cal K}=\zeta$). Since
$b=0$ on each $X_e$, equations (\ref{paction}) show that the projectivising
$U(1)$ action is fiberwise along this set.  It follows that
$Y_e=X_e/U(1)$ is an $S^1$ fibrations over the edge $e$ of
$\Delta$.

It is easy to see from (\ref{Delta}) that the vertices of $\Delta$ lie
along the lines $h_1\dots h_n\subset \R^2$ given by equations
(\ref{h}): $h_j=\{v\in \R^2|v\cdot \nu_j=0\}$.  Each such line
contains precisely two vertices, one on each side of the origin of
$\R^2$; these are mapped into each other by the sign inversion
$a\rightarrow -a$. We shall let $D_j$ denote the diagonal of 
$\Delta$ lying along the line $h_j$; such diagonals will be 
called {\em principal}. 

Since $a=v_1/2$ and since $b=(v_3+iv_2)/2$ vanishes 
along the horizontal locus, we find that $h_j$
coincides with the intersection of the flat $H_j=h_j\times h_j\times h_j$ 
with the two-plane in $\R^6$ given by the first $\R^2$ factor in the
decomposition $\R^6=\R^2\times \R^2\times \R^2$.  The discussion of
Section 3 shows that the $T^2$ fiber of $X$ degenerates to a circle
above each flat.  In particular, the $T^2$ fibers of $X$ above
the edges of $\Delta$ degenerate to circles above each vertex (since the 
vertices of $\Delta$ lie on the lines $h_j\times {0}\times {0}\subset H_j$). 
Upon performing the projectivising quotient, this implies that the $S^1$
fibers of $Y_e\rightarrow e$ degenerate to points above the vertices.
It follows that each
locus $Y_e$ is a two-sphere. The spheres associated with adjacent edges
intersect at a point corresponding to their common vertex.

\subsection{Geometry of the component $Y_V$}

The subspaces $Y_k=\{u\in Y|u_k=0\}=\{w\in Y|w_k^{(+)}=w_k^{(-)}=0\}$
are one-dimensional complex subvarieties of $Y$.  Since $X_j$ are
invariant under the $SU(2)$ action induced by (\ref{Haction}), the
loci $Y_j$ must correspond to the $SU(2)/U(1)=S^2$ orbits of the
induced $SU(2)$ action on $Y$; therefore, each $Y_j$ is a fiber of the
$S^2$ fibration $Y\rightarrow M$. In particular, $Y_j$ are rational 
curves in the twistor space.

\begin{figure}[hbtp]
\begin{center}
\scalebox{0.5}{\input{vsphere.pstex_t}}
\end{center}
\caption{ The locus $X'_k=X_K\cap {\cal M}_\zeta$ is an $S^1$ fibration over 
the two-sphere $\sigma_k$. The figure shows $\sigma_k$ and the two-torus 
obtained by restricting to those $S^1$ fibers of $X'_k$ which lie above 
one of the circles $C_{v_1}$.
\label{vsphere}}
\end{figure}
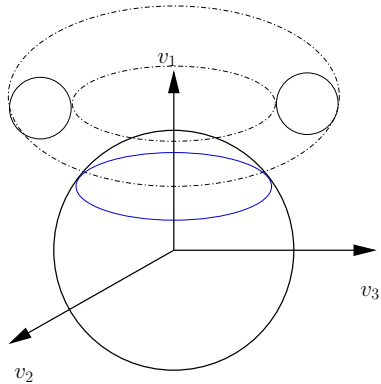

To see directly why $Y_k$ is a two-sphere, notice that substituting 
$w_k^{(+)}=w_k^{(-)}=0$ in equations
(\ref{fibers})  implies 
$\nu_k \cdot a=\nu_k\cdot b=0\Leftrightarrow \nu_k\cdot v=0$, 
which forces the vector $v=(2a,2Im(b),
2 Re(b))$ to lie in the 3-dimensional subspace $H_k\subset \R^6$.
The condition $(a,b)\in \Sigma_\zeta$ further constrains $v$ to lie on a
locus $\sigma_k\subset H_k$, which is topologically a two-sphere. 
The value $b=0\Leftrightarrow v_2=v_3=0$ gives two
opposite points on this sphere (which one can take to be the north and
south pole), which correspond to opposite vertices of
the polygon $\Delta$. Since $\nu_k\cdot a$ vanishes on our locus,
the real component $\pi_r:X\rightarrow \R^2$ of the 
moment map descends to a projection of $X'_k:=X_k\cap {\cal M}_\zeta$ 
onto the principal diagonal $D_k$. The locus $X'_k\subset X$ 
is an $S^1$ fibration over $\sigma_k$, since the
generic $T^2$ fiber of $X\rightarrow \R^6$ 
is collapsed to a circle for $u_k=0$ (figure \ref{vsphere}).  
According to (\ref{paction}),
the projectivising $U(1)$ action on this 
locus fixes the value of $v_1=2a$ while
rotating the vector $(v_2,v_3)=2(Im b, Re b)$ 
in the two-plane defined by this value of $v_1$:
\bea
&&v_1\rightarrow v_1\nn\\
&&v_2\rightarrow v_2 \cos(2\alpha)+v_3 \sin (2\alpha)\\
&&v_3\rightarrow -v_2 \sin(2\alpha)+v_3 \cos (2\alpha)\nn~~,
\eea
where we took $\lambda=e^{2i\alpha}$. 
The orbits are circles $C_{v_1}\subset \sigma_k$ 
lying in the plane defined by $v_1$ (figure \ref{vsphere}). Due to the 
square in the second transformation (\ref{paction}), each orbit covers such
a circle {\em twice}.  In particular, the element $\lambda=-1\Leftrightarrow 
\alpha=\pi$ effects a full rotation along the circle $C_{v_1}$.
Since $v_1$ is invariant under this action, 
we find that $\pi_r=v_1$ descends to a projection of $Y_k=X'_k/U(1)_{eff}$ 
onto $D_k$.

The full projectivising $U(1)$ action on the locus $X'_k$ 
identifies the $S^1$ fibers of $X'_k\rightarrow \sigma_k$
sitting above the circles $C_{v_1}\subset \sigma_k$. 
The precise projection of this action on the $S^1$ fiber 
depends on the restriction of the transformations (\ref{taction}) to
the locus $X_k$. There are two possibilities in this regard:

(a) $U(1)_{eff}=U(1)$ and the $\Z_2$ subgroup generated by $\lambda=-1$ acts
non-trivially on the $S^1$ fiber of 
$X'_k\rightarrow \sigma_k$.

In this case, one must go twice around the
circle $C_{v_1}\subset \sigma_k$ in order to come back to the same point in the
$S^1$ fiber. The projectivising $U(1)$ quotient
gives a copy of $\R\P^1=S^1$ fibered over the segment $\pi_r(X'_k)=D_k$; 
thus the induced map $\pi_r:Y_k\rightarrow D_k$ is an $S^1$ fibration.
Since the circle $C_{v_1}$ collapses to zero size at the poles of $\sigma_k$
(which correspond to the vertices of $\Delta$ connected by $D_k$), we 
find the the $S^1$ fiber of $Y_k$ collapses to a point above the endpoints
of $D_k$. In particular, $Y_k$ is a two-sphere.

(b) $U(1)_{eff}=U(1)/\Z_2$ or $U(1)_{eff}=U(1)$ and 
the $\Z_2$ subgroup of $U(1)$ acts trivially in the direction of the $S^1$
fiber. 

In this case, one comes back to the same point in the fiber
after going once around the circle $C_{v_1}\subset \sigma_k$. 
Once again, the 
projectivising $U(1)$ quotient gives an $S^1$ fibration 
$\pi_r:Y_k\rightarrow D_k$, whose fibers collapse above  the endpoints of 
$D_k$.

We conclude that  each locus $Y_k$ is a two-sphere, which the map 
induced by $\pi_r$ presents as an $S^1$ fibration over $D_k$.  
Such a sphere has
two distinguished points, namely those points sitting above the opposite
vertices of $\Delta$ connected by the principal diagonal $D_k$.  
Each distinguished 
point corresponds to $b=0$ and therefore is shared by $Y_k$ and
two adjacent spheres belonging to the horizontal locus.

\subsection{Relation with the $S^2$ fibration of $Y$ over 
the quaternion-Kahler base}

As discussed above, each vertical sphere $Y_j$ is a fiber of 
$Y\rightarrow M$. For the horizontal locus $Y_H$, we have $b=0$ and 
the $SU(2)$ action (\ref{Aaction}) induces the following transformations 
(\ref{abtfs}):
\bea
\label{abetfs}
&&a\rightarrow (|\lambda|^2-|\kappa|^2)a~\nn\\
&&b\rightarrow -2 \lambda \kappa a~~.
\eea
Since $a\neq 0$, the second equation shows that 
the $SU(2)$ orbit can touch $Y_H$ at another point 
only if $\kappa=0$ or $\lambda=0$. The first case corresponds 
to the diagonal $U(1)$ subgroup 
$A=\left[\begin{array}{cc}\lambda&0\\0&{\overline \lambda}\end{array}\right]$
(with $|\lambda|=1$), which acts trivially on the twistor space (since $Y$ is 
a quotient of ${\cal K}^{-1}(\zeta)\subset X$ through this subgroup).
The case $\lambda=0$ gives 
$A=\left[\begin{array}{cc}0&\kappa\\-{\overline \kappa}&0\end{array}\right]$,
with $|\kappa|=1$. In this case, the transformations (\ref{abetfs}) reduce 
to $a\rightarrow -a$ and $b=0=$fixed, while equation (\ref{Aaction}) gives:
\be
w_k^{(+)}\rightarrow \kappa {\overline w}_k^{(-)}~~,~~
w_k^{(-)}\rightarrow -\kappa {\overline w}_k^{(+)}~~.
\ee
These transformations obviously map the locus $Y_e$ (defined by the equations 
$w_k^{(-\epsilon_k(e))}=0$) into the locus $Y_{-e}$ (defined by the equations 
\footnote{Notice that $\epsilon_k(-e)=-\epsilon_k(e)$.}
$w_k^{(+\epsilon_k(e))}=0$). In particular, the antipodal map 
(which corresponds to the element $\j\in Sp(1)$, for which 
$\lambda=0$ and $\kappa=1$) maps $Y_e$ into $Y_{-e}$ while taking $a$ into 
$-a$. Since $Sp(1)$ acts isometrically on $Y$, 
it follows that the singularity types of $Y$ along $Y_e$ and
$Y_{-e}$ must coincide. In fact, relation (\ref{proj_amap}) shows 
that the antipodal map covers the sign inversion $a\rightarrow -a$ through the 
projection map $\pi_r:Y\rightarrow \R^2$; this relation between the antipodal 
map and sign inversion is valid on the entire distinguished locus 
$Y_D=Y_H\cup Y_V$.

The observations made above show that each sphere $Y_e$ intersects the 
$Sp(1)/U(1)=S^2$ orbit of $Sp(1)$ in  precisely one point. Therefore,
such spheres are horizontal with respect to the fibration $Y\rightarrow M$
(i.e. are lifts of spheres in the ESD base $M$). 

\begin{figure}[hbtp]
\begin{center}
\scalebox{0.6}{\input{DM.pstex_t}}
\end{center}
\caption{Dividing $\Delta$ through the sign inversion $\iota:a\rightarrow -a$ 
gives the polygon $\Delta_M$. The figure shows the case $n=3$.
\label{DM}}
\end{figure}
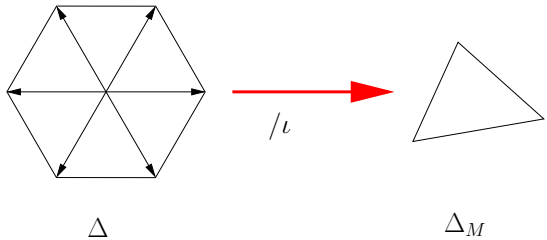

The entire construction can be summarized as follows. One has a system of $n$
spheres in $M$, described by the polygon $\Delta_M$
obtained from $\Delta$ upon quotienting through the sign inversion of
$\R^2$ (figure \ref{DM}). 
Each sphere in $M$ corresponds to an edge of $\Delta_M$, and
two spheres intersect (at a single point) precisely when the
associated edges of $\Delta_M$ touch each other at a vertex. 
Every such sphere has two lifts $Y_e$ and $Y_{-e}$ 
(related by the antipodal map) through the fibration $Y\rightarrow M$. 
These lifts correspond to those opposite edges $e$ and $-e$ of $\Delta$ 
which lie above the associated edge of $\Delta_M$.
The collection of all such lifts gives the
locus $Y_H$. The locus $Y_V$ is the collection of $S^2$ fibers of 
$Y\rightarrow M$ which lie above the vertices of $\Delta_M$.
Comparing with the results of \cite{BGMR,CP} and \cite{CP}, we 
see that the ESD space $M$ is a $T^2$ fibration over the 
compact convex polytope bounded by $\Delta_M$; the $T^2$ fibers 
collapse to circles above the edges of $\Delta_M$ and to points 
above its vertices. The polygon $\Delta_M$ can be identified with the 
polygon extracted in \cite{BGMR} by different methods. It is also 
homeomorphic with the boundary of the (compactified) hyperbolic plane 
appearing in \cite{CP}. 

\subsection{Singularities along $Y_H$ }

Let us fix an edge $e$ of $\Delta$ and consider the non-void branch
$X_e\subset X$ associated with this edge. 
If the projectivising group\footnote{Recall that $U(1)_{eff}=U(1)$ or 
$U(1)/Z_2$ is the group
acting effectively {\em on $X$}.} $U(1)_{eff}$ acts effectively on $X_e$, then
the associated two-sphere $Y_e=X_e/U(1)_{eff}$ is a smooth locus in $Y$,
except possibly for its two points lying above the endpoints of $e$.  If
this action is not effective, then there will be a discrete subgroup
$\Gamma_e$ of $U(1)_{eff}$ which fixes every point of $X_e$; then $Y_e$ 
consists of orbifold singularities of type
$\Gamma_e$, with possible enhancement of the orbifold group at the
endpoints of $e$. The group $\Gamma_e$ can be identified through the 
following argument, a more formal version of which can be found in 
Appendix D. 

Let $\epsilon(e)=(\epsilon_1(e)\dots \epsilon_n(e))$ be the sign
vector associated to the edge $e$ as explained in Subsection 7.1. Then
$X_e$ consists of the tori realized by equations (\ref{Xe}) with the
$T^r$ identifications (\ref{qepsilon}).  Since the complex coordinates 
$w_j^{(-\epsilon_j(e))}$ vanish on $X_e$, the $(n-2)\times n$ 
charge matrix $Q_e$ of
the action induced by (\ref{qepsilon}) on the non-vanishing coordinates 
$w_j^{(\epsilon_j(e))}$ is obtained from $Q$ by changing the signs
of its columns: the $j^{th}$ column of $Q$ is multiplied by
$\epsilon_j(e)$. We have an exact sequence: 
\be
\label{sequence_epsilon}
0\longrightarrow \Z^{n-2} \stackrel{q_e^*}{\longrightarrow} \Z^n
\stackrel{g_e}\longrightarrow \Z^2\longrightarrow 0~~, 
\ee 
where $q^*_e$ is the map defined by the transpose $Q^t_e$, while $g_e$ is
the map whose matrix $G_e$ is obtained from $G$ by multiplying its
columns with the signs $\epsilon_j(e)$. This amounts to replacing
the hyperkahler toric generators $\nu_j\in \Z^2$ with the vectors
$\epsilon_j(e) \nu_j$.

{\bf Observation} 
The middle term of (\ref{sequence_epsilon}) has the following meaning.
Consider the `toric' diagonal $T^{2n}$ action:
\be
\label{m_diag}
w_j^{(+)}\rightarrow \Lambda_j w_j^{(+)}~~,~~
w_j^{(-)}\rightarrow \Lambda_{j+n} w_j^{(-)}~~
\ee
which is relevant to the description of $X$ as an intersection of 
quadrics in a toric variety (Subsection 5.1). 
Restricting this action 
to the non-vanishing coordinates $w_j^{(\epsilon_j(e))}$ on the locus $X_e$, 
we obtain:
\be
\label{Y_e_action}
w_j^{(\epsilon_j(e))}\rightarrow \lambda_j w_j^{(\epsilon_j(e))}~~,
\ee
where $\lambda_j=\Lambda_{j+(1-\epsilon_j(e))\frac{n}{2}}$.
Tensoring the middle term of (\ref{sequence_epsilon}) with $U(1)$ 
gives a torus $T^n$, which we let act on $w_j^{(\epsilon_j(e))}$
according to (\ref{Y_e_action}). This allows us to absorb the signs 
$\epsilon_j(e)$ into the definition of $g_e$.

The projectivising $U(1)$ acts as follows on the locus $X_e$:
\be
\label{pepsilon}
w_j^{(\epsilon_j)}\rightarrow e^{2\pi i \phi}w_j^{(\epsilon_j)}~~.
\ee
This action corresponds to a map form $T^1=S^1$ to $T^n$, described by the
lattice map $\gamma:\Z\rightarrow \Z^n$ which takes the generator $1$ of $\Z$ 
into the vector $\left[\begin{array}{c}1\\ \dots \\1 \end{array}\right]$. 
The induced $U(1)$ action on the $T^2$ fiber $T^n/T^{n-2}$ of $X_e$ 
is described by the composite map 
$\alpha_e=g_e\circ\gamma:\Z\rightarrow \Z^2$, 
which takes $1\in \Z$ into the two-vector:
\be
\label{nue}
\nu_e:=\sum_{j=1}^n{\epsilon_j(e)\nu_j}~~.
\ee
As explained in Appendix D, the vector $\nu_e$ cannot vanish, 
which means the that the map $\alpha_e$ is injective. 
Therefore, we have an exact sequence:
\be
\label{epsilon_sequence}
0\longrightarrow \Z \stackrel{\alpha_e}{\longrightarrow} \Z^2
\stackrel{\beta_e}\longrightarrow A_e\longrightarrow 0~~.
\ee
Let us first assume that $U(1)_{eff}=U(1)$, so that the $\Z_2$ subgroup 
$\{-1,1\}$ acts nontrivially on $X$.
In this case, the group $A_e$ will contain torsion if and only if $\Gamma_e$ is
nontrivial. From the results of Appendix $A$, we have
$\Gamma_e=\Z/\Z\nu_e=\Z_{m_e}$, where $m_e$ is the
greatest common divisor of the coordinates of $\nu_e$.  In particular,
the locus $Y_e$ is smooth if and only if the vector $\nu_e$ is
primitive.  The embedding of $\Gamma_e$ into $T^n$ takes the generator
of $\Z_{m_e}$ into the element
$\lambda=(e^{\frac{2\pi i}{m_e}}\dots e^{\frac{2\pi i}{m_e}})$ of $T^n$. 
When combined with
(\ref{Uproj_action}), 
this describes the action of $\Gamma_e$ on the coordinates 
$w_j^{(-\epsilon_j(e))}$ transverse to the locus $Y_e$:
\be
w_j^{(-\epsilon_j(e))}\rightarrow e^{\frac{2\pi i}{m_e}}
w_j^{(-\epsilon_j(e))}~~.
\ee
If the $\Z_2$ subgroup of the projectivising $U(1)$ acts trivially on 
$X$, then one has $U(1)_{eff}=U(1)/\Z_2$ and the singularity group of 
$Y$ along $Y_e$ is the quotient $\Z_{m_e}/\Z_2$, where $m_e$ is determined 
as above. 

The group $\Gamma_e$ can also be computed 
as the trivially acting subgroup of $T^{n-1}=T^{n-2}\times T^1$ on the locus 
$w_1^{(-\epsilon_1(e))}=\dots =w_n^{(-\epsilon_n(e))}=0$ in $X$. 
This is related to the `toric' approach discussed below in Subsection 7.9.
The equivalence of the two methods is explained in Appendix D.

\subsection{Singularities along $Y_V$}

Let us fix a component $Y_j$ of $Y_V$. To identify the singularity type 
along $Y_j$, we must find the trivially acting subgroup of the 
restriction of the $U(1)^{n-1}_{eff}$ 
action to the locus $w_j^{(+)}=w_j^{(-)}=0$.
This gives the system of equations:
\bea
\label{vsys}
\lambda \prod_{\alpha=1}^{n-2}{\lambda_\alpha^{q_k^{(\alpha)}}}=
\lambda \prod_{\alpha=1}^{n-2}{\lambda_\alpha^{-q_k^{(\alpha)}}}=1~~
{\rm~for}~k\neq j~~,
\eea
which encodes the condition that the non-vanishing coordinates $w_k^{(\pm)}$ 
$(k\neq j)$ must be fixed by the $U(1)^{n-1}_{eff}$ action. 
Here $\lambda, \lambda_1,\dots ,\lambda_{n-2}\in U(1)$. Since an 
element $(\lambda, \lambda_1,\dots ,\lambda_{n-2})$ 
which acts trivially on our locus must fix the vectors $a,b$, 
we find the constraint $\lambda^2=1$ i.e. $\lambda\in \{-1,1\}$ 
(see equation \ref{paction}). Hence it suffices to consider only these 
values of $\lambda$ in the system (\ref{vsys}). 
Following the 
discussion of Subsection 7.1., we distinguish the following cases:

(a) If $U(1)_{eff}=U(1)/\Z_2$, then the $\Z_2$ subgroup of the projectivising 
$U(1)$ acts trivially on $X$ and is eliminated when constructing an effective 
action. Therefore, the system (\ref{vsys}) reduces to:
\be
\label{vsys1}
\prod_{\alpha=1}^{n-2}{\lambda_\alpha^{q_k^{(\alpha)}}}=1~~{\rm~for}~k\neq j~~.
\ee
This coincides with the defining system (\ref{Gamma_j_sys}) for the 
singularity group of the cone $X$ along $X_j-\{0\}$. According to the 
results of Subsection 5.4.3, the multiplicative 
group of solutions to this system is isomorphic with $\Z_{m_j}$. 
Therefore, the singularity group of $Y$ along $Y_j$ coincides with the 
singularity group $\Z_{m_j}$ of $X$ along $X_j-\{0\}$. 

(b) If $U(1)_{eff}=U(1)$, then the $\Z_2$ subgroup of the projectivising 
$U(1)$ acts nontrivially on $X$, and we must consider both values 
$\lambda=1$ and $\lambda=-1$. In this case, (\ref{vsys}) reduces to:
\bea
\label{vsys2}
\prod_{\alpha=1}^{n-2}{\lambda_\alpha^{q_k^{(\alpha)}}}~&=&1
~~{\rm~for}~k\neq j~\nn\\
{\rm ~or~}~\\
\prod_{\alpha=1}^{n-2}{\lambda_\alpha^{q_k^{(\alpha)}}}&=&-1~~
{\rm~for}~k\neq j~~.\nn
\eea
In lattice language, we have the map $q^*_j:\Z^{n-2}\rightarrow \Z^{n-1}$
whose transpose matrix is obtained from $Q$ by deleting the $j^{th}$ column. 
This induces a map of tori $q_j^*:T^{n-2}\rightarrow T^{n-1}$, which we 
denote by the same letter. The group $\Gamma_j$ of solutions to (\ref{vsys2}) 
is the subgroup $(q_j^*)^{-1}(im q_j^*\cap \{-1,1\})$, where $\{-1,1\}$ 
is the multiplicative 
$\Z_2$ subgroup of $T^{n-1}$ formed by the elements 
$1:=(1\dots 1)$ and $-1:=(-1\dots -1)$. 
The kernel of $q_j^*$ 
is the group of solutions to the first set of equations in (\ref{vsys2}), 
which by the results of Subsection 5.4.3 is the singularity group $\Z_{m_j}$
of $X$ along $X_j-\{0\}$. We distinguish the following cases:

(bI) $-1\not \in im q_j^*$, i.e. the second system of equations in 
(\ref{vsys2}) does not admit solutions. In this case, we have 
$im q_j^*\cap \{-1,1\}=\{1\}$, and 
$\Gamma_j=ker (q_j^*:T^{n-2}\rightarrow T^{n-1})$ coincides with the 
singularity group $\Z_{m_j}$ of $X$ along $X_j$.

(bII) $-1\in im q_j^*$, i.e. the second system of equations in (\ref{vsys2}) 
admits solutions. In this case, we have an exact sequence:
\be
\label{extension}
1\rightarrow \Z_{m_j}\rightarrow \Gamma_j\stackrel{q_j^*}{\rightarrow} \Z_2
\rightarrow 1~~,
\ee
where the group on the right is the $\Z_2$ subgroup $\{-1,1\}$ of $T^{n-1}$.
This shows that $\Gamma_j$ is an extension of $\Z_2$ by $\Z_{m_j}$. 
Since $Ext^1(\Z_2,\Z_{m_j})=\Z_c$ \cite{HS}, where $c$ is the greatest 
common divisor of $2$ and $m_j$, we distinguish the following possibilities:

(bII.1) $m_j$ is odd. In this case, $Ext^1(\Z_2,\Z_{m_j})=0$ 
and (\ref{extension}) must be the trivial extension 
$\Gamma_j=\Z_2\times \Z_{m_j}$. Since $m_j$ is odd, we also have 
$\Z_2\times \Z_{m_j}\approx \Z_{2m_j}$, so that $\Gamma_j=\Z_{2m_j}$. 
The isomorphism $\Z_2\times \Z_{m_j}
\rightarrow \Z_{2m_j}$ maps an element $(\alpha,u)$ 
into the element $2u+m_j\alpha$, so that $\{0\}\times \Z_{m_j}$ is mapped 
into the subgroup $\{0,2,4,\dots 2(m_j-1)\}$ of $\Z_{2m_j}$; this 
corresponds to solutions of the first system in (\ref{vsys2}). 

The order two element $m_j$ of $\Z_{2m_j}$ does not belong to this subgroup
(since $m_j$ is odd). Hence this element must correspond to an 
element $\boldlambda$ which satisfies the second system in (\ref{vsys}). 
Since $m_j$ has order two in $\Z_{2m_j}$, we must have 
$\boldlambda^2=1$, so that 
$\lambda_\alpha=\pm 1$. 
It follows that the second system in (\ref{vsys2}) must
admit a solution with $\lambda_\alpha\in \{-1,1\}$, provided that it admits 
any solutions at all. For such values of $\lambda_\alpha$, this system
reduces to the condition that the sum of those rows $\alpha$ 
of $Q_j$ for which $\lambda_\alpha=-1$ is a vector all of whose entries are 
odd. Hence if $m_j$ is odd, 
then a necessary and sufficient condition for the second system 
in (\ref{vsys2}) to admit solutions is that there exist a subcollection 
of rows of $Q_j$ whose sum is a vector having only odd entries.

(bII.2) $m_j$ is even, $m_j=2p_j$. In this case, $Ext^1(\Z_2,\Z_{m_j})=\Z_2$, 
and $\Gamma_j$ is either\footnote{If $m_j=2p_j$ is even, then 
the groups $\Z_2\times \Z_{m_j}$ and $\Z_{2m_j}$ are not isomorphic. 
To see why, notice that the first group has three order two elements, 
namely $(0,p_j),(1,0)$ and $(1,p_j)$, while the second group has 
only one order two element (namely $m_j$).}
the trivial extension $\Z_2\times \Z_{m_j}$ 
or the nontrivial extension $\Z_{2m_j}$. 
We shall show that the first case 
is actually forbidden. For this, let us assume that 
$\Gamma_j=\Z_2\times \Z_{m_j}=\Z_2\times \Z_{2p_j}$. 
Remember from Subsection 5.4.3 that the element $s$ of $\Z_{m_j}=\Z_{2p_j}$
corresponds to an element $\boldlambda:=(\lambda_1\dots \lambda_{n-2})$ 
of $T^{n-2}$ which satisfies the first equations in (\ref{vsys2}) as 
well as the equation:
\be
\prod_{\alpha=1}^{n-2}{\lambda_\alpha^{q_j^{(\alpha)}}}=
e^{2\pi i \frac{s}{m_j}}~~.
\ee
Upon choosing $s=p_j=\frac{m_j}{2}$, we obtain an element 
$\boldlambda^{(1)}$ of $T^{n-2}$ which satisfies:
\bea
\label{sys1}
\prod_{\alpha=1}^{n-2}{(\lambda^{(1)}_\alpha)^{q_k^{(\alpha)}}}~&=&1
~~{\rm~for}~k\neq j~\nn\\
\prod_{\alpha=1}^{n-2}{(\lambda^{(1)}_\alpha)^{q_j^{(\alpha)}}}&=&-1~~.
\eea
This corresponds to the element $(0,p_j)$ of $\Z_2\times \Z_{m_j}$
On the other hand, the element $(1,0)$ of $\Z_2\times \Z_{m_j}$ 
corresponds to an element $\boldlambda^{(2)}\in T^{n-2}$ which satisfies 
$\prod_{\alpha=1}^{n-2}{(\lambda^{(2)}_\alpha)^{q_k^{(\alpha)}}}=-1
~~{\rm~for}~k\neq j~$. Since $(1,0)$ has order two in $\Z_2\times \Z_{m_j}$, 
we have $(\boldlambda^{(2)})^2
=((\lambda^{(2)}_1)^2\dots (\lambda^{(2)}_{n-2})^2)
=(1\dots 1)$, so that $\lambda^{(2)}_\alpha=\pm 1$. 
Therefore, the product 
$\prod_{\alpha=1}^{n-2}{(\lambda^{(2)}_\alpha)^{q_j^{(\alpha)}}}$
equals $+1$ or $-1$.
This product cannot 
equal $-1$, since in that case $\boldlambda^{(2)}$ would be a solution  
of the system (\ref{Z2syst}) of Subsection 6.6, thereby contradicting the 
assumption $U(1)_{eff}=U(1)$. Therefore, we must have:
\bea
\label{sys2}
\prod_{\alpha=1}^{n-2}{(\lambda^{(2)}_\alpha)^{q_k^{(\alpha)}}}&=&-1
~~{\rm~for}~k\neq j~\nn\\
\prod_{\alpha=1}^{n-2}{(\lambda^{(2)}_\alpha)^{q_j^{(\alpha)}}}&=&1~~.
\eea
Upon multiplying (\ref{sys1}) and (\ref{sys2}), we obtain an element 
$\boldlambda^{(3)}=\boldlambda^{(1)}\boldlambda^{(2)}=
(\lambda^{(1)}_1\lambda^{(2)}_1\dots \lambda^{(1)}_{n-2}\lambda^{(2)}_{n-2})$
(corresponding to $(1,p_j)\in \Z_2\times \Z_{m_j}$)
which satisfies equations  (\ref{Z2syst}) of Subsection 6.6, thereby 
contradicting the assumption that the $\Z_2$ subgroup of the projectivising 
$U(1)$ acts nontrivially on $X$. This shows that $\Gamma_j$ cannot be 
the trivial extension of $\Z_2$ by $\Z_{m_j}$. Therefore, we must once 
again have $\Gamma_j=\Z_{2m_j}$. Up to an isomorphism, the 
extension (\ref{extension}) maps $u\in \Z_{m_j}$ into $2u\in \Z_{2m_j}$, so
that $\Z_{m_j}$ corresponds to the subgroup $\{0,2,4\dots 2(m_j-1)\}$
of $\Z_{2m_j}$. 
We conclude that case (bII) always leads to $\Gamma_j=\Z_{2m_j}$. 
Combining everything, we obtain the following:

\

{\bf Proposition} Let $\Gamma_j$ demote the singularity group of $Y$ along
$Y_j$. 

(a)If $U(1)_{eff}=U(1)/\Z_2$, then 
$\Gamma_j$ is isomorphic with $\Z_{m_j}$. 

(b)If $U(1)_{eff}=U(1)$, then $\Gamma_j$ 
coincides with $\Z_{m_j}$ or $\Z_{2m_j}$.

\

Upon combining with the results of Appendix A, this shows that the 
integral Smith form of the matrix ${\tilde Q}_j$ (obtained by deleting 
the $j^{th}$ and $(j+n)^{th}$ rows of ${\tilde Q}$) is always of the type:
\be
{\tilde Q}_j^{ismith}=[diag(1\dots 1, t_j),0]~~,
\ee
where $t_j$ is either $m_j$ or $2m_j$. 
To find which of the two possibilities arises in case (b), it suffices to 
determine $t_j$ by computing this integral Smith form. 

It is also easy to identify the action of $\Gamma_j$ on the transverse 
quaternion coordinate $u_j=w_j^{(+)}+\j w_j^{(-)}$. If $\Gamma_j=\Z_{m_j}$, 
then we recover the transverse action on the locus 
$X_j\subset X$, which was determined in Subsection 5.4.3. Therefore, 
the generator of $\Z_{m_j}$ acts as:
\be
u_j\rightarrow e^{\frac{2\pi i}{m_j}}u_j\Leftrightarrow 
w_j^{(\pm)}\rightarrow e^{\pm \frac{2\pi i}{m_j}}w_j^{(\pm)}~~.
\ee
If $\Gamma_j=\Z_{2m_j}$, then $\Z_{m_j}$ is embedded as the subgroup 
$\{0, 2, 4,\dots 2(m_j-1)\}$ of $\Z_{2m_j}$, acting as above.
The generator of $\Z_{2m_j}$ does not 
belong to this subgroup, and corresponds to an element 
$\boldlambda\in T^{n-2}$
which satisfies the second system in (\ref{vsys2}). Its square
$\boldlambda'=\boldlambda^2=
(\lambda_1^2\dots \lambda_{n-2}^2)$ corresponds to the 
generator of $\Z_{m_j}$ and satisfies the first system. 
Form subsection 5.4.3, we also know that $\boldlambda'$ must satisfy:
\be
\prod_{\alpha=1}^{n-2}{(\lambda'_\alpha)^{q_j^{(\alpha)}}}=
e^{\frac{2\pi i}{m_j}}~~.
\ee
Therefore, we must have:
\be
\prod_{\alpha=1}^{n-2}{(\lambda_\alpha)^{q_j^{(\alpha)}}}=
e^{\frac{\pi i}{m_j}}~~.
\ee
This shows that the generator of 
$\Z_{2m_j}$ acts on the transverse coordinates as:
\be
u_j\rightarrow e^{\frac{\pi i}{m_j}}u_j\Leftrightarrow 
w_j^{(\pm)}\rightarrow e^{\pm \frac{\pi i}{m_j}}w_j^{(\pm)}~~.
\ee

\subsection{Singularities above the vertices of $\Delta$}

As mentioned above, the singularity type on the distinguished locus 
may be enhanced at the intersection points between the vertical and 
horizontal spheres. Each such point corresponds to a vertex of $\Delta$. 
Consider a vertex $A$ lying on the principal diagonal $D_j$ supported
by the line $h_j=\{a|a\cdot \nu_j=0\}$.
If $e$ and $e'$ 
are the edges of $\Delta$ meeting at $A$, then their sign vectors 
$\epsilon=\epsilon(e)$ and $\epsilon'=\epsilon(e')$ 
coincide except in position $j$:
\be
\epsilon_k=\epsilon'_k~~{\rm~for~}k\neq j~~,~~
\epsilon_j=-\epsilon'_j~~.
\ee
Accordingly, the matrices $Q_e$ and $Q_{e'}$ 
(defined as in Subsection 7.5) 
coincide except for their $j^{th}$ columns, which have opposite signs.

\begin{figure}[hbtp]
\begin{center}
\scalebox{0.5}{\input{vertex.pstex_t}}
\end{center}
\caption{ Vertices of $\Delta$ may lead to enhanced singularity types.
\label{vertex}}
\end{figure}
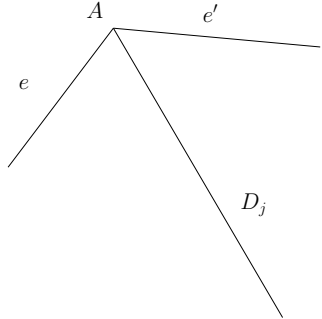

By swapping $e$ and $e'$, we can always assume that $\epsilon_j=+1$ and 
$\epsilon_j=-1$, and we shall do so in what follows.
The spheres $Y_e$ and $Y_{e'}$ are then defined by the constraints 
$w_k^{(-\epsilon_k)}=0$ for $k\neq j$ and $w^{(-)}_j=0$ (for $Y_e$) or 
$w_j^{(+)}=0$ (for $Y_{e'}$). Their intersection point $Y_A$ corresponds to 
$w_k^{(-\epsilon_k)}=0$ for $k\neq j$ 
and $w^{(+)}_j=w^{(-)}_j=0$. Accordingly, the 
$T^{n-1}$ action on the nonvanishing coordinates 
$w_k^{(\epsilon_k)}$ $(k\neq j)$ is described by the $(n-1)\times (n-1)$ 
matrix:
\be
{\bar Q}_A=\left[\begin{array}{cccccc}\epsilon_1col(Q,1)&\dots &
\epsilon_{j-1}col(Q,j-1)~, &\epsilon_{j+1}col(Q,j+1)&\dots&
\epsilon_n col(Q,n)\\1&\dots &1&1&\dots &1\end{array}
\right]~~.
\ee
The singularity type of $Y$ at the point $Y_A$ can now 
be extracted by computing 
the integral Smith form of ${\bar Q}_A$. We note that it is possible for
$Y_A$ to be a singular point even if $Y_e$, $Y_{e'}$ and $Y_j$ are 
smooth in $Y$.
In this case, $Y_A$ is an isolated singular point, 
which induces a codimension {\em six} singularity of ${\cal C}(Y)$.
Such  singularities of the $G_2$ cone are expected to correspond to a 
conformal field theory in five dimensions, whose physics is rather poorly 
understood.

\subsection{Special isometries and good isometries}

Recall from Section 3 that a {\em special isometry} of the $G_2$ cone
${\cal C}(Y)$ is an isometry induced by one of the two-torus 
of triholomorphic isometries
of $X$ (via reduction to $M$, followed by a lift to ${\cal C}(Y)$).
In this subsection, we are interested in characterizing those special 
isometries which fix the cones over $Y_e$ or $Y_j$. Such isometries are 
relevant for understanding the IIA reduction of our models.

\subsubsection{Special isometries which fix ${\cal C}(Y_e)$}

Consider the $U(1)$ subgroup of the two-torus 
of special isometries whose Lie algebra is given by $\R\nu_e\subset \R^2=
Lie(T^2)$, where $\nu_e$ is the vector (\ref{nue}).
The exact sequence (\ref{epsilon_sequence}) shows that this circle group 
corresponds to the specific cycle of the $T^2$ fibers of
$X_e$ which is killed by the projectivising $U(1)$ quotient in order 
to produce the $S^1$ fiber of $Y_e\rightarrow e$.
Therefore, this $U(1)$ subgroup of $T^2$ 
is precisely the group of special isometries which fixes the cone 
${\cal C}(Y_e)$.

\subsubsection{Special isometries which fix ${\cal C}(Y_j)$}

Remember that the $T^2$ fibers of $X\rightarrow \R^6$ 
(i.e. the orbits of the  triholomorphic $T^2$ action on $X$) collapse to 
circles above $X_j-\{0\}$; thus a $U(1)$ subgroup of $T^2$ 
fixes all points $X$. According to Section 5, the Lie algebra of this 
$U(1)$ is spanned by the toric hyperkahler generator $\nu_j$.
It is clear from Section 7.3. that this is the subgroup  which fixes
every point of $Y_j$. Therefore, it also coincides with the 
special isometry subgroup which fixes the cone ${\cal C}(Y_j)$.

\subsubsection{Good special isometries}

When studying the type IIA reduction of $M$-theory on ${\cal C}(Y)$, 
one is interested in 
`good' special isometries of ${\cal C}(Y)$, i.e. special isometries
which fix all of its singular loci.  Let $E_{sing}$ and $V_{sing}$ be
the collections of edges $e$ of $\Delta$ and indices $j=1\dots n$ for
which the associated loci $Y_e$ or $Y_j$ are singular in $Y$.  According to 
the observations made above, `good'
special isometries of $Y$ correspond to the subalgebra of $\R^2=Lie(T^2)$
given by the following intersection of one-dimensional spaces:
\be
G=\left[\cap_{e\in E_{sing}}(\R\nu_e)\right]\cap 
\left[\cap_{j\in V_{sing}}{(\R\nu_j)}\right]~~.
\ee
It is clear that this intersection is generally zero, so that a good 
special isometry is usually impossible to find.

\subsection{On the toric approach to finding the singularities of $Y$}

We end this section with a short discussion of an 
alternate approach to finding the 
singularities of $Y$. This is based on a direct toric analysis starting from 
the embedding of $Y$ in a toric variety, which was discussed in Subsection 6.4.
As we shall see, this somewhat naive method is rather inefficient, which is 
why we prefer the approach discussed above. The content 
of the present subsection is intended 
for readers familiar with toric geometry and 
is not needed for understanding the remainder of the paper. 

If $X$ is a good hyperkahler cone, then one can use an argument similar to that 
of Appendix B to show that the affine variety ${\cal Z}\subset \C^{2n}$ is 
smooth outside the origin. In this case, all singularities of $Y$ must be 
induced from 
singularities of the toric ambient space $\T$ discussed in Subsection 6.4. 
It is well-known that a point $z$ 
can be singular in $\T$ only if some of its homogeneous coordinates 
\cite{Cox}
$z_j=w_j^{(+)}, z_{j+n}=w^{(-)}_j$ vanishes. 
If $V(z)\subset \{1\dots 2n\}$ is the set of indices $j$
associated with the vanishing homogeneous coordinates of $z$, then the
singularity type of $\T$ at $z$ can be computed by considering the map
$ \Z^{r+1}\stackrel{{\tilde q}^*_N}{\longrightarrow}\Z^N$ which is defined as
the projection of ${\tilde q}^*$ of Subsection 6.4. 
onto the sublattice $\Z^N$ of $\Z^{2n}$
associated with the complement $N$ of $V$ in the set $\{1\dots
2n\}$. The associated matrix ${\tilde Q}_N$ is obtained from ${\tilde Q}$ 
by deleting all columns associated with the index set $V$.  
Computing the cokernel
of ${\tilde q}^*_N$ gives a short exact sequence:
\be
0\longrightarrow \Z^{r+1}\stackrel{{\tilde q}^*_N}{\longrightarrow}\Z^N
\stackrel{{\tilde g}_N}{\longrightarrow} A_N\longrightarrow 0~~,
\ee
where the group $A_N$ will generally contain torsion. 
In fact, it follows from Appendix A that the
orbifold group of $\T$ at $z$ coincides with the torsion subgroup
of $A_N$ or with a $\Z_2$ quotient thereof (the second 
possibility arises since the $T^{r+1}$ action on $\C^{2n}$ used to define $\T$ 
may fail to be effective). The former group can be determined by computing 
the integral Smith form of the matrix $Q_N$ of $q_N$, which is 
obtained from ${\tilde Q}$ by keeping only those columns associated with 
the index set $N$, i.e. by deleting all columns associated with $V$. 

Applying this procedure to the toric space $\T$ will typically give a
large collection of singular loci. In general, only a small subset of
these will intersect the twistor space; those which do not are
irrelevant for our purpose. It should be clear from this observation
that the direct toric approach is rather inefficient, since it 
involves a large number of singular loci of $\T$ which do not
intersect $Y$; to find which of them do, one must check existence of 
solutions for a system of quadratic equations obtained by restricting 
(\ref{quadrics}) to each of these singular loci.

The procedure of the previous subsections avoids this problem by using 
information
about the $T^d$ fibration $Y\rightarrow \R^{3d}$ in order to implicitly
solve the complex moment map constraints (\ref{quadrics}).  Since in
this paper we are mainly interested in $G_2$ cones, we have presented this
procedure for the case $d=2$ only; it is not hard to see that this
approach generalizes to higher dimensions.

For the case $d=2$, the analysis of the previous subsections tells us 
precisely which singular loci of $\T$ can intersect the twistor
space:

(a) Singularities along loci given by simultaneous vanishing of
$w_j^{(+)}=z_j$ and $w_j^{(-)}=z_{j+n}$ for some fixed $j$;

(b) Singularities along loci given by the simultaneous vanishing of
precisely one coordinate $w^{(-\epsilon_j)}$ in {\em each} of the 
pairs $w^{(+)}_j, w_j^{(-)}$.

Loci of the first type intersect $Y$ along $Y_j$, while loci of the
second type intersect $Y$ along $Y_\epsilon$. According to our results, 
the second case can still lead to a void intersection; indeed, $Y_\epsilon$ 
is non-void if and only if the sign vector $\epsilon$ equals $\epsilon(e)$ for 
some edge $e$ of $\Delta$. Combining this with the previous discussion allows 
us to extract a `simplified toric approach', which tests for singularities 
only along the distinguished locus. In fact, (a) and (b) tell us that it 
suffices to consider index sets $V,N$ of the following forms:

(1) $V=\{j,j+n\}~~,~~N=\{1\dots 2n\}-V$, for the vertical loci $Y_j$

and

(2) $N=\{j+(1-\epsilon_j(e))\frac{n}{2} |j=1\dots n\}~~,~~V=\{1\dots 2n\}-N$, 
for the horizontal loci $Y_e$.

Upon restricting to these loci, one obtains matrices 
${\tilde Q}_N$ of the form ${\tilde Q}_j=delcols ({\tilde Q}, j, j+n)$
(for $Y_j$) and ${\bar Q}_e=\left[\begin{array}{c}
\epsilon_1(e)col(Q,1)~\dots ~\epsilon_n(e)
col(Q,n)\\1~~~\dots~~~ 1\end{array}\right]$. The singularity type along $Y_e$ 
and $Y_j$ results upon computing the integral Smith forms of these matrices. 
This approach still has the disadvantage that it is unclear 
why the singularity type is always a cyclic group, a fact which is
easy to see in the alternate approach of the previous subsections. 
For the horizontal loci $Y_e$, a 
direct explanation of this fact in `toric' language can be found 
in Appendix D.

\section{Examples}

\subsection{A model with three charges and a good isometry}

Let $X=\H^5///_{0}U(1)^3$, with (good and torsion-free) 
quaternion charge matrix:
\be
Q=\left[ \begin {array}{ccccc} 1&0&0&-2&1\\0&1&0&1&1
\\0&0&1&1&-2\end {array} \right]~~.
\ee
The toric hyperkahler generators $\nu_j$ $(j=1..5)$ 
are given by the columns of:
\be
G=\left[ \begin {array}{ccccc} 1&1&-2&0&-1\\0&3&-3&-1
&-2\end {array} \right]~~. 
\ee
Since all generators are primitive, the hyperkahler cone $X$ is smooth
outside the apex. It is also clear from 
the form of $Q$ that the projectivising $U(1)$ acts effectively on this cone.

\begin{figure}[hbtp]
\begin{center}
\mbox{\epsfxsize=7truecm \epsffile{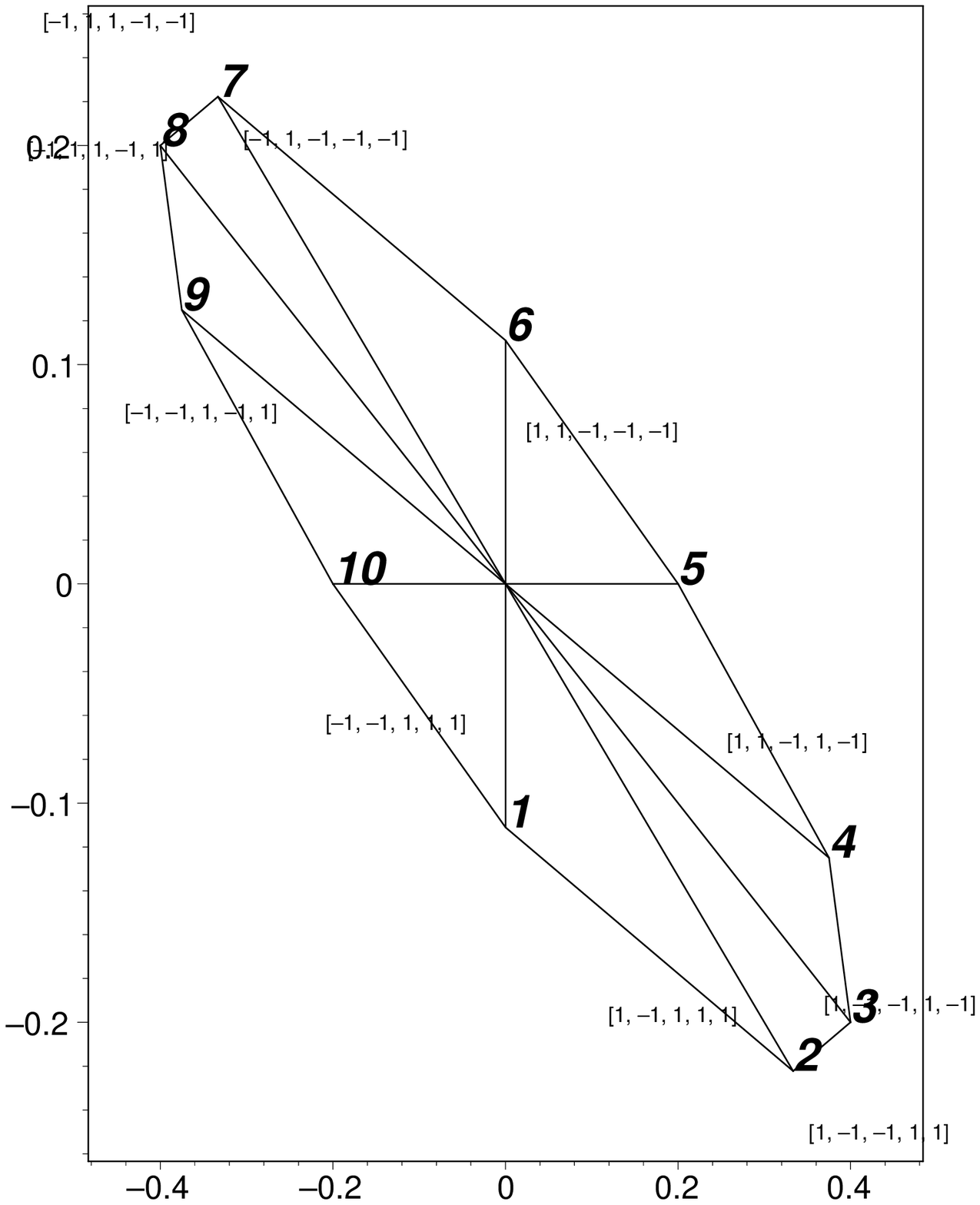}}
\end{center}
\caption{The polygon $\Delta$.\label{e1}}
\end{figure}

With the choice $\zeta=1$, the 
vertices of $\Delta$ are given by the columns of the following matrix: 
\be
V=\left[ \begin {array}{cccccccccc} 0&1/3&2/5&3/8&1/5&0&-1/3&-2/5&-3/8&
-1/5\\-1/9&-2/9&-1/5&-1/8&0&1/9&2/9&1/5&1/8&0
\end {array} \right]~~. 
\ee
This is shown in figure \ref{e1} with the associated signs for its edges.
The principal diagonals $D_j\subset h_j=\{a\cdot \nu_j=0\}$ 
are given by the segments 
$D_1=[1,6]~,~D_2=[4,9]~,~ D_3=[2,7]~,~ D_4=[5,10]~,~ D_5=[3,8]$.

One obtains two horizontal spheres of 
$\Z_3$ singularities associated with the opposite edges 
$e=[1,2]$ and $-e=[7,6]$ ($\nu_e=\left[\begin{array}{c}-3\\-9\end{array}
\right]$), and one vertical sphere\footnote{Since $X$ has no 
singularities outside of its apex, we see that the singularity type 
along $Y_j$ coincides with $\Z_{m_j}=\Z_1=\{0\}$ for $j=1,3$ and 
with $\Z_{2m_j}=\Z_2$ for $j=2$, where $m_1=m_2=m_3=1$ characterize 
the smooth loci $X_j\subset X$. This illustrates the two possibilities 
in case (b) of the Proposition of Subsection 7.6.} of $\Z_2$ singularities 
from the principal diagonal $D_2$. The horizontal singularities are 
related by the antipodal map of the $S^2$ fibration $Y\rightarrow M$. 
One also obtains $\Z_9,\Z_6,\Z_7$ and $\Z_5$ singularities at the pairs 
of opposite vertices $\{1,6\}$, $\{2,7\}$, $\{3,8\}$ and $\{4,9\}$ 
respectively.
Since
$\nu_e=-3\nu_2=
\left[ \begin {array}{c} -3\\-9\end {array}\right]$, 
the $G_2$ cone admits a $U(1)$ of good isometries with Lie algebra 
$\R\nu_2$.

\subsection{An example with three charges and without a good isometry }

Let us consider $X=\H^5///_{0}U(1)^3$, with:
\bea
Q=\left[ \begin {array}{ccccc} 1&0&0&3&1\\0&1&0&1&2
\\0&0&1&3&3\end {array} \right]~~,~~ 
G= \left[ \begin {array}{ccccc} 1&2&3&0&-1\\0&5&6&1&-3
\end {array} \right]~~. 
\eea 
Since the third column $\nu_3$ of $G$ fails to be primitive, the toric 
hyperkahler cone $X$ has $\Z_3$ singularities along the locus $u_3=0$. 
The projectivising $U(1)$ acts effectively on $X$.

\begin{figure}[hbtp]
\begin{center}
\mbox{\epsfxsize=7truecm \epsffile{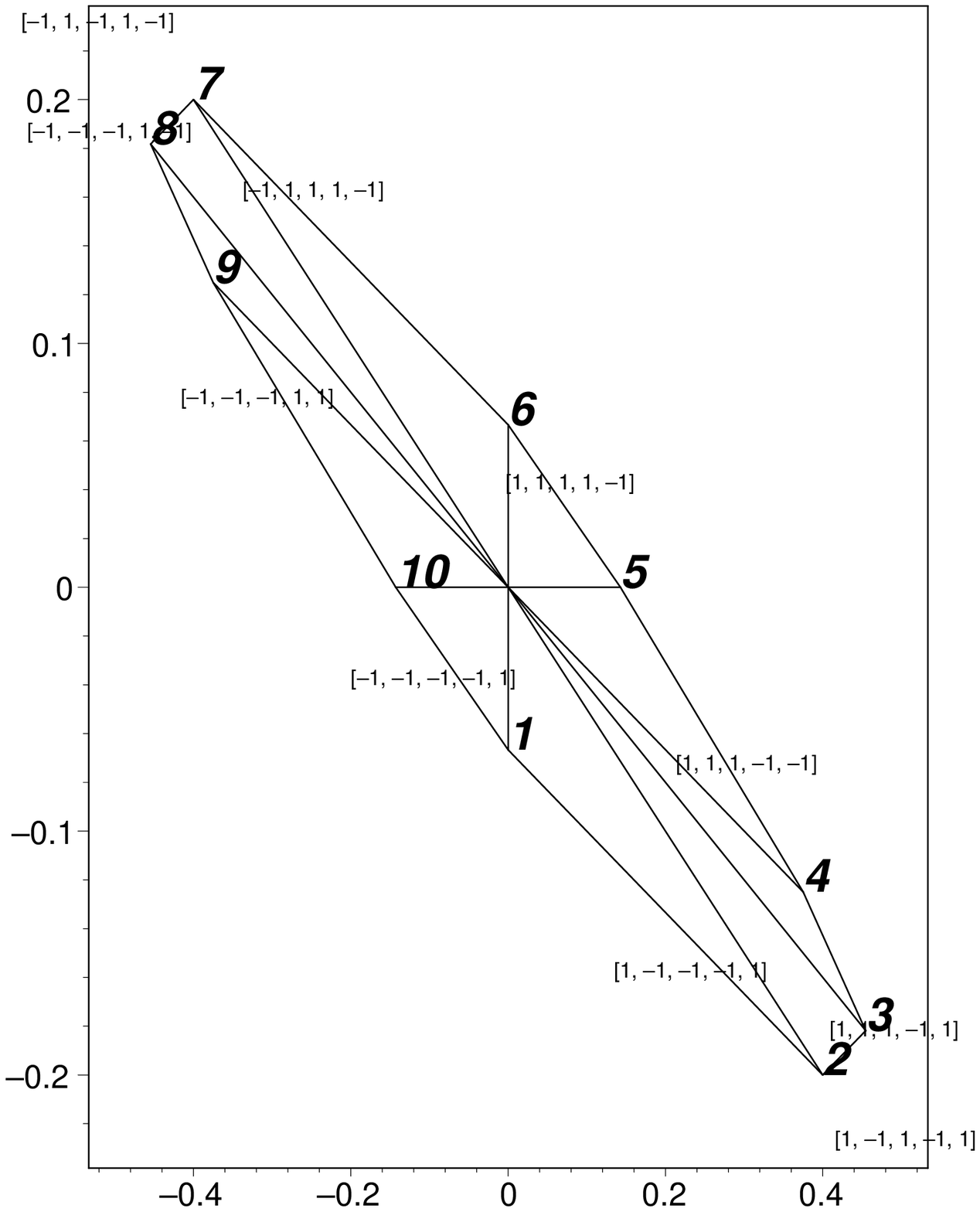}}
\end{center}
\caption{The polygon $\Delta$.\label{e4}}
\end{figure}

The vertices of $\Delta$ (for $\zeta=1$) are given by the columns of the 
matrix:
\be
V=\left[ \begin {array}{cccccccccc} 0&2/5&5/11&3/8&1/7&0&-2/5
&-5/11&-3/8&-1/7\\-1/15&-1/5&-2/11&-1/8&
0&1/15&1/5&2/11&1/8&0\end {array} \right]~~.
\ee
This is drawn in figure \ref{e4}, together with the sign vectors of its edges. 
The principal diagonals $D_1\dots D_5$ are given by the segments
$[1,6],[3,8],[2,7],[5,10],[4,9]$, in this order.

One obtains a pair of horizontal spheres of $\Z_5$ singularities from the 
opposite edges $e=[1,2]$ and $-e=[6,7]$ (with $\nu_e=5\nu_5=
\left[ \begin {array}{c} -5\\-15
\end {array} \right]$)
as well as vertical spheres of $\Z_3$
and $\Z_2$ singularities\footnote{Notice that the singularity type of $Y$ 
along $Y_j$ coincides with that of $X$ along $X_j$ for $j\neq 5$, 
but it is `doubled' (from $\Z_1=\{0\}$ to $\Z_2$) for $j=5$.}
from the diagonals $D_3=[2,7]$ and $D_5=[4,9]$.
One also has $\Z_{15},\Z_{11},\Z_9,\Z_7$ and $\Z_6$ 
singularities at the points 
associated with the pairs of vertices $\{1,6\},\{ 2,7\},\{3,8\}, \{4,9\}$ 
and $\{5,10\}$. The model has no good isometries.

\subsection{A model with two families of singularities}
For this example, we take $X=\H^4///_{0}U(1)^2$, with:
\be
Q=\left[ \begin {array}{cccc} 1&0&2&1\\0&1&1&2
\end {array} \right]~~,~~G=\left[ \begin {array}{cccc} 1&2&0&-1\\0&3&1&-2
\end {array} \right]~~.
\ee  
The toric hyperkahler cone is smooth outside its apex. 
For this model, the $\Z_2$ subgroup $\{-1,1\}$ of the projectivising 
$U(1)$ acts trivially on $X$. Indeed, the sum of the two rows of $Q$ 
is a vector all of whose entries are odd. As explained above,
this must be taken into account 
when computing the singularities of the twistor space. 

\begin{figure}[hbtp]
\begin{center}
\mbox{\epsfxsize=7truecm \epsffile{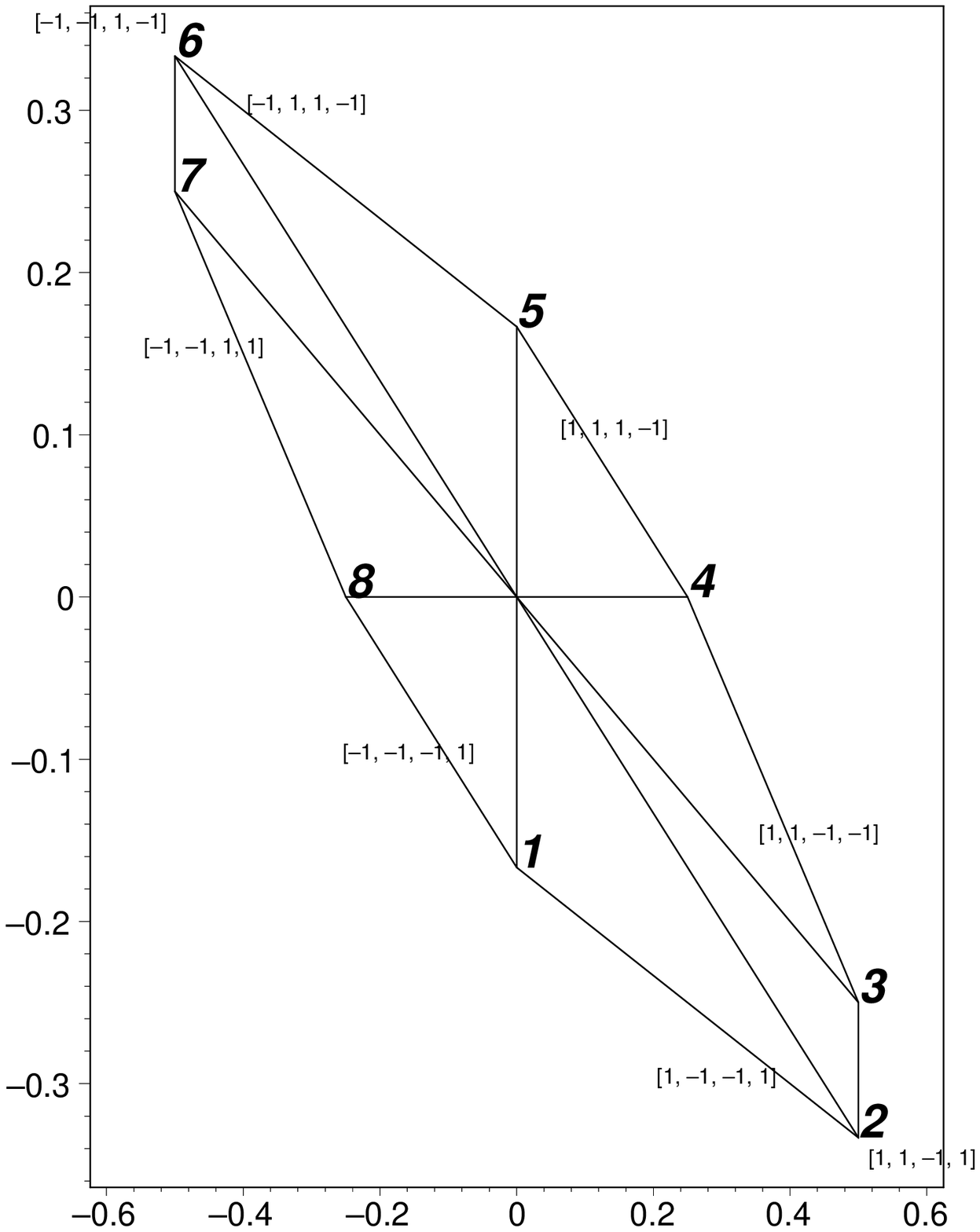}}
\end{center}
\caption{The polygon $\Delta$.\label{2brane}}
\end{figure}

With the choice $\zeta=1$, the 
vertices of $\Delta$ are given by the columns of the matrix:
\be
V=\left[ \begin {array}{cccccccc} 0&1/2&1/2&1/4&0&-1/2&-1/2&-1/4
\\-1/6&-1/3&-1/4&0&1/6&1/3&1/4&0\end {array}
 \right]~~.
\ee
This polygon is shown in figure \ref{2brane}, together with the sign
vectors of its edges. 
The principal diagonals $D_1\dots D_4$ 
correspond to the segments $[1,5],[2,6],[4,8],[3,7]$. 

One obtains two $S^2$'s worth of $\Z_2$ singularities on the horizontal 
loci associated with the opposite edges $e=[3,4]$ and $-e=[7,8]$. 
The associated vector $\nu_e$ equals 
$\left[ \begin {array}{c} 4
\\4\end {array} \right]$. 
The model has no 
spheres of singularities on the vertical locus, but has 
isolated $\Z_3$ singularities at the points corresponding to the 
pairs of opposite vertices $\{1,5\}$ and $\{2,6\}$. The vertices $3,7$ 
correspond to $\Z_2$ singularities whose transverse action matches that 
of a generic point along the $S^2$'s associated with $[3,4]$ and $[7,8]$.
The points associated with the opposite vertices $\{4,8\}$ are smooth. 
This model has an $S^1$ of good isometries, with Lie algebra 
generated by the vector $\left[ \begin {array}{c} 1\\1\end {array} \right]$. 

\section{Physical interpretation}

The physical interpretation of our models is very similar to that 
of \cite{Witten_Acharya}, and can be extracted by using similar arguments.
Let us start with the Abelian symmetries originating from the 
supergravity 3-form.  
Using the results of \cite{BGMR,Bielawski}, 
we find that $b_2({\cal C}(Y))=n-1$,
where $n$ is the number appearing in the 
construction  $X=\H^n///_0U(1)^{n-2}$ of the associated hyperkahler cone.
This gives a 
$U(1)^{n-1}$ Abelian symmetry produced by the reduction of 
the M-theory 3-form; as in \cite{Witten_Acharya}, this symmetry is 
broken to $U(1)^{n-2}$ for the deformed $G_2$ metric (\ref{G2deformed}).
Exactly as in \cite{Witten_Acharya}, we can apply the argument 
of \cite{Witten_anom} to conclude 
that $M$-theory on our $G_2$ cones produces chiral
fermions localized at the apex of the $G_2$ cone and charged under the 
Abelian $U(1)^{n-1}$ component of the resulting symmetry group.

Our models also produce nonabelian symmetries originating from 
codimension four singularities of the $G_2$ metric. 
As is well-known (see \cite{KKV} and references therein) quantum field
theories can be geometrically engineered using local description of
singularities of type IIA compactification spaces. 
As discussed in \cite{Achar}, the same idea
extends to M-theory compactified on singular spaces of $G_2$ holonomy. 
In the paper cited, it was argued that
$A-D-E$ singularities in codimension four lead to an $N=1$
super Yang-Mills theory carrying corresponding $A-D-E$ gauge group 
and localized at the singularity.
As explained in Section 7 and illustrated 
in Section 8, 
codimension four singularities in 
our models are always cyclic,  and therefore 
of type $A_{n-1}$. This gives an $SU(n)$ gauge theory 
living at every codimension four singular locus with 
$\Z_n$ singularity type. The location of 
such singularities and the their 
singularity type can be extracted with the methods of Section 7. 

Codimension six singularities of the $G_2$ metric, 
if present, are less well understood. 
They correspond to isolated singular points of 
the twistor space,  
and are presumably related to certain conformal field theories 
in five dimensions. We have encountered such singularities in the 
examples of Section 8.  

Since our $G_2$ spaces have a two-torus of
isometries, we can choose a $U(1) \subset T^2$ and dimensionally
reduce M-theory to IIA. The resulting type IIA  background
contains a nonvanishing RR one-form and hence will correspond to a
configuration of D6-branes. As in \cite{Witten_Acharya, AW,
Townsend} the D-brane  worldvolumes are comprised of
$\R^{3,1}$, the radial direction $r$ and a singular locus $Y_e$ or $Y_j$ 
in $Y$, which is preserved (as a set) by the $U(1)$ isometry. 
From Subsection 7.8.3, we know that 
good isometries (namely isometries which fix all points of every
singular locus) are very rare. If such isometries do not exist, then the
IIA background is strongly coupled along those six-branes associated with 
singular loci which are not pointwise invariant under the isometry chosen 
for the reduction. If a good isometry exists, then reducing through it 
leads to a IIA solution which describes a weakly coupled system of 
D6-branes. In fact we did
find such models in the examples of Subsections 8.1 and 8.3 (and many more 
such models can be constructed upon using our criteria). 
For the example of Subsection 8.1., the vector 
$\nu_e$ associated with the edge $e = [1,2]$ was equal to
$-3 \nu_2$. Reducing along the isometry generated by $\nu_2$ we
obtain weakly-coupled D6-branes associated with both the horizontal and the
vertical loci. In the example of Subsection 8.3, one 
has a good isometry generated by the vector $\nu_e$ with 
$e = [3,4]$, since there are no singularities on the vertical locus.

After reduction to IIA, the residual $U(1)$ isometry 
of the background can be used to obtain a T-dual IIB description. 
From the discussion of Section 5, we know that the $T^2$ fibers 
describing the triholomorphic orbits in the hyperkahler cone $X$
are collapsed to circles on the loci $X_j$. When reducing $X_j$ to $Y_j$ 
through the projectivising $U(1)$ action, these $S^1$ fibers become 
circles along the sphere $Y_j$ (namely the circle fibers of the 
fibration of $Y_j$ over the principal diagonal $D_j$ of $\Delta$.)
Therefore, the T-duality mapping our system to IIB acts along a 
worldvolume direction of the vertical branes. This converts a von Neumann 
direction into a Dirichlet direction, thereby leading to a IIB fivebrane. 
Since we are dualizing a von Neumann into a Dirichlet direction, we 
expect that T-dualization 
of the vertical  6-branes will produce {\em delocalized} 5-branes in the 
IIB background; the delocalization should occur along the T-dual of the 
$S^1$ fiber of $Y_j\rightarrow D_j$. A similar argument applies to those 
IIA 6-branes which are associated with horizontal loci $Y_e$. In this case, 
we know from Section 5 that the triholomorphic $T^2$ orbits of $X$ 
are not collapsed on the loci $X_e$. However, the projectivising $U(1)$ 
now acts along the $T^2$ fibers of $X_e$, which therefore descend to circles 
upon reduction to $Y_e$; these are the $S^1$ fibers of $Y_e\rightarrow e$. 
Once again, we find that $T$-dualization is performed along a worldvolume 
direction, and therefore vertical IIA 6-branes will correspond to 
delocalized five-branes in IIB. If the IIA reduction is performed through a 
good isometry, then the weak coupling arguments given above show that the 
IIA solution describes a system of D6-branes, 
while the T-dual 
IIB background describes a system of delocalized D5-branes. 
In this case, the IIA description 
provides an alternate explanation of the origin of chiral fermions.
If the isometry used in the reduction is not good (which is necessarily the 
case for models which do not admit a good isometry) then we obtain 
a set of strongly coupled `branes' in IIA and we expect a set of strongly 
coupled dual loci in IIB. The physical interpretation of these loci 
is less clear, though the associated supergravity solutions 
can be analyzed explicitly \cite{metrics}. However, it is natural to 
extend our conclusions to these cases, and propose that the resulting 
M-theory backgrounds lead to nonabelian gauge theories whose gauge groups 
are constructed in the manner discussed above.

\section{Conclusions}

By using the relation between an Einstein self-dual orbifold $M$, its twistor 
space $Y$ and its hyperkahler cone $X$, we developed methods to identify the 
location and type of singularities of the twistor space of a 
compact, Einstein self-dual orbifold of positive scalar curvature which
admits a two-torus of isometries. This amounts to considering the class 
of models for which the hyperkahler cone $X$ of $M$ is `toric hyperkahler' 
in the sense of \cite{BD}, i.e. admits a presentation as a toral hyperkahler quotient $X=\H^n///_0U(1)^{n-2}$. This is precisely the class of ESD spaces 
for which one expects a simplification in view of the work of \cite{HKLR}.
Upon combining our methods with the construction 
of \cite{BS,GP}, we obtained an algorithm for analyzing the associated $G_2$ 
cones ${\cal C}(M)$, 
which allows us to extract the low energy gauge group
produced by  $M$-theory on such backgrounds. 
This identifies the basic physics of such models, which form 
a vast generalization of those considered in \cite{Witten_Acharya}. 
Since the $G_2$ cones belonging to this family admit a $T^2$ of 
isometries, such $M$-theory 
backgrounds have a two-torus of T-dual type IIA and IIB 
descriptions, whose physics describes systems of strongly and/or weakly 
coupled 6-branes, T-dual to systems of delocalized  fivebranes. 
By using abstract geometric arguments, we showed that the generic model in 
this family does not allow for a  description in terms of weakly coupled 
D-branes only, which means that its type II reduction is not amenable to  
conformal field theory techniques.
Those (non-generic) models which {\em do} admit such a description can be 
identified by a simple criterion, and we presented some examples of this type.
The techniques developed in this paper are computationally quite effective
and they allow for an explicit analysis of any representative 
of this large family of models (the complexity of the computations involved 
depends on the number $n$). 

Since the ESD spaces of interest admit a two-torus of isometries, 
they belong to the class recently considered in the work of Calderbank 
and Pedersen, who wrote down the explicit ESD metric for the most general space
of this type. When combined with the construction of \cite{BS,GP}, this allows
for an explicit description of the associated $G_2$ metrics, and therefore 
of the IIA backgrounds obtained by performing the KK reduction, together with
their type IIB duals. This analysis provides an independent confirmation 
of the conclusions of the present paper, as we will 
show in \cite{metrics}.

\

\acknowledgments{The authors thank M. Rocek for collaboration 
in the initial stages of this project and for reading the manuscript.
C.~I.~L thanks the CIT-USC Center for Theoretical Physics 
for hospitality during part of the preparation of this work. He is indebted
to I. Bars, R. Corrado, C. Romelsberger, P. Berglund and 
A.~Brandhuber for interesting conversations 
and for providing a stimulating atmosphere.
This work was supported by the Research Foundation under NSF 
grant PHY-0098527. }

\appendix

\section{On maps of lattices and associated maps of tori}

This appendix collects a few basic facts about maps of lattices.
Some of the properties described below are well-known and used intensively, for 
example, in toric geometry \cite{Oda, Fulton,Danilov, Cox, Cox_review}. 
By definition, a lattice 
is a free, finitely generated Abelian group, which is the same as a free, 
finitely generated $\Z$-module. A map of lattices is a group morphism 
between such objects, which is the same as a morphism of $\Z$-modules.
We remind the reader that the ring $\Z$ of 
integers is a PID (principal ideal domain)\footnote{Recall that a 
PID is a commutative ring without divisors of zero, 
for which every ideal is principal, i.e. generated by a single element.}, 
which is why the homological
algebra of $\Z$-modules has a particularly simple form \cite{HS}. 
In particular, a finitely generated 
$\Z$-module is projective if and only if it is free, which is equivalent with 
it being torsion free. 

\subsection{The structure theorem of lattice maps}

The following result is well-known:

\begin{Theorem} 
Given a lattice map $f:\Z^r\rightarrow \Z^n$, one can always 
find integral 
bases $v_1\dots v_r$ of $\Z^r$ and $u_1\dots u_n$ of $\Z^n$ such that 
$f(v_\alpha)=t_\alpha u_\alpha$ for all $\alpha=1\dots r$, where $t_\alpha$ 
are non-negative integers 
satisfying the division relations $t_1|t_2|\dots |t_r$. 
\end{Theorem}

This is simply the structure theorem of finitely-generated Abelian groups, 
adapted to our context. 
If $e'_\alpha$ and $e_j$ are the canonical bases of $\Z^r$ and $\Z^n$, then 
$f$ can be described by an $n\times r$ matrix $F_{j\alpha}=f_j^{(\alpha)}$, 
whose entries are defined through:
\be
f(e'_\alpha)=\sum_{j=1}^n{f_j^{(\alpha)}e_j}~~.
\ee
The theorem says that there exist matrices $U\in GL(n,\Z)$ and 
$V\in GL(r,\Z)$ such that $U^{-1}FV$ is in {\em integral Smith form}:
\be
U^{-1}FV=F^{ismith}=\left[\begin{array}{cccc}t_1&0&0&0\\0&t_2&0&0\
\\\cdots&\cdots&\cdots&\cdots\end{array}\right]~~.
\ee
Indeed, $V$ is the matrix whose columns are $v_\alpha$ and $U$ is the matrix 
whose columns are $u_j$.
We shall mostly be interested in the case $r\leq n$.
Then the product $\prod_{j=\alpha}^r{t_\alpha}$ equals the so-called 
$r^{th}$ {\em discriminantal divisor} $\g (F)$ of the matrix $F$, 
i.e. the greatest common divisor of all $r\times r$ minor determinants of $F$.

The integers $t_1\dots t_r$ are independent of the choice of bases 
$v_\alpha$, $u_j$ with properties as in the theorem, and are called 
the {\em torsion coefficients} of the lattice map $f$ (also known 
as the {\em invariant factors} of the associated matrix $F$). 
They coincide with the torsion coefficients of the torsion subgroup of the 
cokernel of $f$, i.e. one has a group 
isomorphism $\Z^n/f(\Z^r)=\Z_{t_1}\oplus \cdots \oplus 
\Z_{t_r}\oplus \Z^{n-r}$, where $\Z_t:=\Z/{t\Z}$ and 
$\Z^{n-r}$ is defined to be zero if $n\leq r$. If $t=0$, then $\Z_0$ is 
a copy of $\Z$. If $t=1$, then $\Z_1=\Z/\Z=0$.  
It is clear that the map $f$ is injective if and only if $t_r> 0$, 
in which case all torsion coefficients are non-vanishing. 

The following is an immediate consequence of the structure theorem: 

\begin{Proposition} Let $f:\Z^r\rightarrow \Z^n$ be an {\em injective} map 
of lattices 
(thus $r\leq n$ and 
$f(\Z^r)$ is a sublattice of $\Z^n$). Then the following are equivalent:

(a) $coker(f)=\Z^n/f(\Z^r)$ is torsion free (and thus free, i.e. a 
copy of the lattice $\Z^d$, where $d=n-r$)

(b) $f(\Z^r)$ is a direct summand in $\Z^n$

(c) All torsion coefficients of $f$ equal one

(d) The $r^{th}$ discriminantal divisor of the matrix $F$ equals one.

In this case, one has a short exact sequence:
\be
\label{sex}
0\longrightarrow \Z^r\stackrel{f}{\longrightarrow}\Z^n\stackrel{g}{\longrightarrow}\Z^d\longrightarrow 0~~.
\ee
\end{Proposition}

Another basic result is:

\begin{Proposition} Let $g:\Z^n\rightarrow \Z^d$ be a map of lattices. 
Then the kernel of $g$ is a lattice.
\end{Proposition}

{\bf Proof} The ring of integers is a PID, so that all projective 
$\Z$-modules are free, and any submodule of 
a projective module is free \cite{HS}.

\paragraph{Observation} Note that any short exact sequence of lattices 
(\ref{sex}) is split exact. Indeed, $Ext^1(\Z^r, \Z^d)$ vanishes since 
$\Z^r$ is a projective module. In particular, the dual sequence 
\be
\label{sexdual}
0\longrightarrow (\Z^d)^*\stackrel{g^*}{\longrightarrow}(\Z^n)^*
\stackrel{f^*}{\longrightarrow}(\Z^r)^*\longrightarrow 0~~,
\ee
where $(\Z^n)^*:=Hom_\Z(\Z^n,\Z)$ etc. is also exact. 

\subsection{Some properties of torus maps}

A map of lattices $f:\Z^r\rightarrow \Z^n$ induces a map of tori
$f:T^r\rightarrow T^n$ upon tensoring with the Abelian group 
$U(1)$. If $f$ is injective, then the short exact sequence:
\be
\label{Asex}
0\longrightarrow \Z^r
\stackrel{f}{\longrightarrow}\Z^n\stackrel{g}{\longrightarrow}A\longrightarrow 0~~
\ee
(where $A=coker(f)$) induces a long exact Tor sequence \cite{HS} which 
collapses to four terms since $\Z^n$ is flat so that 
$Tor^{\Z}(\Z^n, U(1))$ vanishes:
\be
\label{torseq}
0\longrightarrow \Gamma 
\longrightarrow T^r\stackrel{f}{\longrightarrow}T^n\stackrel{g}{\longrightarrow}A\otimes_{\Z} U(1)\longrightarrow 0~~,
\ee
where $\Gamma= Tor^{\Z}(A,U(1))$.
The sequence (\ref{torseq}) collapses to a short exact sequence if and 
only if $A$ is torsion-free; by the result above, this happens precisely
when $f$ has trivial torsion coefficients, in which case one has $A=\Z^d$ 
and $A\otimes_\Z U(1)=T^d$. Otherwise, (\ref{torseq}) shows that 
the induced map of tori $f:T^r\rightarrow T^n$ will fail to be injective; 
its (finite) kernel $\Gamma$ is the torsion group $Tor^{\Z}(A,U(1))$.
Let $t_1\dots t_r$ be the torsion coefficients of $f$ and $v_\alpha,u_j$ 
be bases of $\Z^r$ and $\Z^n$ such that $f(v_\alpha)=t_\alpha u_\alpha$.
We let $T(A)$ denote the torsion subgroup of $A$.

\begin{Proposition} Assume that $f$ is injective. 
Then $\Gamma$ and the torsion subgroup $T(A)$ are both 
isomorphic with $\Z_{t_1}\times \dots \times \Z_{t_r}$. An isomorphism 
$\psi:\Z_{t_1}\times \dots \times \Z_{t_r}\rightarrow \Gamma$ is given by:
\be
\label{Gamma_emb}
\psi(s_1\dots s_r)=
(e^{2\pi i\sum_{\alpha=1}^r{v^{(\beta)}_\alpha \frac{s_\alpha}{t_\alpha}}})_
{\beta=1\dots r}~~,
\ee
where $v_\alpha=\sum_{\beta=1}^r{v_\alpha^{(\beta)} e'_\beta}$, 
with $(e'_\beta)$ the canonical basis of $\Z^r$. Here $\Gamma$ is 
viewed as a subgroup of $T^r=U(1)^r$, embedded by inclusion. 
\label{fptor}
\end{Proposition}

{\bf Proof:} Writing $f(e'_\alpha)=\sum_{j=1}^{n}{f_j^{(\alpha)}e_j}$, 
the embedding $T^r\rightarrow T^n$ is given by:
\be
\label{action}
(\lambda_\alpha)_{\alpha =1\dots r} \rightarrow 
(\prod_{\alpha=1}^{r}{\lambda_\alpha^{f_j^{(\alpha)}}})_{j=1\dots n}~~,
\ee
where $\lambda_\alpha\in U(1)$. 
The group $\Gamma$  consists of the solutions $\lambda\in T^r$ of the system:
\be
\label{system}
\prod_{\alpha=1}^{r}{\lambda_\alpha^{f_j^{(\alpha)}}}=1 {\rm~for~all~}~
j=1\dots n~~.  
\ee 
Since $f$ is injective, the solution set is finite. 
Writing $\lambda_\beta=e^{2 \pi i \phi_\beta}$ (with $\phi_\beta\in \R/\Z$) 
reduces this system to: 
\be
\label{system1}
f(\phi)=0 ~~{\rm~~in~~}(\R/\Z)^n~~.  
\ee 
It is clear that all solutions 
$\phi=(\phi_1 \dots \phi_r)$ belong to $(\Q/\Z)^r$. With the structure 
given by addition, 
these form a group isomorphic with $\Gamma$. Using the structure theorem for
$f$, we write 
$\phi=\sum_{\alpha=1}^r{\phi^{(\alpha)}v_\alpha}$, which reduces (\ref{system1}) to the form:
\be
t_\alpha\phi^{(\alpha)}=0~~{\rm~~in~~}\R/\Z~{\rm~~for~all~}\alpha~~.
\ee
The general solution is 
$\phi^{(\alpha)}=\frac{s_\alpha}{t_\alpha}$, 
with $s_\alpha\in \Z_{t_\alpha}$ (this is well-defined 
as an element of $\Q/\Z$). Since $\phi_\beta=\sum_{\alpha}{\phi^{(\alpha)}
v_\alpha^{(\beta)}}$, we find that
$\Gamma$ coincides with $\Z_{t_1}\times \dots \times \Z_{t_r}$ 
via the isomorphism $\psi$. 
The remaining statement follows from $T(A)=T(\Z^n/f(\Z^r))=
\langle u_1\dots u_r\rangle_{\Z}
/\langle t_1u_1\dots t_ru_r\rangle_{\Z}=
\Z_{t_1}\times \dots \times \Z_{t_r}$.

\

Consider the short exact sequence of lattices (\ref{sex})
and a partition of $\{1\dots n\}$  into subsets $V$ and $N$, 
where $V$ has at most $d$ 
(and hence $N$ has at least $r$) elements.  Associated to this 
partition is a direct sum decomposition 
$\Z^n=\Z^V\oplus \Z^N$, with $\Z^V=\oplus_{j\in V}{\Z e_j}$ and 
$\Z^N=\oplus_{j\in N}{\Z e_j}$, where 
$(e_j)$ is the canonical basis of $\Z^n$. 
We let $j_V$ and $j_N$ denote the injections 
of $\Z^V$ and $\Z^N$ into $\Z^n$ and 
$p_V,p_N$ the projections of $\Z^n$ onto these sublattices. 
We have a split short exact sequence:
\be
0\longrightarrow \Z^V\stackrel{j_V}{\longrightarrow}\Z^n
\stackrel{p_N}{\longrightarrow} \Z^N\longrightarrow 0~~,
\ee
with left and right splittings given by $p_V$ and $j_N$: 
\be
p_V\circ j_V=id_{\Z^V}~~,~~p_N\circ j_N=id_{\Z^N}~~.
\ee
Let us define $f_N:=p_N\circ f$ and $g_V:=g\circ j_V$. 
Computing the cokernel of $f_N$ gives an exact sequence: 
\be
\label{horizontal_sequence2}
\Z^r\stackrel{f_N}{\longrightarrow}\Z^N\stackrel{\beta}
{\longrightarrow}A\longrightarrow 0~~,
\ee
where the group $A=\Z^N/im(f_N)$ may have torsion, even though the cokernel 
of $f$ is torsion-free. The situation is summarized in figure \ref{proj}.

\begin{figure}[hbtp]
\begin{center}
\scalebox{0.4}{\input{proj.pstex_t}}
\end{center}
\caption{Exact sequences.\label{proj}}
\end{figure}

\begin{Proposition} \label{Prop} The following are equivalent: 

(a) The map $f_N$ is injective

(b) The map $g_V$ is injective

In this case, 
there exists a unique morphism $\alpha$ from $\Z^d$ to $A$ which closes 
figure \ref{proj} to a commutative diagram. 
Moreover, the resulting vertical sequence:
\be
\label{vertical_sequence}
0\longrightarrow \Z^V\stackrel{g_V}{\longrightarrow}\Z^d\stackrel{\alpha}{\longrightarrow}A\longrightarrow 0
\ee
is exact. This allows us to compute $A=coker(f_N)$ as the cokernel of $g_V$.
\label{goodseq}
\end{Proposition}

{\bf Proof:} Assume that $f_N$ is injective. If $x$ belongs to $ker g_V$, then 
$g(j_V(x))=g_V(x)=0$ so $j_V(x)\in ker g=im f$, which gives an element 
$y\in \Z^r$ such that $j_V(x)=f(y)$. Now $f_N(y)=p_N(f(y))=p_N(j_V(x))=0$
since $p_N\circ j_V$ vanishes. Hence $f_N(y)=0$ which implies $y=0$ 
by injectivity of $f_N$. Thus $j_V(x)=0$ and so $x=0$ since $j_V$ is injective.
Therefore $ker g_V=0$ and $g_V$ is injective. 

To prove the converse implication, it suffices to notice that the diagram 
in figure \ref{proj} is symmetric with respect to the northwest-southeast 
diagonal; hence the proof of $(b)\Rightarrow (a)$ is formally identical 
to that of $(a)\Rightarrow (b)$.

Let us now assume that (a) (and thus (b)) hold.
Then existence and uniqueness of $\alpha$ follows by applying the 
3-lemma to the two horizontal exact sequences.
Surjectivity of $\alpha$ follows trivially from surjectivity of 
$\beta$ and $p_N$ and commutativity of 
the lower right square.

It remains to prove exactness of (\ref{vertical_sequence}) at the
middle. For this, consider an element $x$ in the kernel of $\alpha$.
Then $x=g(y)$ for some $y$ in $\Z^n$, and
$\beta(p_N(y))=\alpha(g(y))=\alpha(x)=0$, which shows that $p_N(y)$ lies
in $ker \beta=im f_N$.  Thus $p_N(y)=f_N(z)$ for some $z$ in
$\Z^r$. Since $f_N=p_N\circ f$, this gives $y=f(z)+t$, 
with $t\in ker p_N=im j_V$, so that $y=f(z)+j_V(s)$ 
for some $s$ in $\Z^V$. Hence
$x=g(y)=g(j_V(s))=g_V(s)$, where we used exactness of the upper
horizontal sequence as well as commutativity of the upper right
square. It follows that $ker\alpha\subset im g_V$. The opposite
inclusion follows from $\alpha\circ g_V=\alpha\circ g\circ
j_V=\beta\circ p_N\circ j_V=0$, since $p_N\circ j_V=0$.

\begin{figure}[hbtp]
\begin{center}
\scalebox{0.4}{\input{proj_tor.pstex_t}}
\end{center}
\caption{The associated exact sequences of tori.\label{proj_tor}}
\end{figure}
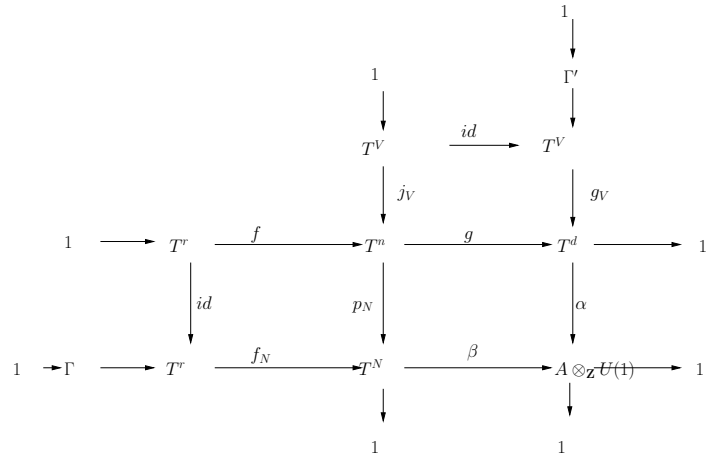

\paragraph{Observation} Assume that the hypothesis (a) 
(and thus (b)) of Proposition \ref{Prop} holds. 
Upon tensoring the diagram of figure \ref{proj} 
with $U(1)$, we obtain the diagram shown in figure \ref{proj_tor}, 
where the groups $\Gamma$ and $\Gamma'$ are isomorphic.
Computing $\Gamma$ from the bottom horizontal 
sequence, one obtains a presentation of type (\ref{Gamma_emb}),
where this time $v^{(N)}_\alpha$ is a basis 
of $\Z^r$ such that 
$f_N(v^{(N)}_\alpha)=t^{(N)}_\alpha u^{(N)}_\alpha$ for some basis 
$u^{(N)}_j$ of $\Z^N$.
Notice that $\Gamma$ maps to $T^V$ through the composite $p_V\circ f$.
Upon using expression (\ref{Gamma_emb}), 
we find that this map is given by:
\be
\label{Gamma_V}
\lambda_j=
\prod_{\alpha=1}^{r}{e^{\frac{2\pi i}{t^{(N)}_\alpha}s_\alpha 
\sum_{\beta=1}^r{
f_j^{(\beta)}v^{(N,\beta)}_\alpha}}}~~(j\in V)~~,
\ee
so that the generator of $\Z_{t^{(N)}_\alpha}$ embeds into $T^V$ as 
$(e^{\frac{2\pi i}{t^{(N)}_\alpha}\sum_{\beta=1}^r{
f_j^{(\beta)}v^{(N,\beta)}_\alpha}})_{j\in V}$. 
In matrix language, we have $F_NV_N=U_NF_N^{ismith}$, where $V_N$, $U_N$ are 
invertible $r\times r$ and $|N|\times |N|$ 
integral matrices whose columns are $v^{(N)}_\alpha$ and 
$u^{(N)}_j$ ($|N|$ denotes the 
cardinality of the set $N$, and $F_N$ is the $|N|\times r$ matrix of 
the map $f_N$, which results from the matrix $F$ of $f$ by deleting the rows 
associated with the index set $V$). 
Then (\ref{Gamma_V}) says that the generator
of $\Z_{t^{(N)}_\alpha}$ embeds into 
$T^V$ according to the element $(FV_N)_{j\alpha}$
of the matrix $FV_N$ (where $j\in V$). The collection of such elements 
forms the matrix obtained from $FV_N$ by deleting the rows associated with the 
index set $N$.

If one uses the vertical rightmost sequence instead, then one 
determines $\Gamma$ from the invariant factors $t'_k$ of the matrix 
$G_V$. In this case, $\Gamma$ embeds into $T^V$ by the uppermost vertical 
arrow, which according to (\ref{Gamma_emb}) is given by:
\be
\lambda_j=
\prod_{k\in V}{e^{\frac{2\pi i}{t'_k}s_kv^{'(j)}_k}}~~(j\in V)~~,
\ee
where this time $v'_k, u'_j$ are bases of $\Z^V$ and $\Z^d$ such that 
$g_V(v'_k)=t'_ku'_k$. These are the columns of invertible 
integral matrices $U',V'$ which bring the matrix $G_V$ 
(obtained from $G$ by deleting the columns associated with $N$) 
to its integral Smith form.

From figure \ref{proj_tor}, we also obtain:
\be
\label{GGprime}
j_V(\Gamma')=j_V(ker g_V)=
ker g\cap im j_v=im f \cap ker p_N=f(ker f_N)=f(\Gamma)~~.
\ee
Since $j_V$ and $f$ are injective maps of tori, this allows us 
to determine one of the isomorphic presentations $\Gamma,\Gamma'$ given the 
other. The meaning of this is as follows. The group $\Gamma$ is 
defined as the kernel of the map $f_N:T^r\rightarrow T^N$. Proposition
\ref{Prop} shows that this group is isomorphic with 
$\Gamma':=ker (g_V:T^V\rightarrow T^d)$, while relation (\ref{GGprime}) 
shows that $\Gamma=f^{-1}(j_V(\Gamma'))$. 
In many applications, it is 
easier to determine $\Gamma'$ and reconstruct its embedding 
$\Gamma$ into $T^r$ by using the later relation. 

\paragraph{A particular case} If the set $V$ has only one element, say 
$V=\{j\}$, then $g_V(\Z^V)=\Z\nu_j$ (where $\nu_j=g(e'_j)$ is the $j^{th}$ 
column of the matrix $G$ of $g$) and 
$A=\Z^d/\Z\nu_j$. If $m_j = gcd(\nu_j^1\dots \nu_j^d)$,
then we can write $\nu_j=m_j w_1$, where $w_1$ is a primitive
integral vector in $\Z^d$. This vector can be completed to an
integral basis $w_1\dots w_d$ of $\Z^d$, which gives
$A\approx \Z_{m_j}\times \Z^{d-1}$; in particular, $G_V$ has a single 
invariant factor, given by $t'_1=m$. 
Hence $\Gamma=T(A)=\Z_{m_j}$. In this case, the matrix $G_V$ reduces to 
the column vector $\nu_j$ and there is a single basis vector $v_1=e_j$ for the 
lattice $\Z^V=\Z e_j\approx \Z$.
The embedding of $\Z_{m_j}$ into $T_V=U(1)$
takes the generator of $\Z_{m_j}$ into $e^{\frac{2\pi i}{m_j}}$, while 
the embedding of $\Z_{m_j}$ into $T^n$ effected by $j_V$ 
takes this generator into the element
$\Lambda=(1\dots 1,e^{\frac{2\pi i}{m_j}}, 1\dots 1)\in T^n$, where  
$e^{\frac{2\pi i}{m_j}}$ sits in position $j$. If we let
$\boldlambda=
(\lambda_1\dots\lambda_r)\in U(1)^r$ parameterize elements of $T^r$,
then relation (\ref{GGprime}) tells us that $\Gamma$ is the multiplicative 
group of solutions to the system of equations:
\bea
\label{Gamma_sys}
\prod_{\alpha=1}^r{\lambda_\alpha^{f_k^{(\alpha)}}}&=&
1~~{\rm~for~}~~k\neq j\nn\\
\prod_{\alpha=1}^r{\lambda_\alpha^{f_j^{(\alpha)}}}&=&e^{\frac{2\pi i s}{m_j}}
~~,~~s\in \Z_{m_j}~~.
\eea
Since $f:T^r\rightarrow T^n$ is an injective map of tori, this makes it obvious
why the group of such solutions is isomorphic with 
$\Z_{m_j}$, and gives the embedding of the latter group into $T^r$. Note that 
$\Gamma$ is defined by the first $n-1$ equations of 
(\ref{Gamma_sys}), which are associated with the non-injective map 
$f_N:T^r\rightarrow T^N$. 
Knowledge of the last equation (which follows automatically
from the first under our assumptions) supplements these $n-1$ conditions 
by the last constraint in (\ref{Gamma_sys}). That is, any solution 
$\boldlambda=(\lambda_1\dots \lambda_{n-2})$ 
of the first $n-1$ equations will automatically satisfy 
the last equation of (\ref{Gamma_sys}) for some uniquely determined 
element $s=s_{\boldlambda}\in \Z_{m_j}$, 
and the map $\boldlambda\rightarrow s_{\boldlambda}$ 
gives the isomorphism $\Gamma\approx \Z_{m_j}$. 
This allows us to determine 
the structure of $\Gamma$ without explicitly solving the original $n-1$ 
equations.

\section{Singularities of toric hyperkahler cones}

\subsection{Good toric hyperkahler cones}

Consider a toric hyperkahler cone $X=\H^n///_0T^r$, 
defined by a short exact sequence: 
\be
\label{qtsequence}
0\longrightarrow \Z^r\stackrel{q^*}{\longrightarrow}\Z^n
\stackrel{g}{\longrightarrow} \Z^d\longrightarrow 0~~,
\ee
where $q^*$ is the transpose of a map $q:\Z^n\rightarrow \Z^r$.
Let $Q,G$ be the matrices of $q$ and $g$ with respect to the canonical bases 
of the appropriate lattices. 
Let ${\vec \mu}:\H^n\rightarrow \R^r\otimes \R^3$ 
be the hyperkahler moment map of the associated $T^r$ 
action and ${\cal N}={\vec \mu}^{-1}(0)\subset \H^n$ be its zero 
level set. We recall the following:

{\bf Fact} The image of the differential $d_u{\vec \mu}:\H^n\rightarrow 
\R^r\otimes \R^3$ at a point 
$u\in {\cal N}$ coincides with $stab_{T^r}(u)^\perp\otimes \R^3$, where 
$stab_{T^r}(u)\subset \R^r$ is the Lie algebra of the stabilizer 
$Stab_{T^r}(u)$ of $u$ in $T^r$.

It follows that the point $u$ is smooth in ${\cal N}$ 
if and only if $stab_{T^r}(u)$ vanishes, i.e. 
$Stab_{T^r}(u)$ is a finite group. Given a point $u\in {\cal N}$, we define 
$V(u)=\{j\in \{1\dots n\}|u_j=0\}$ and $N(u)=\{j\in \{1\dots n\}|u_j\neq 0\}$.
This gives a partition $\{1\dots n\}=V(u)\cup N(u)$ and associated 
decompositions $\R^n=\R^V\oplus \R^N$, $\Z^n=\Z^V\oplus \Z^N$ and 
$q^*=q^*_V\oplus q^*_N$. Here $q^*_V=p_V\circ q^*$ and $q^*_N=p_N\circ q^*$, 
where $p_N,p_V$ are the projections of $\Z^n$ onto $\Z^N$ and $\Z^V$.
It is clear that $stab_{T^n}(u)$ (the Lie algebra of the stabilizer 
of $u$ in the `diagonal' torus $T^n$) equals $\R^V$ (indeed, only the 
vanishing coordinates of $u$ are invariant under the diagonal $T^n$ action
$u_j\rightarrow \Lambda_j u_j$ on $\H^n$). Therefore,
$stab_{T^r}(u)$ equals the preimage of the subspace $\R^V\subset \R^n$ 
through (the real extension of) the  map $q^*$, i.e. the 
kernel of the map $q^*_N:\R^r\rightarrow \R^N$. Thus $u$ is a singular point 
of ${\cal N}$ if and only if this map is not injective. We have 
the diagram of figure \ref{proj}, with $f$ replaced by $q^*$ and $f_N$ 
replaced by $q^*_N$.

\begin{Lemma} Suppose that any $d$ of the toric hyperkahler generators 
$\nu_1\dots \nu_n$ are linearly independent over $\R$, i.e. all $d\times d$ 
minor determinants of the matrix $G$ are non-vanishing.
Then for  any point $u\in {\cal N}-\{0\}$, the set $V(u)$ has at 
most $d-1$ elements (hence the set $N(u)$ has at least $r+1$ elements). 
\label{Lemma}
\end{Lemma}
{\bf Proof:} Write $u_k=w_k^{(+)}+\j w_k^{(-)}$. 
As explained in Section 5, the fact that $u$ belongs to ${\cal N}$ means: 
\be
\label{pp}
|w_k^{(+)}|^2-|w_k^{(-)}|^2=2\nu_k \cdot a~~,~~
w_k^{(+)} w_k^{(-)}=\nu_k \cdot b
\ee
for some $a\in \R^2$ and $b\in \C^2$. 
If $k$ is an element of $V$, 
then $w_k^{(+)}=w_k^{(-)}=0$ and we obtain $\nu_k \cdot a=0$ and 
$\nu_k \cdot b=0$. If $V$ has more than $d-1$ elements, this implies 
$a=b=0$ and thus $u=0$ by eqs (\ref{pp}), since in this case 
the vectors $\nu_k$ $(k\in V)$ generate $\R^d$ by the assumption 
of the Lemma. Hence $V$ has at most $d-1$ elements provided that $u\neq 0$.

\begin{Proposition} The following are equivalent: 

(a) All $r\times r$ minor determinants  of $Q$ are nonzero 

(b) Any $d$ of the vectors $\nu_1\dots \nu_n$ are 
linearly independent over $\R$, i.e. all $d\times d$ minor determinants of $G$ 
are nonzero.

In this case, the origin of $\H^n$ is the only singular point of 
${\cal N}$. 
A toric hyperkahler cone is called {\em good} if it  satisfies condition 
$(a)$ (and thus $(b)$).
\label{goodcones}
\end{Proposition}

{\bf Proof:} Consider the diagram in figure \ref{proj} with $f$ replaced by 
$q^*$. 
It is clear that conditions (a),
(b) amount respectively to injectivity of $f=q^*_N$ or of $g_V$ for 
all partitions $\{1\dots n\}=V\cup N$ with $|V|=d$ and $|N|=r$. Hence  
equivalence of (a) and (b) follows immediately from Proposition \ref{goodseq}. 

Let us now assume that (a) (and thus (b)) holds. Fix a point 
$u\in {\cal N}-\{0\}$. 
As mentioned above, $u$ is singular in ${\cal N}$ 
if and only if  the map $q^*_N:\R^r\rightarrow \R^n$ (where $N=N(u)$)
is not injective. This amounts to requiring 
that all maximal minor determinants of the matrix 
$Q_N$ (obtained from $Q$ by deleting the columns associated with $V(u)$ and 
keeping the columns associated with $N(u)$)
vanish. Since $u\neq 0$ 
and $(b)$ holds, Lemma \ref{Lemma} 
assures us that $N(u)$ has at least $r+1$ elements, 
so that $Q_N$ has $r$ rows and at least $r+1$ columns. This means that 
its maximal minors coincide with certain $r\times r$ minors of $Q$, whose 
determinant 
cannot vanish since (a) holds. Therefore, $u$ cannot be a singular point of 
${\cal N}$.

\

\begin{Corollary} Let $X=\H^n///_0 T^{n-2}$ be an 
{\em eight-dimensional} toric hyperkahler cone. Then the following 
are equivalent:

(a) The cone $X$ is good.

(b) No two of the three-dimensional flats $H_1\dots H_n$ coincide in $\R^6$.

(c) No two of the lines $h_1\dots h_n$ coincide in $\R^2$.
\label{goodconescorr}
\end{Corollary}

{\bf Proof:} For $X$ an eight-dimensional toric hyperkahler cone,
the hyperplanes $h_j$ are lines in $\R^2$ which pass through the origin. 
Two such lines intersect
outside of the origin if and only if they coincide. Since $H_j=h_j\times 
h_j\times h_j$, it follows that  
two flats can intersect outside of the origin if and only 
if they coincide. Now $h_j=\{a\in \R^2|a\cdot \nu_j=0\}$, so two lines 
$h_i$ and $h_j$ coincide if and only if $\nu_i$ and $\nu_j$ are 
linearly dependent. Since $d=2$, the conclusion follows from Proposition 
\ref{goodcones}. 

\

\subsection{Singularities of good toric hyperkahler cones}

Let $X$ be a good, $d$-dimensional toric hyperkahler cone. Since 
${\cal N}$ is smooth except at the origin, all singularities of 
$X={\cal N}/T^r$ outside its apex arise from points $u\in {\cal N}-\{0\}$ 
which have nontrivial finite stabilizers with respect to 
the $T^r$ action. Since 
we assume this action to be effective (as follows from the fact that 
$coker(q^*)$ is torsion-free), this can only happen if some of the 
coordinates of $u$ vanish\footnote{Indeed, the 
map of tori $T^r\rightarrow T^n$ is injective.}. 
Considering the sets $N(u)$ and $V(u)$ 
as above, the torus $T^n$ decomposes as $T^V\times T^N$, with $T^V$ acting 
(trivially) on the vanishing coordinates and $T^N$ acting on the 
nonvanishing components of $u$. It is clear that the $T^r$-stabilizer 
$\Gamma_u=Stab_{T^r}(u)$ of $u$
coincides with the kernel of  the map $T^r\rightarrow T^N$
induced from $T^r\rightarrow T^n$ by composing with the projection 
$T^n\rightarrow T^N$. 
Since $X$ is good and $u\neq 0$, we have $|N|\geq r+1$ and 
$q^*_N$ is injective.
Hence the situation is precisely that considered in figure \ref{proj}, 
with $f$ replaced by $q^*$.  
Now Proposition \ref{fptor} shows that $\Gamma_u=Tor^{\Z}(A,U(1))$, where 
$A=coker(q^*_N)$.  Since $q^*_N$ is injective, 
Proposition \ref{goodseq} allows 
us to also compute $A$ as the cokernel of the map $g_V:\Z^V\rightarrow \Z^d$.

In the case $d=2$, Lemma \ref{Lemma} 
shows that at most
one quaternion 
coordinate $u_j$ can vanish unless $u$ coincides with the apex of $X$. 
It follows that singularities of $X$ can occur only along the loci
$X_j=\{u\in X|u_j=0\}$ discussed in Section 5. For $u\in X_j-\{0\}$,
one has $V(u)=\{j\}$ and 
the observation following Proposition \ref{goodseq} shows that 
$\Gamma_u$ must be a cyclic group, determined  by the greatest common divisor 
of the two components of the toric hyperkahler generator $\nu_j$.
This gives the following:

\begin{Proposition} Let $X=\H^n///_0U(1)^{n-2}$ be a good toric hyperkahler cone of real dimension eight. Then:

(1) All singularities of $X$ lie in one of the four-dimensional loci 
$X_j=\{u\in X|u_j=0\}$. Two such loci intersect 
at precisely one point, namely the apex of $X$.

(2) The locus $X_j-\{0\}$ is smooth if and only if the associated toric hyperkahler generator $\nu_j\in \Z^2$ 
is a primitive vector.

(3) If $\nu_j$ is not primitive, then each point on the locus
$X_j-\{0\}$ is a $\Z_m$ quotient singularity of $X$, where $m$ is the
greatest common divisor of the coordinates $\nu_j^1$ and $\nu_j^2$ of
$\nu_j$. The action of the generator of $\Z_{m_j}$ on the quaternion
coordinate $u_j$ transverse to the singular locus $X_j$ is given by:
\be
u_j\rightarrow e^{\frac{2\pi i}{m_j}}u_j~~.
\ee

\label{goodconesing}
\end{Proposition}

Recall from Appendix A that the singularity group $\Gamma_j$ along $X_j-\{0\}$ 
is given by the multiplicative group of solutions to the following system:
\be
\prod_{\alpha=1}^{n-2}{\lambda^{q_k^{(\alpha)}}}=1~~{\rm~for~}~~k\neq j~~.
\ee
As explained in Appendix A, any solution $\boldlambda=
(\lambda_1\dots \lambda_{n-2})$ 
of this system will automatically 
satisfy the equation:
\be
\prod_{\alpha=1}^{n-2}{\lambda^{q_j^{(\alpha)}}}=e^{2\pi i \frac{s}{m_j}}
~~{\rm~for~}~~k\neq j~~,
\ee
for some element $s=s_{\boldlambda} \in \Z_{m_j}$, which is uniquely determined 
by $\boldlambda$. The isomorphism $\Gamma_j\rightarrow \Z_{m_j}$
is given by the map $\boldlambda\rightarrow s_{\boldlambda}$.

{\bf Observation} If the toric hyperkahler cone is not good, then 
some of its singularities outside the apex are induced from singularities 
of the zero level set ${\cal N}={\vec \mu}^{-1}(0)$. Such cases can 
be analyzed by algebraic geometry methods, upon using the toric embedding 
of $X$ given in Subsection 5.1.

\section{Criterion for effectiveness of the projectivising $U(1)$ action}

Consider a $d$-dimensional toric hyperkahler cone $X=\H^n///_0 U(1)^r$ $(d=n-r)$.
Let $Q^{ismith}$ be the integral Smith 
form of the $r\times n$ charge matrix $Q$ and $U\in SL(r, \Z)$, 
$V\in SL(n,\Z)$ such that $Q^{ismith}=U^{-1}QV$. 
Since $Q$ has trivial invariant factors, the matrix 
$Q_{ismith}$ has the form $[I_r, 0]$, where $I_r$ is the $r\times r$ identity 
matrix. Remember that 
$X$ carries an action of $Sp(1)$ induced by:
\be
(u_1\dots u_n) \rightarrow (u_1 t^{-1}\dots u_n t^{-1})~~,
\ee
where $t$ is a unit norm quaternion (viewed as an element of $Sp(1)$).
Consider the integral $n$-vector 
$F:=V^t\left[\begin{array}{c}1\\1\\\dots \\1\end{array}\right]$ (i.e. 
$F$ is the sum of all rows of $V$).

\begin{Proposition}

The $\Z_2$ subgroup $\{1,-1\}$ of  $Sp(1)$ acts trivially on $X$ if and only 
if the components $F_{r+1}\dots F_n$  are even. Equivalently,
$\{-1,1\}$ acts trivially on $X$ if and only if there exist $1\leq m\leq r$
and $1\leq \alpha_1<\alpha_2<\cdots \alpha_m\leq r$ such that all components 
of the $n$-vector $w$ defined as the sum of the rows 
$\alpha_1\dots \alpha_m$ of $Q$ are odd. In this case, the indices $\alpha_k$
with this property are uniquely determined, and the the action of 
the $\Z_2$ subgroup of $Sp(1)$ on $\H^n$ 
coincides with the action on $\H^n$ of the $\Z_2$ subgroup of $T^r$ 
generated by: 
\bea
&&\lambda_\alpha=+1,~~{\rm~for~}~~\alpha\neq \alpha_j\nn\\
&&\lambda_{\alpha_j}=-1~~{\rm~for~~all~}~~j=1\dots m~~.
\eea
\end{Proposition}

{\bf Proof:} The generator of the given $\Z_2$ subgroup of $Sp(1)$ 
acts on quaternion coordinates 
through sign inversion, $u_j\rightarrow -u_j$. This descends to a trivial 
action on $X$ if and only if this transformation can be realized through the 
action of $T^r$, i.e. precisely when the following system admits a solution 
$\boldlambda=(\lambda_1\dots \lambda_r)\in U(1)^r$:
\be
\label{Z2sys1}
\prod_{\alpha=1}^{r}{\lambda_\alpha^{q_j^{(\alpha)}}}=-1~{\rm~for~all~}
j=1\dots n~~.
\ee
Upon writing $\lambda_\alpha=e^{2\pi i \phi_\alpha}$ 
(with $\phi_\alpha\in \R/\Z$), we can 
express (\ref{Z2sys1}) in the equivalent form:
\be
Q^t\phi=\frac{1}{2}E~~({\rm mod}~(\Z^n))~~,
\ee
where 
$\phi=\left[\begin{array}{c}\phi_1\\\phi_2\\\dots \\\phi_{r}
\end{array}\right]$ and $E$ is the $n$-vector
$\left[\begin{array}{c}1\\1\\\dots \\1\end{array}\right]$. Since
$Q=UQ^{ismith}V^{-1}$, this becomes:
\be
(Q^{ismith})^t\psi=F~~({\rm mod}~(2\Z)^n)~~,
\ee
where $\psi=2U^t\phi\in (\R/(2\Z))^n$. 
Using the form $Q^{ismith}=[I_r,0]$, this gives 
$\psi=\left[\begin{array}{c}F_1\\F_2\\\dots \\F_r\end{array}\right]\in 
\Z^r/(2\Z)^r=\Z_2^r$
and: 
\be
\left[\begin{array}{c}F_{r+1}\\F_{r+2}\\\dots \\F_n\end{array}\right]=0 
~~({\rm mod}~(2\Z)^d)~~,
\ee
which proves the first part of the proposition. Note that we have 
$\phi=\frac{1}{2}U^{-t}\psi\in (\frac{1}{2}\Z)^r/\Z^r\approx (\Z_2)^r$, 
if we identify the subgroup $\{0,1/2\}$ of $\Q/\Z$ with $\Z_2$. This means that
the solution $\boldlambda$ of (\ref{Z2sys1}) (when it exists) has the form
$\lambda_\alpha=(-1)^{\sum_{\beta=1}^r{(U^{-1})_{\beta\alpha}F_\beta}}$, 
i.e. $\lambda_\alpha$ must be $+1$ or $-1$. Using this fact directly in 
(\ref{Z2sys1}), we see that a solution exists if and only if there exists
a subcollection $\alpha_1\dots \alpha_m$ 
of the rows of $Q$ whose sum $w$ is a row vector all of whose 
entries are odd ($\alpha_j$ are those indices $\alpha$ for which 
$\lambda_{\alpha_j}=-1$). To see why the 
rows $\alpha_1\dots \alpha_m$ are uniquely determined, suppose that there 
exists another choice $\alpha'_1\dots \alpha'_{m'}$ 
of rows with this property, and let $w'$ be the integral 
$n$-vector obtained by adding all these rows (by assumption, both 
$w$ and $w'$ have only odd entries). Eliminating the common rows 
between these two collections, we obtain two disjoint 
sets $S_1$ and $S_2$ of rows of $Q$ 
which have the property that 
all entries of the vector 
$\sum_{k\in S_1}{row(Q,k)}-\sum_{k\in S_2}{row(Q,k)}=
\sum_{j=1}^m{row(Q,\alpha_j)}-\sum_{j=1}^{m'}{row(Q,\alpha'_j)}=w-w'$ 
are even; but this immediately implies that all $r\times r$ minor determinants
of $Q$ are even, so that the discriminantal divisor $\g(Q)$ would be even.
This contradicts the assumption $\g(Q)=1\Leftrightarrow Q^{ismith}=[I_r,0]$ 
made in the definition of toric hyperkahler spaces. 
The second part of the proposition follows.

\section{Singularities of the twistor space along $Y_e$ }

Let $Y$ be the twistor space $Y$ associated with a good, eight-dimensional 
toric hyperkahler cone. Consider the locus $Y_e=Y_{\epsilon(e)}$ 
associated with an edge $e$ of the characteristic polygon $\Delta$. 
Remember that $Y_e=X_e/T^1$, where $X_e={\vec \pi}^{-1}
(e\times \{0_{\R^2}\}\times \{0_{\R^2}\})$ is 
a $T^2$ fibration over $e\subset \R^2$. 
We want to describe the quotient of the $T^2$ fiber
of $X_e$ by $T^1$. For this, we shall use the fact that the $T^2$ fiber 
itself is a quotient of $T^n$ by $T^r=T^{n-2}$.

To describe this in terms of lattices, 
consider the decomposition $\Z^{n-1}=\Z^{n-2}\times \Z$ associated to 
the product $T^{n-1}=T^{n-2}\times T^1$ and 
the corresponding projections 
$p_1:\Z^{n-1}\rightarrow \Z^{n-2}$, $p_2:\Z^{n-1}\rightarrow \Z$ and 
injections 
$j_1:\Z^{n-2}\rightarrow \Z^{n-1}$, 
$j_2:\Z\rightarrow \Z^{n-1}$. 

We have a split exact sequence: 
\be
0\longrightarrow \Z^{n-2}\stackrel{j_1}{\longrightarrow} \Z^{n-1}\stackrel{p_2}{\longrightarrow} \Z\longrightarrow 0~~.
\ee
Restricting to the locus $X_e$, we are left with the non-vanishing variables 
$w^{(\epsilon_j(e))}_j$, and with the exact sequence  (\ref{sequence_epsilon})
of Subsection 7.5:
\be
\label{seps}
0\longrightarrow \Z^{n-2} \stackrel{q_e^*}{\longrightarrow} \Z^n
\stackrel{g_e}\longrightarrow \Z^2\longrightarrow 0~~. 
\ee
As in Subsection 7.5, 
the middle term of (\ref{seps}) corresponds to the $T^n$ action 
obtained by restricting the `toric' diagonal $T^{2n}$ action:
\be
\label{mod_diag}
w_j^{(+)}\rightarrow \Lambda_j w_j^{(+)}~~,~~
w_j^{(-)}\rightarrow \Lambda_{j+n} w_j^{(-)}~~
\ee
to the non-vanishing coordinates $w_j^{(\epsilon_j(e))}$:
\be
w_j^{(\epsilon_j(e))}\rightarrow \lambda_j w_j^{(\epsilon_j(e))}~~,
\ee
where $\lambda_j=\Lambda_{j+(1-\epsilon_j(e))\frac{n}{2}}$.

The projectivising $U(1)$ 
action on $w^{(\epsilon_j)}_j$ 
results from the map $\gamma:\Z\rightarrow \Z^n$ defined through:
\be
\label{gamma}
\gamma(1)=t:=\left[\begin{array}{c}1\\\dots \\1\\\end{array}\right]~~.
\ee
The action of $T^{n-1}=T^{n-2}\times T^1$ results from the map 
${\bar q}^*_e=q^*_e\circ p_1+\gamma\circ p_2:\Z^{n-1}\rightarrow \Z^n$.
If $u=(v,m)$ is an element of $\Z^{n-1}=\Z^{n-2}\times \Z$ (with 
$v\in \Z^{n-2}$ and $m\in \Z$), then 
${\bar q}^*_e(u)=q^*_e(v)+\gamma(m)$. The $(n-1)\times n$ 
matrix of the transpose map ${\bar q}_e$ is given by
${\bar Q}_e=\left[\begin{array}{c}~Q_e~\\1\dots 1\end{array}\right]$. 

\paragraph{Observation} The matrix ${\bar Q}_e$ is related to the toric
approach to the singularities of $Y$ outlined in Subsection 7.9. 
Indeed, ${\bar Q_e}$ is obtained from the toric charge matrix 
${\tilde Q}=\left[\begin{array}{c}Q~,-Q\\1 \dots 1\end{array}\right]$
of the ambient space $\T$ upon keeping only those columns associated with the 
coordinates $w_j^{(\epsilon_j(e))}$ which do not vanish on the locus $Y_e$. 
From this perspective, the singularity type along $Y_e$ can be determined 
by using the results of Appendix A, i.e. by determining the integral Smith 
form of the matrix ${\bar Q}_e$. In this approach, it is not clear that 
the singularity group along $Y_e$ is always a cyclic group. This facts
only becomes clear when considering the alternate description used in 
Subsection 7.5. The purpose of this appendix is to explain the equivalence of 
these two methods, and to provide a rigorous justification of the latter.

The injection $\gamma$ induces a map 
$\alpha=g_e\circ \gamma:\Z\rightarrow \Z^2$, 
which describes the embedding of the 
projectivising $U(1)$ in the $T^2$ fiber of $X_e$. This map is specified by:
\be
\alpha(1)=g_e(t)=\sum_{j=1}^n{\epsilon_j(e)\nu_j}=\nu_e~~.
\ee

\begin{Lemma}

(a) The vector $\nu_e$ does not vanish. Therefore, the map $\alpha_e$ is 
nonzero (and injective). 

(b) The map ${\bar q}^*_e$ is injective, i.e. the matrix ${\bar Q}_e$
has maximal rank.

\end{Lemma}

{\bf Proof:} To show $(a)$, consider the vector $p_e$ 
(associated with the middle 
point of the edge $e$) as in Subsection 7.5. Since 
$\epsilon_j(e)=sign(p_e\cdot \nu_j)$, we have: 
\be
p_e\cdot \nu_e=\sum_{j=1}^n{\epsilon_j(e)p_e\cdot \nu_j}=
\sum_{j=1}^n{|p_e\cdot \nu_j|}~~.
\ee
Therefore, vanishing of $\nu_e$ would imply $p_e\cdot \nu_j=0$ i.e. 
$p_e\in D_j$ for all $j$ ($j$ are the principal diagonals of the 
characteristic polygon $\Delta$). This is impossible, since $p_e$ 
belongs to the interior of the edge $e$.

To show (b), consider an element 
$u\in \Z^{n-1}$ such 
that ${\bar g}^*_e(u)=0$. Writing $u=(v,m)$ with $v\in \Z^{n-2}$ and 
$m\in \Z$, we have $0={\bar g}^*_e(u)=q^*_e(v)+\gamma(m)$, so that 
$\alpha(m)=g_e(\gamma(m))=-g_e(q^*_e(v))=0$ since 
$g_e\circ q^*_e$ vanishes. Since $\alpha$ is injective, this gives $m=0$
and $q^*_e(v)=0$. This implies $v=0$ since $q^*_e$ is injective. 
Thus $u=0$ and ${\bar q}^*_e$ is injective.

\

In view of the Lemma, one has short exact sequences:
\be
0\longrightarrow \Z^{n-1}\stackrel{{\bar q}^*_e}{\longrightarrow} 
\Z^n\stackrel{{\bar g}_e}{\longrightarrow} 
A_e\longrightarrow 0~~.
\ee
and
\be
\label{Zsequence}
0\longrightarrow \Z\stackrel{\alpha}{\longrightarrow} 
\Z^2\stackrel{\beta}{\longrightarrow} A'_e\longrightarrow 0~~.
\ee
Both $A_e$ and $A'_e$ may contain torsion; 
this corresponds to possible fixed points 
of the $T^{n-1}$ and $T^1$ actions on $X_e$. The situation is summarized 
in figure \ref{doublesequence}.

\begin{figure}[hbtp]
\begin{center}
\scalebox{0.4}{\input{doublesequence.pstex_t}}
\end{center}
\caption{Lattice diagram for the embedding of $T^1$ into the 
$T^2$ fiber on the locus $X_\epsilon$. \label{doublesequence}}
\end{figure}

\begin{Proposition} The square formed by the maps ${\bar q}^*_e, g_e, p_2$ 
and $\alpha$ commutes. 
Moreover, there exists a unique morphism $\Phi:A_e\rightarrow A'_e$ which
produces a commutative square on the right, and this morphism is an
isomorphism.
\end{Proposition}

{\bf Proof:} (1) We first show that the square indicated commutes. If
$u=(v,m)$ is an element of $\Z^{n-1}=\Z^{n-2}\times \Z$, then
$p_1(u)=v$ and $p_2(u)=m$. We have $g_e\circ
{\bar q}^*_e(u)=g_e(q^*_e(v)+\gamma(m))= g_e\circ
\gamma(m)=\alpha(m)=\alpha\circ p_2(u)$, where we used the definition
of ${\bar g}_e$ and $\alpha$ as well as exactness of the vertical sequence. Thus
$g_e\circ {\bar q}^*_e=\alpha\circ p_2$, which is the desired
statement.

(2) Forgetting the morphism $j_2$, we have maps $p_2$ and
$g_e$ which close the first square. Since the horizontal
sequences are short exact, applying the 3-lemma shows that there
exists a unique morphism $\Phi:A_e\rightarrow A'_e$ which closes the last
square to a commutative diagram. Surjectivity of $\Phi$ follows from the 
commutative square on the right (since $\beta$ and $g_e$ are surjective).
To show injectivity of $\Phi$, consider
an element $x\in A_e$ such that $\Phi(x)=0$.  Then $x={\bar g}_e(y)$
with $y\in \Z^n$ and $g_e(y)\in ker \beta=im\alpha$. Thus
$g_e(y)=\alpha(m)$ for some $m\in \Z$. Since
$\alpha=g_e\circ \gamma$, this means that $y=\gamma(m)+w$, with
$w$ an element of $kerg_e=imq^*_e$. Hence there exists a
$v\in \Z^{n-2}$ such that $y=\gamma(m)+q^*_e(v)=
{\bar q}^*_e(v,m)$. This implies $y\in im {\bar q}^*_e=ker
{\bar g}_e$, and thus $x={\bar g}_e(y)=0$.

\

Consider the horizontal sequences in figure \ref{doublesequence}.
Tensoring with $U(1)$ gives two exact sequences on four
terms (the associated long exact Tor sequences collapse because
$\Z^n$ and $\Z^2$ are flat.). The leftmost terms of the resulting
sequences are $\Gamma_e=Tor^{\Z}(A_e,U(1))$ and
$\Gamma'_e=Tor^{\Z}(A'_e,U(1))$, the kernels of the maps
$T^{n-1}\rightarrow T^n$ and $T^1\rightarrow T^2$ (figure \ref{tor}).

\begin{figure}[hbtp]
\begin{center}
\scalebox{0.4}{\input{tor.pstex_t}}
\end{center}
\caption{The associated diagram of tori.\label{tor}}
\end{figure}

\begin{Corollary} The groups $\Gamma_e$ and $\Gamma'_e$ 
are isomorphic.
\end{Corollary}

{\bf Proof:} Follows from naturality of $Tor$. 

\

This shows that one obtains the same result by computing $\Gamma_e$ from 
the maps $T^1\rightarrow T^2$ or $T^{n-2}\rightarrow T^n$ on the locus $Y_e$.

It is now easy to describe the orbifold group of the twistor space 
along $Y_e$, as well as its action on the transverse 
coordinates $w_j^{(-\epsilon_j(e))}$. For this, we have to remember that 
the group which acts effectively on $X$ is $U(1)_{eff}=U(1)$ or $U(1)/\Z_2$, 
according to the criterion of Appendix C. This means that 
the singularity group coincides with $\Gamma_e$ or $\Gamma_e/\Z_2$.

\begin{Corollary} Assume that $Y_e$ is nondegenerate. 
Then the group $\Gamma_e$ is isomorphic with $\Z_{m_e}$, 
where $m_e$ is the greatest common divisor 
of the coordinates of $\nu_e$.
The generator of  $\Z_{m_e}$ acts on the coordinates $w_j^{(-\epsilon_j(e))}$
transverse to $Y_e$ through:
\be
\label{tvs_action}
w_j^{(-\epsilon_j(e))}\rightarrow e^{\frac{2\pi i}{m_e}}~~
w_j^{(-\epsilon_j(e))}~~.
\ee
If the projectivising $U(1)$ acts effectively on $X$, then the singularity 
group of $Y$ along $Y_e$ coincides with $\Gamma_e$, with the transverse 
action given above. Otherwise, the singularity group is $\Gamma_e/\Z_2$, with 
transverse action induced by (\ref{tvs_action}).

\end{Corollary}

{\bf Proof:} The sequence (\ref{Zsequence}) presents $A'_e=A_e$ 
as:
\be
A'_e=\Z^2/\alpha(\Z)=\Z^2/\Z\nu_e~~,
\ee
and thus the group $\Gamma_e$ as:
\be
\Gamma_e=Tor^{\Z}(A'_e,U(1))=T(A'_e)=\Z_{m_e}~~.
\ee
The embedding of 
$\Gamma_e$ into the $T^n$ factor of $T^{2n}$ which acts diagonally 
on the transverse coordinates $w_j^{(-\epsilon_j(e))}$ is  
induced by a copy of the map 
$\gamma$ given in (\ref{gamma}). This takes the generator of 
$\Z_{m_e}$ into the element 
$\lambda=(e^{\frac{2\pi i}{m_e}}\dots e^{\frac{2\pi i}{m_e}})$.
This immediately gives the transverse action.
The remaining statements are obvious.

\paragraph{Observation 1} 
Recall that the torus $T^n$ appearing in the presentation 
$X=\H^n///_0T^n$ acts on $w_j^{(\pm)}$ as $w_j^{(\pm)}\rightarrow 
\Lambda_j^{\pm 1} w_j^{(\pm)}$. Therefore, the embedding of $\Z_{m_e}$
into this torus takes the generator of $\Z_{m_e}$ into 
$(e^{-\epsilon_1(e)\frac{2\pi i}{m_e}}\dots e^{-\epsilon_n(e)
\frac{2\pi i}{m_e}})$. 

\paragraph{Observation 2} When combined  with the results of Appendix A,
the two Corollaries show 
that the integral Smith form of the matrix ${\bar Q}_e$ 
is always of the type 
${\bar Q}_e^{ismith}=[diag(1\dots 1, m_e),0]$, i.e. at most 
one invariant factor of ${\bar Q}_e$ can be nontrivial. This explains 
why the two methods for identifying the singularities of $Y$ along $Y_e$ 
lead to the same result. 

\section{Embedding $X$ in a toric variety}

Consider a toric hyperkahler cone described by the sequence 
(\ref{quat_sequence}). Recall the  
embedding $X\subset \S$, where $\S=\C^{2n}/(\C^*)^r$.
The toric variety $\S$ results upon decomposing the 
quaternion coordinates into complex coordinates as in 
(\ref{complex_coords}). Since $w^{(+)}_j$ and $w^{(-)}_j$ acquire opposite 
charges $q_j^{(\alpha)}$ and $-q_j^{(\alpha)}$, 
this is described by the map $j=(id, -id):\Z^n\rightarrow \Z^{2n}$.
The complex coordinates have been arranged as 
$z_1=w^{(+)}_1\dots z_n=w^{(+)}_n, z_{n+1}=w^{(-)}_1\dots z_{2n}=w^{(-)}_n$, 
and are acted on by $T^{2n}$ through the diagonal transformations 
$z_\rho\rightarrow \Lambda_\rho z_\rho$. 
The quotienting torus $T^r$ maps to $T^{2n}$ through the composite map 
${\hat q}^*=j\circ q^*=(q^*, -q^*)$, which is obviously injective (since 
$q^*$ is). 

{\bf Proposition} The cokernel of ${\hat q}^*$ is torsion free. 
In particular, we have an exact sequence:
\be
\label{qsequence}
0\longrightarrow \Z^r\stackrel{{\hat q}^*}{\longrightarrow}\Z^{2n}
\stackrel{{\hat g}}{\longrightarrow} 
\Z^{2d+r}\longrightarrow 0~~;
\ee

{\bf Proof:} By the structure theorem of lattice maps, we have bases 
$v_\alpha$ of $\Z^r$ and $u_j$ of $\Z^n$ such that 
$q^*(v_\alpha)=u_\alpha$ (remember that we assume $Q^{ismith}=[I_r,0]$). 
It is easy to check that the vectors $U_\alpha:=(u_\alpha, -u_\alpha), 
U_{-\alpha}=(u_\alpha,0)$ ($\alpha=1\dots r$) 
and $U_j:=(u_j,0)$, $U_{-j}=(0,u_j)$ ($j=r+1\dots n$) 
form a basis of $\Z^{2n}$. Since ${\hat q}^*(v_\alpha)=U_\alpha$, 
the map ${\hat q}^*$ has integral Smith form with respect to the bases 
$v$ and $U$, with unit torsion coefficients. 
This implies that its cokernel is torsion-free.

\end{document}

%% file: deltaM0.pstex_t
\begin{picture}(0,0)%
\includegraphics{deltaM0.pstex}%
\end{picture}%
\setlength{\unitlength}{4144sp}%
\begingroup\makeatletter\ifx\SetFigFont\undefined%
\gdef\SetFigFont#1#2#3#4#5{%
  \reset@font\fontsize{#1}{#2pt}%
  \fontfamily{#3}\fontseries{#4}\fontshape{#5}%
  \selectfont}%
\fi\endgroup%
\begin{picture}(2250,2475)(2431,-2626)
\put(2431,-421){\makebox(0,0)[lb]{\smash{\SetFigFont{17}{20.4}{\rmdefault}{\bfdefault}{\updefault}\special{ps: gsave 0 0 0 setrgbcolor}$A_1$\special{ps: grestore}}}}
\put(4681,-1276){\makebox(0,0)[lb]{\smash{\SetFigFont{17}{20.4}{\rmdefault}{\bfdefault}{\updefault}\special{ps: gsave 0 0 0 setrgbcolor}$A_2$\special{ps: grestore}}}}
\put(2836,-2626){\makebox(0,0)[lb]{\smash{\SetFigFont{17}{20.4}{\rmdefault}{\bfdefault}{\updefault}\special{ps: gsave 0 0 0 setrgbcolor}$A_3$\special{ps: grestore}}}}
\end{picture}

%% file: geometries.pstex_t
\begin{picture}(0,0)%
\includegraphics{geometries.pstex}%
\end{picture}%
\setlength{\unitlength}{4144sp}%
\begingroup\makeatletter\ifx\SetFigFont\undefined%
\gdef\SetFigFont#1#2#3#4#5{%
  \reset@font\fontsize{#1}{#2pt}%
  \fontfamily{#3}\fontseries{#4}\fontshape{#5}%
  \selectfont}%
\fi\endgroup%
\begin{picture}(1935,2175)(2161,-2311)
\put(4096,-1636){\makebox(0,0)[lb]{\smash{\SetFigFont{17}{20.4}{\rmdefault}{\bfdefault}{\updefault}\special{ps: gsave 0 0 0 setrgbcolor}$Y$\special{ps: grestore}}}}
\put(2971,-376){\makebox(0,0)[lb]{\smash{\SetFigFont{17}{20.4}{\rmdefault}{\bfdefault}{\updefault}\special{ps: gsave 0 0 0 setrgbcolor}$X$\special{ps: grestore}}}}
\put(2161,-1231){\makebox(0,0)[lb]{\smash{\SetFigFont{17}{20.4}{\rmdefault}{\bfdefault}{\updefault}\special{ps: gsave 0 0 0 setrgbcolor}$S$\special{ps: grestore}}}}
\put(3061,-2311){\makebox(0,0)[lb]{\smash{\SetFigFont{17}{20.4}{\rmdefault}{\bfdefault}{\updefault}\special{ps: gsave 0 0 0 setrgbcolor}$M$\special{ps: grestore}}}}
\end{picture}

%% file: double.pstex_t
\begin{picture}(0,0)%
\includegraphics{double.pstex}%
\end{picture}%
\setlength{\unitlength}{4144sp}%
\begingroup\makeatletter\ifx\SetFigFont\undefined%
\gdef\SetFigFont#1#2#3#4#5{%
  \reset@font\fontsize{#1}{#2pt}%
  \fontfamily{#3}\fontseries{#4}\fontshape{#5}%
  \selectfont}%
\fi\endgroup%
\begin{picture}(1845,2835)(2251,-2311)
\put(4096,-1636){\makebox(0,0)[lb]{\smash{\SetFigFont{14}{16.8}{\rmdefault}{\bfdefault}{\updefault}\special{ps: gsave 0 0 0 setrgbcolor}$Y$\special{ps: grestore}}}}
\put(3061,-2311){\makebox(0,0)[lb]{\smash{\SetFigFont{14}{16.8}{\rmdefault}{\bfdefault}{\updefault}\special{ps: gsave 0 0 0 setrgbcolor}$M$\special{ps: grestore}}}}
\put(3196,-466){\makebox(0,0)[lb]{\smash{\SetFigFont{14}{16.8}{\rmdefault}{\bfdefault}{\updefault}\special{ps: gsave 0 0 0 setrgbcolor}$X$\special{ps: grestore}}}}
\put(3151,344){\makebox(0,0)[lb]{\smash{\SetFigFont{14}{16.8}{\rmdefault}{\bfdefault}{\updefault}\special{ps: gsave 0 0 0 setrgbcolor}$X'$\special{ps: grestore}}}}
\put(2251,-601){\makebox(0,0)[lb]{\smash{\SetFigFont{14}{16.8}{\rmdefault}{\bfdefault}{\updefault}\special{ps: gsave 0 0 0 setrgbcolor}$S'$\special{ps: grestore}}}}
\put(2251,-1231){\makebox(0,0)[lb]{\smash{\SetFigFont{14}{16.8}{\rmdefault}{\bfdefault}{\updefault}\special{ps: gsave 0 0 0 setrgbcolor}$S$\special{ps: grestore}}}}
\end{picture}

%% file: 3lemma.pstex_t
\begin{picture}(0,0)%
\includegraphics{3lemma.pstex}%
\end{picture}%
\setlength{\unitlength}{4144sp}%
\begingroup\makeatletter\ifx\SetFigFont\undefined%
\gdef\SetFigFont#1#2#3#4#5{%
  \reset@font\fontsize{#1}{#2pt}%
  \fontfamily{#3}\fontseries{#4}\fontshape{#5}%
  \selectfont}%
\fi\endgroup%
\begin{picture}(9810,2241)(1,-1951)
\put(7651,-1951){\makebox(0,0)[lb]{\smash{\SetFigFont{17}{20.4}{\familydefault}{\mddefault}{\updefault}\special{ps: gsave 0 0 0 setrgbcolor}$\Z^{2d+r}$\special{ps: grestore}}}}
\put(7696,-106){\makebox(0,0)[lb]{\smash{\SetFigFont{17}{20.4}{\rmdefault}{\bfdefault}{\updefault}\special{ps: gsave 0 0 0 setrgbcolor}$\Z^d$\special{ps: grestore}}}}
\put(4816,-106){\makebox(0,0)[lb]{\smash{\SetFigFont{17}{20.4}{\rmdefault}{\bfdefault}{\updefault}\special{ps: gsave 0 0 0 setrgbcolor}$\Z^n$\special{ps: grestore}}}}
\put(1891,-106){\makebox(0,0)[lb]{\smash{\SetFigFont{17}{20.4}{\rmdefault}{\bfdefault}{\updefault}\special{ps: gsave 0 0 0 setrgbcolor}$\Z^r$\special{ps: grestore}}}}
\put(1621,-961){\makebox(0,0)[lb]{\smash{\SetFigFont{17}{20.4}{\rmdefault}{\bfdefault}{\updefault}\special{ps: gsave 0 0 0 setrgbcolor}$id$\special{ps: grestore}}}}
\put(  1,-106){\makebox(0,0)[lb]{\smash{\SetFigFont{17}{20.4}{\rmdefault}{\bfdefault}{\updefault}\special{ps: gsave 0 0 0 setrgbcolor}$0$\special{ps: grestore}}}}
\put(  1,-1951){\makebox(0,0)[lb]{\smash{\SetFigFont{17}{20.4}{\rmdefault}{\bfdefault}{\updefault}\special{ps: gsave 0 0 0 setrgbcolor}$0$\special{ps: grestore}}}}
\put(1846,-1951){\makebox(0,0)[lb]{\smash{\SetFigFont{17}{20.4}{\rmdefault}{\bfdefault}{\updefault}\special{ps: gsave 0 0 0 setrgbcolor}$\Z^r$\special{ps: grestore}}}}
\put(4726,-1951){\makebox(0,0)[lb]{\smash{\SetFigFont{17}{20.4}{\rmdefault}{\bfdefault}{\updefault}\special{ps: gsave 0 0 0 setrgbcolor}$\Z^{2n}$\special{ps: grestore}}}}
\put(4636,-961){\makebox(0,0)[lb]{\smash{\SetFigFont{17}{20.4}{\rmdefault}{\bfdefault}{\updefault}\special{ps: gsave 0 0 0 setrgbcolor}$j$\special{ps: grestore}}}}
\put(7471,-961){\makebox(0,0)[lb]{\smash{\SetFigFont{17}{20.4}{\rmdefault}{\bfdefault}{\updefault}\special{ps: gsave 0 0 0 setrgbcolor}$f$\special{ps: grestore}}}}
\put(9811,-106){\makebox(0,0)[lb]{\smash{\SetFigFont{17}{20.4}{\rmdefault}{\bfdefault}{\updefault}\special{ps: gsave 0 0 0 setrgbcolor}$0$\special{ps: grestore}}}}
\put(9766,-1951){\makebox(0,0)[lb]{\smash{\SetFigFont{17}{20.4}{\rmdefault}{\bfdefault}{\updefault}\special{ps: gsave 0 0 0 setrgbcolor}$0$\special{ps: grestore}}}}
\put(6301, 74){\makebox(0,0)[lb]{\smash{\SetFigFont{17}{20.4}{\rmdefault}{\bfdefault}{\updefault}\special{ps: gsave 0 0 0 setrgbcolor}$ g$\special{ps: grestore}}}}
\put(3106, 74){\makebox(0,0)[lb]{\smash{\SetFigFont{17}{20.4}{\rmdefault}{\bfdefault}{\updefault}\special{ps: gsave 0 0 0 setrgbcolor}$q^*$\special{ps: grestore}}}}
\put(3106,-1726){\makebox(0,0)[lb]{\smash{\SetFigFont{17}{20.4}{\rmdefault}{\bfdefault}{\updefault}\special{ps: gsave 0 0 0 setrgbcolor}${\hat q}^*$\special{ps: grestore}}}}
\put(6346,-1681){\makebox(0,0)[lb]{\smash{\SetFigFont{17}{20.4}{\rmdefault}{\bfdefault}{\updefault}\special{ps: gsave 0 0 0 setrgbcolor}${\hat g}$\special{ps: grestore}}}}
\end{picture}

%% file: moment.pstex_t
\begin{picture}(0,0)%
\includegraphics{moment.pstex}%
\end{picture}%
\setlength{\unitlength}{4144sp}%
\begingroup\makeatletter\ifx\SetFigFont\undefined%
\gdef\SetFigFont#1#2#3#4#5{%
  \reset@font\fontsize{#1}{#2pt}%
  \fontfamily{#3}\fontseries{#4}\fontshape{#5}%
  \selectfont}%
\fi\endgroup%
\begin{picture}(7875,1926)(46,-2347)
\put( 46,-691){\makebox(0,0)[lb]{\smash{\SetFigFont{17}{20.4}{\rmdefault}{\bfdefault}{\updefault}\special{ps: gsave 0 0 0 setrgbcolor}$X$\special{ps: grestore}}}}
\put( 46,-1996){\makebox(0,0)[lb]{\smash{\SetFigFont{17}{20.4}{\rmdefault}{\bfdefault}{\updefault}\special{ps: gsave 0 0 0 setrgbcolor}$0$\special{ps: grestore}}}}
\put(2026,-1996){\makebox(0,0)[lb]{\smash{\SetFigFont{17}{20.4}{\rmdefault}{\bfdefault}{\updefault}\special{ps: gsave 0 0 0 setrgbcolor}${\vec \R}^d$\special{ps: grestore}}}}
\put(5896,-1951){\makebox(0,0)[lb]{\smash{\SetFigFont{17}{20.4}{\rmdefault}{\bfdefault}{\updefault}\special{ps: gsave 0 0 0 setrgbcolor}${\vec \R}^r$\special{ps: grestore}}}}
\put(7921,-1951){\makebox(0,0)[lb]{\smash{\SetFigFont{17}{20.4}{\rmdefault}{\bfdefault}{\updefault}\special{ps: gsave 0 0 0 setrgbcolor}$0$\special{ps: grestore}}}}
\put(4006,-1996){\makebox(0,0)[lb]{\smash{\SetFigFont{17}{20.4}{\rmdefault}{\bfdefault}{\updefault}\special{ps: gsave 0 0 0 setrgbcolor}${\vec \R}^n$\special{ps: grestore}}}}
\put(4006,-691){\makebox(0,0)[lb]{\smash{\SetFigFont{17}{20.4}{\rmdefault}{\bfdefault}{\updefault}\special{ps: gsave 0 0 0 setrgbcolor}$\H^n$\special{ps: grestore}}}}
\put(2026,-691){\makebox(0,0)[lb]{\smash{\SetFigFont{17}{20.4}{\rmdefault}{\bfdefault}{\updefault}\special{ps: gsave 0 0 0 setrgbcolor}${\cal N}$\special{ps: grestore}}}}
\put(2971,-2221){\makebox(0,0)[lb]{\smash{\SetFigFont{17}{20.4}{\rmdefault}{\bfdefault}{\updefault}\special{ps: gsave 0 0 0 setrgbcolor}${\vec g}^*$\special{ps: grestore}}}}
\put(1891,-1276){\makebox(0,0)[lb]{\smash{\SetFigFont{17}{20.4}{\rmdefault}{\bfdefault}{\updefault}\special{ps: gsave 0 0 0 setrgbcolor}$ {\vec \pi}_0$\special{ps: grestore}}}}
\put(3736,-1276){\makebox(0,0)[lb]{\smash{\SetFigFont{17}{20.4}{\rmdefault}{\bfdefault}{\updefault}\special{ps: gsave 0 0 0 setrgbcolor}${\vec \eta}$\special{ps: grestore}}}}
\put(766,-1276){\makebox(0,0)[lb]{\smash{\SetFigFont{17}{20.4}{\rmdefault}{\bfdefault}{\updefault}\special{ps: gsave 0 0 0 setrgbcolor}$ {\vec \pi}$\special{ps: grestore}}}}
\put(5401,-1276){\makebox(0,0)[lb]{\smash{\SetFigFont{17}{20.4}{\rmdefault}{\bfdefault}{\updefault}\special{ps: gsave 0 0 0 setrgbcolor}${\vec \mu}$\special{ps: grestore}}}}
\put(4951,-2266){\makebox(0,0)[lb]{\smash{\SetFigFont{17}{20.4}{\rmdefault}{\bfdefault}{\updefault}\special{ps: gsave 0 0 0 setrgbcolor}${\vec q}$\special{ps: grestore}}}}
\end{picture}

%% file: flats.pstex_t
\begin{picture}(0,0)%
\includegraphics{flats.pstex}%
\end{picture}%
\setlength{\unitlength}{4144sp}%
\begingroup\makeatletter\ifx\SetFigFont\undefined%
\gdef\SetFigFont#1#2#3#4#5{%
  \reset@font\fontsize{#1}{#2pt}%
  \fontfamily{#3}\fontseries{#4}\fontshape{#5}%
  \selectfont}%
\fi\endgroup%
\begin{picture}(3215,4233)(1474,-3517)
\put(1756,-3436){\makebox(0,0)[lb]{\smash{\SetFigFont{17}{20.4}{\rmdefault}{\bfdefault}{\updefault}\special{ps: gsave 0 0 0 setrgbcolor}$h_i$\special{ps: grestore}}}}
\put(4096,-3436){\makebox(0,0)[lb]{\smash{\SetFigFont{17}{20.4}{\rmdefault}{\bfdefault}{\updefault}\special{ps: gsave 0 0 0 setrgbcolor}$h_j$\special{ps: grestore}}}}
\put(1756,-2086){\makebox(0,0)[lb]{\smash{\SetFigFont{17}{20.4}{\rmdefault}{\bfdefault}{\updefault}\special{ps: gsave 0 0 0 setrgbcolor}$X_i$\special{ps: grestore}}}}
\put(3781,-2131){\makebox(0,0)[lb]{\smash{\SetFigFont{17}{20.4}{\rmdefault}{\bfdefault}{\updefault}\special{ps: gsave 0 0 0 setrgbcolor}$X_j$\special{ps: grestore}}}}
\end{picture}

%% file: 3lemma2.pstex_t
\begin{picture}(0,0)%
\includegraphics{3lemma2.pstex}%
\end{picture}%
\setlength{\unitlength}{4144sp}%
\begingroup\makeatletter\ifx\SetFigFont\undefined%
\gdef\SetFigFont#1#2#3#4#5{%
  \reset@font\fontsize{#1}{#2pt}%
  \fontfamily{#3}\fontseries{#4}\fontshape{#5}%
  \selectfont}%
\fi\endgroup%
\begin{picture}(9810,2286)(1,-3751)
\put(1846,-1951){\makebox(0,0)[lb]{\smash{\SetFigFont{17}{20.4}{\rmdefault}{\bfdefault}{\updefault}\special{ps: gsave 0 0 0 setrgbcolor}$\Z^r$\special{ps: grestore}}}}
\put(  1,-1951){\makebox(0,0)[lb]{\smash{\SetFigFont{17}{20.4}{\rmdefault}{\bfdefault}{\updefault}\special{ps: gsave 0 0 0 setrgbcolor}$0$\special{ps: grestore}}}}
\put(  1,-3751){\makebox(0,0)[lb]{\smash{\SetFigFont{17}{20.4}{\rmdefault}{\bfdefault}{\updefault}\special{ps: gsave 0 0 0 setrgbcolor}$0$\special{ps: grestore}}}}
\put(1666,-2806){\makebox(0,0)[lb]{\smash{\SetFigFont{17}{20.4}{\rmdefault}{\bfdefault}{\updefault}\special{ps: gsave 0 0 0 setrgbcolor}$s$\special{ps: grestore}}}}
\put(4591,-2896){\makebox(0,0)[lb]{\smash{\SetFigFont{17}{20.4}{\rmdefault}{\bfdefault}{\updefault}\special{ps: gsave 0 0 0 setrgbcolor}$id$\special{ps: grestore}}}}
\put(4726,-1951){\makebox(0,0)[lb]{\smash{\SetFigFont{17}{20.4}{\rmdefault}{\bfdefault}{\updefault}\special{ps: gsave 0 0 0 setrgbcolor}$\Z^{2n}$\special{ps: grestore}}}}
\put(4681,-3751){\makebox(0,0)[lb]{\smash{\SetFigFont{17}{20.4}{\rmdefault}{\bfdefault}{\updefault}\special{ps: gsave 0 0 0 setrgbcolor}$\Z^{2n}$\special{ps: grestore}}}}
\put(7471,-2851){\makebox(0,0)[lb]{\smash{\SetFigFont{17}{20.4}{\rmdefault}{\bfdefault}{\updefault}\special{ps: gsave 0 0 0 setrgbcolor}$p$\special{ps: grestore}}}}
\put(7651,-1951){\makebox(0,0)[lb]{\smash{\SetFigFont{17}{20.4}{\rmdefault}{\bfdefault}{\updefault}\special{ps: gsave 0 0 0 setrgbcolor}$\Z^{2d+r}$\special{ps: grestore}}}}
\put(9766,-1951){\makebox(0,0)[lb]{\smash{\SetFigFont{17}{20.4}{\rmdefault}{\bfdefault}{\updefault}\special{ps: gsave 0 0 0 setrgbcolor}$0$\special{ps: grestore}}}}
\put(9811,-3751){\makebox(0,0)[lb]{\smash{\SetFigFont{17}{20.4}{\rmdefault}{\bfdefault}{\updefault}\special{ps: gsave 0 0 0 setrgbcolor}$0$\special{ps: grestore}}}}
\put(6346,-1681){\makebox(0,0)[lb]{\smash{\SetFigFont{17}{20.4}{\rmdefault}{\bfdefault}{\updefault}\special{ps: gsave 0 0 0 setrgbcolor}${\hat g}$\special{ps: grestore}}}}
\put(3106,-1726){\makebox(0,0)[lb]{\smash{\SetFigFont{17}{20.4}{\rmdefault}{\bfdefault}{\updefault}\special{ps: gsave 0 0 0 setrgbcolor}${\hat q}^*$\special{ps: grestore}}}}
\put(3151,-3436){\makebox(0,0)[lb]{\smash{\SetFigFont{17}{20.4}{\rmdefault}{\bfdefault}{\updefault}\special{ps: gsave 0 0 0 setrgbcolor}${\tilde q}^*$\special{ps: grestore}}}}
\put(6436,-3436){\makebox(0,0)[lb]{\smash{\SetFigFont{17}{20.4}{\rmdefault}{\bfdefault}{\updefault}\special{ps: gsave 0 0 0 setrgbcolor}${\tilde g}$\special{ps: grestore}}}}
\put(7561,-3751){\makebox(0,0)[lb]{\smash{\SetFigFont{17}{20.4}{\rmdefault}{\bfdefault}{\updefault}\special{ps: gsave 0 0 0 setrgbcolor}$A$\special{ps: grestore}}}}
\put(1801,-3751){\makebox(0,0)[lb]{\smash{\SetFigFont{17}{20.4}{\rmdefault}{\bfdefault}{\updefault}\special{ps: gsave 0 0 0 setrgbcolor}$\Z^{r+1}$\special{ps: grestore}}}}
\end{picture}

%% file: Delta_gen.pstex_t
\begin{picture}(0,0)%
\includegraphics{Delta_gen.pstex}%
\end{picture}%
\setlength{\unitlength}{4144sp}%
\begingroup\makeatletter\ifx\SetFigFont\undefined%
\gdef\SetFigFont#1#2#3#4#5{%
  \reset@font\fontsize{#1}{#2pt}%
  \fontfamily{#3}\fontseries{#4}\fontshape{#5}%
  \selectfont}%
\fi\endgroup%
\begin{picture}(4448,3040)(2010,-3358)
\put(3421,-781){\makebox(0,0)[lb]{\smash{\SetFigFont{17}{20.4}{\rmdefault}{\bfdefault}{\updefault}\special{ps: gsave 0 0 0 setrgbcolor}$D_1$\special{ps: grestore}}}}
\put(4321,-1231){\makebox(0,0)[lb]{\smash{\SetFigFont{17}{20.4}{\rmdefault}{\bfdefault}{\updefault}\special{ps: gsave 0 0 0 setrgbcolor}$D_2$\special{ps: grestore}}}}
\put(5131,-1906){\makebox(0,0)[lb]{\smash{\SetFigFont{17}{20.4}{\rmdefault}{\bfdefault}{\updefault}\special{ps: gsave 0 0 0 setrgbcolor}$D_3$\special{ps: grestore}}}}
\put(5086,-2851){\makebox(0,0)[lb]{\smash{\SetFigFont{17}{20.4}{\rmdefault}{\bfdefault}{\updefault}\special{ps: gsave 0 0 0 setrgbcolor}$D_4$\special{ps: grestore}}}}
\put(5851,-1051){\makebox(0,0)[lb]{\smash{\SetFigFont{17}{20.4}{\rmdefault}{\bfdefault}{\updefault}\special{ps: gsave 0 0 0 setrgbcolor}$p_e$\special{ps: grestore}}}}
\end{picture}

%% file: dist.pstex_t
\begin{picture}(0,0)%
\includegraphics{dist.pstex}%
\end{picture}%
\setlength{\unitlength}{4144sp}%
\begingroup\makeatletter\ifx\SetFigFont\undefined%
\gdef\SetFigFont#1#2#3#4#5{%
  \reset@font\fontsize{#1}{#2pt}%
  \fontfamily{#3}\fontseries{#4}\fontshape{#5}%
  \selectfont}%
\fi\endgroup%
\begin{picture}(4600,3343)(1936,-3487)
\put(5986,-826){\makebox(0,0)[lb]{\smash{\SetFigFont{17}{20.4}{\rmdefault}{\bfdefault}{\updefault}\special{ps: gsave 0 0 0 setrgbcolor}$Y_e$\special{ps: grestore}}}}
\put(5131,-2626){\makebox(0,0)[lb]{\smash{\SetFigFont{17}{20.4}{\rmdefault}{\bfdefault}{\updefault}\special{ps: gsave 0 0 0 setrgbcolor}$Y_j$\special{ps: grestore}}}}
\end{picture}

%% file: vsphere.pstex_t
\begin{picture}(0,0)%
\includegraphics{vsphere.pstex}%
\end{picture}%
\setlength{\unitlength}{4144sp}%
\begingroup\makeatletter\ifx\SetFigFont\undefined%
\gdef\SetFigFont#1#2#3#4#5{%
  \reset@font\fontsize{#1}{#2pt}%
  \fontfamily{#3}\fontseries{#4}\fontshape{#5}%
  \selectfont}%
\fi\endgroup%
\begin{picture}(4454,4461)(1779,-3301)
\put(1891,-3301){\makebox(0,0)[lb]{\smash{\SetFigFont{17}{20.4}{\rmdefault}{\bfdefault}{\updefault}\special{ps: gsave 0 0 0 setrgbcolor}$v_2$\special{ps: grestore}}}}
\put(6031,-2311){\makebox(0,0)[lb]{\smash{\SetFigFont{17}{20.4}{\rmdefault}{\bfdefault}{\updefault}\special{ps: gsave 0 0 0 setrgbcolor}$v_3$\special{ps: grestore}}}}
\put(3601,479){\makebox(0,0)[lb]{\smash{\SetFigFont{17}{20.4}{\rmdefault}{\bfdefault}{\updefault}\special{ps: gsave 0 0 0 setrgbcolor}$v_1$\special{ps: grestore}}}}
\end{picture}

%% file: DM.pstex_t
\begin{picture}(0,0)%
\includegraphics{DM.pstex}%
\end{picture}%
\setlength{\unitlength}{4144sp}%
\begingroup\makeatletter\ifx\SetFigFont\undefined%
\gdef\SetFigFont#1#2#3#4#5{%
  \reset@font\fontsize{#1}{#2pt}%
  \fontfamily{#3}\fontseries{#4}\fontshape{#5}%
  \selectfont}%
\fi\endgroup%
\begin{picture}(5379,2308)(1339,-2536)
\put(2161,-2536){\makebox(0,0)[lb]{\smash{\SetFigFont{17}{20.4}{\rmdefault}{\bfdefault}{\updefault}\special{ps: gsave 0 0 0 setrgbcolor}$\Delta$\special{ps: grestore}}}}
\put(5716,-2491){\makebox(0,0)[lb]{\smash{\SetFigFont{17}{20.4}{\rmdefault}{\bfdefault}{\updefault}\special{ps: gsave 0 0 0 setrgbcolor}$\Delta_M$\special{ps: grestore}}}}
\put(3961,-1501){\makebox(0,0)[lb]{\smash{\SetFigFont{17}{20.4}{\rmdefault}{\bfdefault}{\updefault}\special{ps: gsave 0 0 0 setrgbcolor}$/\iota$\special{ps: grestore}}}}
\end{picture}

%% file: vertex.pstex_t
\begin{picture}(0,0)%
\includegraphics{vertex.pstex}%
\end{picture}%
\setlength{\unitlength}{4144sp}%
\begingroup\makeatletter\ifx\SetFigFont\undefined%
\gdef\SetFigFont#1#2#3#4#5{%
  \reset@font\fontsize{#1}{#2pt}%
  \fontfamily{#3}\fontseries{#4}\fontshape{#5}%
  \selectfont}%
\fi\endgroup%
\begin{picture}(3759,3852)(1564,-4168)
\put(4366,-2851){\makebox(0,0)[lb]{\smash{\SetFigFont{17}{20.4}{\rmdefault}{\bfdefault}{\updefault}\special{ps: gsave 0 0 0 setrgbcolor}$D_j$\special{ps: grestore}}}}
\put(1711,-1411){\makebox(0,0)[lb]{\smash{\SetFigFont{17}{20.4}{\rmdefault}{\bfdefault}{\updefault}\special{ps: gsave 0 0 0 setrgbcolor}$e$\special{ps: grestore}}}}
\put(3916,-601){\makebox(0,0)[lb]{\smash{\SetFigFont{17}{20.4}{\rmdefault}{\bfdefault}{\updefault}\special{ps: gsave 0 0 0 setrgbcolor}$e'$\special{ps: grestore}}}}
\put(2521,-556){\makebox(0,0)[lb]{\smash{\SetFigFont{17}{20.4}{\rmdefault}{\bfdefault}{\updefault}\special{ps: gsave 0 0 0 setrgbcolor}$A$\special{ps: grestore}}}}
\end{picture}

%% file: proj.pstex_t
\begin{picture}(0,0)%
\includegraphics{proj.pstex}%
\end{picture}%
\setlength{\unitlength}{4144sp}%
\begingroup\makeatletter\ifx\SetFigFont\undefined%
\gdef\SetFigFont#1#2#3#4#5{%
  \reset@font\fontsize{#1}{#2pt}%
  \fontfamily{#3}\fontseries{#4}\fontshape{#5}%
  \selectfont}%
\fi\endgroup%
\begin{picture}(10047,6087)(394,-6001)
\put(5941,-2131){\makebox(0,0)[lb]{\smash{\SetFigFont{17}{20.4}{\rmdefault}{\bfdefault}{\updefault}\special{ps: gsave 0 0 0 setrgbcolor}$j_V$\special{ps: grestore}}}}
\put(6886,-1231){\makebox(0,0)[lb]{\smash{\SetFigFont{17}{20.4}{\rmdefault}{\bfdefault}{\updefault}\special{ps: gsave 0 0 0 setrgbcolor}$id$\special{ps: grestore}}}}
\put(8596,-3796){\makebox(0,0)[lb]{\smash{\SetFigFont{17}{20.4}{\rmdefault}{\bfdefault}{\updefault}\special{ps: gsave 0 0 0 setrgbcolor}$\alpha$\special{ps: grestore}}}}
\put(5446,-2941){\makebox(0,0)[lb]{\smash{\SetFigFont{17}{20.4}{\rmdefault}{\bfdefault}{\updefault}\special{ps: gsave 0 0 0 setrgbcolor}$\Z^n$\special{ps: grestore}}}}
\put(5356,-4786){\makebox(0,0)[lb]{\smash{\SetFigFont{17}{20.4}{\rmdefault}{\bfdefault}{\updefault}\special{ps: gsave 0 0 0 setrgbcolor}$\Z^N$\special{ps: grestore}}}}
\put(2476,-4786){\makebox(0,0)[lb]{\smash{\SetFigFont{17}{20.4}{\rmdefault}{\bfdefault}{\updefault}\special{ps: gsave 0 0 0 setrgbcolor}$\Z^r$\special{ps: grestore}}}}
\put(631,-4786){\makebox(0,0)[lb]{\smash{\SetFigFont{17}{20.4}{\rmdefault}{\bfdefault}{\updefault}\special{ps: gsave 0 0 0 setrgbcolor}$0$\special{ps: grestore}}}}
\put(631,-2941){\makebox(0,0)[lb]{\smash{\SetFigFont{17}{20.4}{\rmdefault}{\bfdefault}{\updefault}\special{ps: gsave 0 0 0 setrgbcolor}$0$\special{ps: grestore}}}}
\put(2521,-2941){\makebox(0,0)[lb]{\smash{\SetFigFont{17}{20.4}{\rmdefault}{\bfdefault}{\updefault}\special{ps: gsave 0 0 0 setrgbcolor}$\Z^r$\special{ps: grestore}}}}
\put(8281,-4786){\makebox(0,0)[lb]{\smash{\SetFigFont{17}{20.4}{\rmdefault}{\bfdefault}{\updefault}\special{ps: gsave 0 0 0 setrgbcolor}$A$\special{ps: grestore}}}}
\put(10396,-4786){\makebox(0,0)[lb]{\smash{\SetFigFont{17}{20.4}{\rmdefault}{\bfdefault}{\updefault}\special{ps: gsave 0 0 0 setrgbcolor}$0$\special{ps: grestore}}}}
\put(10441,-2941){\makebox(0,0)[lb]{\smash{\SetFigFont{17}{20.4}{\rmdefault}{\bfdefault}{\updefault}\special{ps: gsave 0 0 0 setrgbcolor}$0$\special{ps: grestore}}}}
\put(8326,-2941){\makebox(0,0)[lb]{\smash{\SetFigFont{17}{20.4}{\rmdefault}{\bfdefault}{\updefault}\special{ps: gsave 0 0 0 setrgbcolor}$\Z^d$\special{ps: grestore}}}}
\put(2926,-3796){\makebox(0,0)[lb]{\smash{\SetFigFont{17}{20.4}{\rmdefault}{\bfdefault}{\updefault}\special{ps: gsave 0 0 0 setrgbcolor}$id$\special{ps: grestore}}}}
\put(5401,-1501){\makebox(0,0)[lb]{\smash{\SetFigFont{17}{20.4}{\rmdefault}{\bfdefault}{\updefault}\special{ps: gsave 0 0 0 setrgbcolor}$\Z^V$\special{ps: grestore}}}}
\put(3736,-2761){\makebox(0,0)[lb]{\smash{\SetFigFont{17}{20.4}{\rmdefault}{\bfdefault}{\updefault}\special{ps: gsave 0 0 0 setrgbcolor}$f$\special{ps: grestore}}}}
\put(3736,-4561){\makebox(0,0)[lb]{\smash{\SetFigFont{17}{20.4}{\rmdefault}{\bfdefault}{\updefault}\special{ps: gsave 0 0 0 setrgbcolor}$f_N$\special{ps: grestore}}}}
\put(6931,-2761){\makebox(0,0)[lb]{\smash{\SetFigFont{17}{20.4}{\rmdefault}{\bfdefault}{\updefault}\special{ps: gsave 0 0 0 setrgbcolor}$g$\special{ps: grestore}}}}
\put(8821,-2131){\makebox(0,0)[lb]{\smash{\SetFigFont{17}{20.4}{\rmdefault}{\bfdefault}{\updefault}\special{ps: gsave 0 0 0 setrgbcolor}$g_V$\special{ps: grestore}}}}
\put(6976,-4516){\makebox(0,0)[lb]{\smash{\SetFigFont{17}{20.4}{\rmdefault}{\bfdefault}{\updefault}\special{ps: gsave 0 0 0 setrgbcolor}$\beta$\special{ps: grestore}}}}
\put(5896,-3751){\makebox(0,0)[lb]{\smash{\SetFigFont{17}{20.4}{\rmdefault}{\bfdefault}{\updefault}\special{ps: gsave 0 0 0 setrgbcolor}$p_N$\special{ps: grestore}}}}
\put(8326,-1501){\makebox(0,0)[lb]{\smash{\SetFigFont{17}{20.4}{\rmdefault}{\bfdefault}{\updefault}\special{ps: gsave 0 0 0 setrgbcolor}$\Z^V$\special{ps: grestore}}}}
\put(5581,-376){\makebox(0,0)[lb]{\smash{\SetFigFont{17}{20.4}{\rmdefault}{\bfdefault}{\updefault}\special{ps: gsave 0 0 0 setrgbcolor}$0$\special{ps: grestore}}}}
\put(5626,-5956){\makebox(0,0)[lb]{\smash{\SetFigFont{17}{20.4}{\rmdefault}{\bfdefault}{\updefault}\special{ps: gsave 0 0 0 setrgbcolor}$0$\special{ps: grestore}}}}
\put(8371,-6001){\makebox(0,0)[lb]{\smash{\SetFigFont{17}{20.4}{\rmdefault}{\bfdefault}{\updefault}\special{ps: gsave 0 0 0 setrgbcolor}$0$\special{ps: grestore}}}}
\put(8461,-376){\makebox(0,0)[lb]{\smash{\SetFigFont{17}{20.4}{\rmdefault}{\bfdefault}{\updefault}\special{ps: gsave 0 0 0 setrgbcolor}$0$\special{ps: grestore}}}}
\end{picture}

%% file: proj_tor.pstex_t
\begin{picture}(0,0)%
\includegraphics{proj_tor.pstex}%
\end{picture}%
\setlength{\unitlength}{4144sp}%
\begingroup\makeatletter\ifx\SetFigFont\undefined%
\gdef\SetFigFont#1#2#3#4#5{%
  \reset@font\fontsize{#1}{#2pt}%
  \fontfamily{#3}\fontseries{#4}\fontshape{#5}%
  \selectfont}%
\fi\endgroup%
\begin{picture}(10260,6765)(181,-5956)
\put(5941,-2131){\makebox(0,0)[lb]{\smash{\SetFigFont{17}{20.4}{\rmdefault}{\bfdefault}{\updefault}\special{ps: gsave 0 0 0 setrgbcolor}$j_V$\special{ps: grestore}}}}
\put(6886,-1231){\makebox(0,0)[lb]{\smash{\SetFigFont{17}{20.4}{\rmdefault}{\bfdefault}{\updefault}\special{ps: gsave 0 0 0 setrgbcolor}$id$\special{ps: grestore}}}}
\put(8596,-3796){\makebox(0,0)[lb]{\smash{\SetFigFont{17}{20.4}{\rmdefault}{\bfdefault}{\updefault}\special{ps: gsave 0 0 0 setrgbcolor}$\alpha$\special{ps: grestore}}}}
\put(5266,-3796){\makebox(0,0)[lb]{\smash{\SetFigFont{17}{20.4}{\rmdefault}{\bfdefault}{\updefault}\special{ps: gsave 0 0 0 setrgbcolor}$p_N$\special{ps: grestore}}}}
\put(2926,-3796){\makebox(0,0)[lb]{\smash{\SetFigFont{17}{20.4}{\rmdefault}{\bfdefault}{\updefault}\special{ps: gsave 0 0 0 setrgbcolor}$id$\special{ps: grestore}}}}
\put(3736,-2761){\makebox(0,0)[lb]{\smash{\SetFigFont{17}{20.4}{\rmdefault}{\bfdefault}{\updefault}\special{ps: gsave 0 0 0 setrgbcolor}$f$\special{ps: grestore}}}}
\put(3736,-4561){\makebox(0,0)[lb]{\smash{\SetFigFont{17}{20.4}{\rmdefault}{\bfdefault}{\updefault}\special{ps: gsave 0 0 0 setrgbcolor}$f_N$\special{ps: grestore}}}}
\put(6931,-2761){\makebox(0,0)[lb]{\smash{\SetFigFont{17}{20.4}{\rmdefault}{\bfdefault}{\updefault}\special{ps: gsave 0 0 0 setrgbcolor}$g$\special{ps: grestore}}}}
\put(8821,-2131){\makebox(0,0)[lb]{\smash{\SetFigFont{17}{20.4}{\rmdefault}{\bfdefault}{\updefault}\special{ps: gsave 0 0 0 setrgbcolor}$g_V$\special{ps: grestore}}}}
\put(946,-2896){\makebox(0,0)[lb]{\smash{\SetFigFont{17}{20.4}{\rmdefault}{\bfdefault}{\updefault}\special{ps: gsave 0 0 0 setrgbcolor}$1$\special{ps: grestore}}}}
\put(8371,569){\makebox(0,0)[lb]{\smash{\SetFigFont{17}{20.4}{\rmdefault}{\bfdefault}{\updefault}\special{ps: gsave 0 0 0 setrgbcolor}$1$\special{ps: grestore}}}}
\put(5536,-376){\makebox(0,0)[lb]{\smash{\SetFigFont{17}{20.4}{\rmdefault}{\bfdefault}{\updefault}\special{ps: gsave 0 0 0 setrgbcolor}$1$\special{ps: grestore}}}}
\put(10441,-2941){\makebox(0,0)[lb]{\smash{\SetFigFont{17}{20.4}{\rmdefault}{\bfdefault}{\updefault}\special{ps: gsave 0 0 0 setrgbcolor}$1$\special{ps: grestore}}}}
\put(10396,-4786){\makebox(0,0)[lb]{\smash{\SetFigFont{17}{20.4}{\rmdefault}{\bfdefault}{\updefault}\special{ps: gsave 0 0 0 setrgbcolor}$1$\special{ps: grestore}}}}
\put(5536,-5956){\makebox(0,0)[lb]{\smash{\SetFigFont{17}{20.4}{\rmdefault}{\bfdefault}{\updefault}\special{ps: gsave 0 0 0 setrgbcolor}$1$\special{ps: grestore}}}}
\put(8326,-5956){\makebox(0,0)[lb]{\smash{\SetFigFont{17}{20.4}{\rmdefault}{\bfdefault}{\updefault}\special{ps: gsave 0 0 0 setrgbcolor}$1$\special{ps: grestore}}}}
\put(8281,-4786){\makebox(0,0)[lb]{\smash{\SetFigFont{17}{20.4}{\rmdefault}{\bfdefault}{\updefault}\special{ps: gsave 0 0 0 setrgbcolor}$A\otimes_\Z U(1)$\special{ps: grestore}}}}
\put(8326,-2941){\makebox(0,0)[lb]{\smash{\SetFigFont{17}{20.4}{\rmdefault}{\bfdefault}{\updefault}\special{ps: gsave 0 0 0 setrgbcolor}$T^d$\special{ps: grestore}}}}
\put(8101,-1456){\makebox(0,0)[lb]{\smash{\SetFigFont{17}{20.4}{\rmdefault}{\bfdefault}{\updefault}\special{ps: gsave 0 0 0 setrgbcolor}$T^V$\special{ps: grestore}}}}
\put(5401,-1501){\makebox(0,0)[lb]{\smash{\SetFigFont{17}{20.4}{\rmdefault}{\bfdefault}{\updefault}\special{ps: gsave 0 0 0 setrgbcolor}$T^V$\special{ps: grestore}}}}
\put(5446,-2941){\makebox(0,0)[lb]{\smash{\SetFigFont{17}{20.4}{\rmdefault}{\bfdefault}{\updefault}\special{ps: gsave 0 0 0 setrgbcolor}$T^n$\special{ps: grestore}}}}
\put(5356,-4786){\makebox(0,0)[lb]{\smash{\SetFigFont{17}{20.4}{\rmdefault}{\bfdefault}{\updefault}\special{ps: gsave 0 0 0 setrgbcolor}$T^N$\special{ps: grestore}}}}
\put(2476,-4786){\makebox(0,0)[lb]{\smash{\SetFigFont{17}{20.4}{\rmdefault}{\bfdefault}{\updefault}\special{ps: gsave 0 0 0 setrgbcolor}$T^r$\special{ps: grestore}}}}
\put(2521,-2941){\makebox(0,0)[lb]{\smash{\SetFigFont{17}{20.4}{\rmdefault}{\bfdefault}{\updefault}\special{ps: gsave 0 0 0 setrgbcolor}$T^r$\special{ps: grestore}}}}
\put(6976,-4516){\makebox(0,0)[lb]{\smash{\SetFigFont{17}{20.4}{\rmdefault}{\bfdefault}{\updefault}\special{ps: gsave 0 0 0 setrgbcolor}$\beta$\special{ps: grestore}}}}
\put(181,-4786){\makebox(0,0)[lb]{\smash{\SetFigFont{17}{20.4}{\rmdefault}{\bfdefault}{\updefault}\special{ps: gsave 0 0 0 setrgbcolor}$1$\special{ps: grestore}}}}
\put(946,-4786){\makebox(0,0)[lb]{\smash{\SetFigFont{17}{20.4}{\rmdefault}{\bfdefault}{\updefault}\special{ps: gsave 0 0 0 setrgbcolor}$\Gamma$\special{ps: grestore}}}}
\put(8416,-421){\makebox(0,0)[lb]{\smash{\SetFigFont{17}{20.4}{\rmdefault}{\bfdefault}{\updefault}\special{ps: gsave 0 0 0 setrgbcolor}$\Gamma'$\special{ps: grestore}}}}
\end{picture}

%% file: doublesequence.pstex_t
\begin{picture}(0,0)%
\includegraphics{doublesequence.pstex}%
\end{picture}%
\setlength{\unitlength}{4144sp}%
\begingroup\makeatletter\ifx\SetFigFont\undefined%
\gdef\SetFigFont#1#2#3#4#5{%
  \reset@font\fontsize{#1}{#2pt}%
  \fontfamily{#3}\fontseries{#4}\fontshape{#5}%
  \selectfont}%
\fi\endgroup%
\begin{picture}(9810,5820)(631,-5956)
\put(3511,-1906){\makebox(0,0)[lb]{\smash{\SetFigFont{17}{20.4}{\rmdefault}{\bfdefault}{\updefault}\special{ps: gsave 0 0 0 setrgbcolor}$p_1$\special{ps: grestore}}}}
\put(5536,-5956){\makebox(0,0)[lb]{\smash{\SetFigFont{17}{20.4}{\rmdefault}{\bfdefault}{\updefault}\special{ps: gsave 0 0 0 setrgbcolor}$0$\special{ps: grestore}}}}
\put(4276,-4021){\makebox(0,0)[lb]{\smash{\SetFigFont{17}{20.4}{\rmdefault}{\bfdefault}{\updefault}\special{ps: gsave 0 0 0 setrgbcolor}$\gamma$\special{ps: grestore}}}}
\put(5401,-1501){\makebox(0,0)[lb]{\smash{\SetFigFont{17}{20.4}{\rmdefault}{\bfdefault}{\updefault}\special{ps: gsave 0 0 0 setrgbcolor}$\Z^{n-2}$\special{ps: grestore}}}}
\put(8101,-3796){\makebox(0,0)[lb]{\smash{\SetFigFont{17}{20.4}{\rmdefault}{\bfdefault}{\updefault}\special{ps: gsave 0 0 0 setrgbcolor}$\Phi$\special{ps: grestore}}}}
\put(6976,-4516){\makebox(0,0)[lb]{\smash{\SetFigFont{17}{20.4}{\rmdefault}{\bfdefault}{\updefault}\special{ps: gsave 0 0 0 setrgbcolor}$\beta$\special{ps: grestore}}}}
\put(10396,-4786){\makebox(0,0)[lb]{\smash{\SetFigFont{17}{20.4}{\rmdefault}{\bfdefault}{\updefault}\special{ps: gsave 0 0 0 setrgbcolor}$0$\special{ps: grestore}}}}
\put(10441,-2941){\makebox(0,0)[lb]{\smash{\SetFigFont{17}{20.4}{\rmdefault}{\bfdefault}{\updefault}\special{ps: gsave 0 0 0 setrgbcolor}$0$\special{ps: grestore}}}}
\put(5446,-2941){\makebox(0,0)[lb]{\smash{\SetFigFont{17}{20.4}{\rmdefault}{\bfdefault}{\updefault}\special{ps: gsave 0 0 0 setrgbcolor}$\Z^n$\special{ps: grestore}}}}
\put(2521,-2941){\makebox(0,0)[lb]{\smash{\SetFigFont{17}{20.4}{\rmdefault}{\bfdefault}{\updefault}\special{ps: gsave 0 0 0 setrgbcolor}$\Z^{n-1}$\special{ps: grestore}}}}
\put(2926,-3796){\makebox(0,0)[lb]{\smash{\SetFigFont{17}{20.4}{\rmdefault}{\bfdefault}{\updefault}\special{ps: gsave 0 0 0 setrgbcolor}$p_2$\special{ps: grestore}}}}
\put(2116,-3796){\makebox(0,0)[lb]{\smash{\SetFigFont{17}{20.4}{\rmdefault}{\bfdefault}{\updefault}\special{ps: gsave 0 0 0 setrgbcolor}$j_2$\special{ps: grestore}}}}
\put(631,-2941){\makebox(0,0)[lb]{\smash{\SetFigFont{17}{20.4}{\rmdefault}{\bfdefault}{\updefault}\special{ps: gsave 0 0 0 setrgbcolor}$0$\special{ps: grestore}}}}
\put(631,-4786){\makebox(0,0)[lb]{\smash{\SetFigFont{17}{20.4}{\rmdefault}{\bfdefault}{\updefault}\special{ps: gsave 0 0 0 setrgbcolor}$0$\special{ps: grestore}}}}
\put(2476,-4786){\makebox(0,0)[lb]{\smash{\SetFigFont{17}{20.4}{\rmdefault}{\bfdefault}{\updefault}\special{ps: gsave 0 0 0 setrgbcolor}$\Z$\special{ps: grestore}}}}
\put(3736,-4561){\makebox(0,0)[lb]{\smash{\SetFigFont{17}{20.4}{\rmdefault}{\bfdefault}{\updefault}\special{ps: gsave 0 0 0 setrgbcolor}$\alpha$\special{ps: grestore}}}}
\put(5356,-4786){\makebox(0,0)[lb]{\smash{\SetFigFont{17}{20.4}{\rmdefault}{\bfdefault}{\updefault}\special{ps: gsave 0 0 0 setrgbcolor}$\Z^2$\special{ps: grestore}}}}
\put(5536,-376){\makebox(0,0)[lb]{\smash{\SetFigFont{17}{20.4}{\rmdefault}{\bfdefault}{\updefault}\special{ps: gsave 0 0 0 setrgbcolor}$0$\special{ps: grestore}}}}
\put(3736,-2761){\makebox(0,0)[lb]{\smash{\SetFigFont{17}{20.4}{\rmdefault}{\bfdefault}{\updefault}\special{ps: gsave 0 0 0 setrgbcolor}${\bar q}^*_e$\special{ps: grestore}}}}
\put(6931,-2761){\makebox(0,0)[lb]{\smash{\SetFigFont{17}{20.4}{\rmdefault}{\bfdefault}{\updefault}\special{ps: gsave 0 0 0 setrgbcolor}${\bar g}_e$\special{ps: grestore}}}}
\put(5941,-2131){\makebox(0,0)[lb]{\smash{\SetFigFont{17}{20.4}{\rmdefault}{\bfdefault}{\updefault}\special{ps: gsave 0 0 0 setrgbcolor}$q^*_e$\special{ps: grestore}}}}
\put(8326,-2941){\makebox(0,0)[lb]{\smash{\SetFigFont{17}{20.4}{\rmdefault}{\bfdefault}{\updefault}\special{ps: gsave 0 0 0 setrgbcolor}$A_e$\special{ps: grestore}}}}
\put(8281,-4786){\makebox(0,0)[lb]{\smash{\SetFigFont{17}{20.4}{\rmdefault}{\bfdefault}{\updefault}\special{ps: gsave 0 0 0 setrgbcolor}$A'_e$\special{ps: grestore}}}}
\put(6031,-3751){\makebox(0,0)[lb]{\smash{\SetFigFont{17}{20.4}{\rmdefault}{\bfdefault}{\updefault}\special{ps: gsave 0 0 0 setrgbcolor}$ g_e$\special{ps: grestore}}}}
\end{picture}

%% file: tor.pstex_t
\begin{picture}(0,0)%
\includegraphics{tor.pstex}%
\end{picture}%
\setlength{\unitlength}{4144sp}%
\begingroup\makeatletter\ifx\SetFigFont\undefined%
\gdef\SetFigFont#1#2#3#4#5{%
  \reset@font\fontsize{#1}{#2pt}%
  \fontfamily{#3}\fontseries{#4}\fontshape{#5}%
  \selectfont}%
\fi\endgroup%
\begin{picture}(11115,2241)(46,-4786)
\put(1396,-3841){\makebox(0,0)[lb]{\smash{\SetFigFont{17}{20.4}{\rmdefault}{\bfdefault}{\updefault}\special{ps: gsave 0 0 0 setrgbcolor}$\Phi_*$\special{ps: grestore}}}}
\put(1621,-2941){\makebox(0,0)[lb]{\smash{\SetFigFont{17}{20.4}{\rmdefault}{\bfdefault}{\updefault}\special{ps: gsave 0 0 0 setrgbcolor}$\Gamma_e$\special{ps: grestore}}}}
\put(1576,-4786){\makebox(0,0)[lb]{\smash{\SetFigFont{17}{20.4}{\rmdefault}{\bfdefault}{\updefault}\special{ps: gsave 0 0 0 setrgbcolor}$\Gamma'_e$\special{ps: grestore}}}}
\put(4456,-4561){\makebox(0,0)[lb]{\smash{\SetFigFont{17}{20.4}{\rmdefault}{\bfdefault}{\updefault}\special{ps: gsave 0 0 0 setrgbcolor}$\alpha$\special{ps: grestore}}}}
\put(7696,-4516){\makebox(0,0)[lb]{\smash{\SetFigFont{17}{20.4}{\rmdefault}{\bfdefault}{\updefault}\special{ps: gsave 0 0 0 setrgbcolor}$\beta$\special{ps: grestore}}}}
\put(8821,-3796){\makebox(0,0)[lb]{\smash{\SetFigFont{17}{20.4}{\rmdefault}{\bfdefault}{\updefault}\special{ps: gsave 0 0 0 setrgbcolor}$\Phi$\special{ps: grestore}}}}
\put(2971,-3796){\makebox(0,0)[lb]{\smash{\SetFigFont{17}{20.4}{\rmdefault}{\bfdefault}{\updefault}\special{ps: gsave 0 0 0 setrgbcolor}$p_2$\special{ps: grestore}}}}
\put(4456,-2761){\makebox(0,0)[lb]{\smash{\SetFigFont{17}{20.4}{\rmdefault}{\bfdefault}{\updefault}\special{ps: gsave 0 0 0 setrgbcolor}${\bar q}^*_e$\special{ps: grestore}}}}
\put(7651,-2761){\makebox(0,0)[lb]{\smash{\SetFigFont{17}{20.4}{\rmdefault}{\bfdefault}{\updefault}\special{ps: gsave 0 0 0 setrgbcolor}${\bar g}_e$\special{ps: grestore}}}}
\put(6076,-4786){\makebox(0,0)[lb]{\smash{\SetFigFont{17}{20.4}{\rmdefault}{\bfdefault}{\updefault}\special{ps: gsave 0 0 0 setrgbcolor}$T^2$\special{ps: grestore}}}}
\put(6166,-2941){\makebox(0,0)[lb]{\smash{\SetFigFont{17}{20.4}{\rmdefault}{\bfdefault}{\updefault}\special{ps: gsave 0 0 0 setrgbcolor}$T^n$\special{ps: grestore}}}}
\put(3196,-4786){\makebox(0,0)[lb]{\smash{\SetFigFont{17}{20.4}{\rmdefault}{\bfdefault}{\updefault}\special{ps: gsave 0 0 0 setrgbcolor}$T^1$\special{ps: grestore}}}}
\put( 46,-2941){\makebox(0,0)[lb]{\smash{\SetFigFont{17}{20.4}{\rmdefault}{\bfdefault}{\updefault}\special{ps: gsave 0 0 0 setrgbcolor}$1$\special{ps: grestore}}}}
\put( 46,-4786){\makebox(0,0)[lb]{\smash{\SetFigFont{17}{20.4}{\rmdefault}{\bfdefault}{\updefault}\special{ps: gsave 0 0 0 setrgbcolor}$1$\special{ps: grestore}}}}
\put(11161,-2941){\makebox(0,0)[lb]{\smash{\SetFigFont{17}{20.4}{\rmdefault}{\bfdefault}{\updefault}\special{ps: gsave 0 0 0 setrgbcolor}$1$\special{ps: grestore}}}}
\put(11116,-4786){\makebox(0,0)[lb]{\smash{\SetFigFont{17}{20.4}{\rmdefault}{\bfdefault}{\updefault}\special{ps: gsave 0 0 0 setrgbcolor}$1$\special{ps: grestore}}}}
\put(9001,-4786){\makebox(0,0)[lb]{\smash{\SetFigFont{17}{20.4}{\rmdefault}{\bfdefault}{\updefault}\special{ps: gsave 0 0 0 setrgbcolor}$A'_e\otimes_\Z U(1)$\special{ps: grestore}}}}
\put(9091,-2896){\makebox(0,0)[lb]{\smash{\SetFigFont{17}{20.4}{\rmdefault}{\bfdefault}{\updefault}\special{ps: gsave 0 0 0 setrgbcolor}$A_e\otimes_\Z U(1)$\special{ps: grestore}}}}
\put(3286,-2941){\makebox(0,0)[lb]{\smash{\SetFigFont{17}{20.4}{\rmdefault}{\bfdefault}{\updefault}\special{ps: gsave 0 0 0 setrgbcolor}$T^{n-1}$\special{ps: grestore}}}}
\put(6031,-3796){\makebox(0,0)[lb]{\smash{\SetFigFont{17}{20.4}{\rmdefault}{\bfdefault}{\updefault}\special{ps: gsave 0 0 0 setrgbcolor}$ g_e$\special{ps: grestore}}}}
\end{picture}